\documentclass[11pt,a4paper]{article}
\pdfoutput=1
\usepackage{amssymb,amsmath,amsfonts, mathtools, mathrsfs}
\usepackage[utf8]{inputenc} 
\usepackage{relsize}
\usepackage[dvipsnames]{xcolor}
\usepackage{stmaryrd}
\usepackage{comment}
\usepackage{soul}
\SetSymbolFont{stmry}{bold}{U}{stmry}{m}{n}
\usepackage[noblocks]{authblk}

\usepackage{graphicx}
\usepackage{caption}
\usepackage{tensor}
\usepackage{subfigure}
\usepackage{enumerate}
\usepackage{dsfont}
\usepackage{verbatim}
\usepackage{amsfonts,euscript,amssymb,stmaryrd,braket}
\usepackage{tikz}
\usetikzlibrary{arrows,decorations.markings,patterns}

\usepackage{slashed}

\def\clock{{\count0=\time
		\divide\count0 60
		\ifnum\count0<10 0\fi\the\count0
		\multiply\count0 -60 \advance\count0 \time
		:\ifnum\count0<10 0\fi \the\count0
}}
\newcommand{\timestamp}{{\small\vbox{\hbox{\tt\jobname.tex}
			\hbox{\the\day/\the\month/\the\year, \clock}}}}

\newcommand{\eps}{\varepsilon}

\newcommand{\nn}{\nonumber}
\newcommand{\bea}{\begin{eqnarray}}
\newcommand{\eea}{\end{eqnarray}}

\DeclareMathOperator{\tr}{\textrm{Tr}}

\newcommand{\be}{\begin{equation}}
\newcommand{\ee}{\end{equation}}

\def\ri {{\rm i}}
\makeatletter
\let\old@startsection=\@startsection
\let\oldl@section=\l@section
\renewcommand{\@startsection}[6]{\old@startsection{#1}{#2}{#3}{#4}{#5}{#6\mathversion{bold}}}
\renewcommand{\l@section}[2]{\oldl@section{\mathversion{bold}#1}{#2}}
\makeatother

\makeatother

\numberwithin{equation}{section}

\usepackage{color}

\def\be{\begin{equation}}
\def\ee{\end{equation}}
\def\ba{\begin{eqnarray}}
\def\ea{\end{eqnarray}}

\definecolor{princetonorange}{rgb}{1.0, 0.56, 0.0}
\definecolor{WildStrawberry}{rgb}{1.0, 0.26, 0.64}
\definecolor{rossocorsa}{rgb}{0.83, 0.0, 0.0}
\definecolor{navyblue}{rgb}{0.0, 0.0, 0.5}

\newif\ifnatbibsort\natbibsorttrue

\relax

\ifnatbibsort\RequirePackage[numbers,sort&compress]{natbib}\else\RequirePackage[numbers,compress]{natbib}\fi
\RequirePackage[colorlinks=true
,urlcolor=blue
,anchorcolor=blue
,citecolor=blue
,filecolor=blue
,linkcolor=blue
,menucolor=blue
,pagecolor=blue
,linktocpage=true
,pdfproducer=medialab
,pdfa=true
]{hyperref}

\begin{document}
\graphicspath{ {./Figures/} }

\rule{0pt}{1cm}

\begin{center}

	{\Large\bf \mathversion{bold}
	Entanglement negativity for a free scalar chiral current
	}

	\vskip  0.8cm
	{
	Malen Arias$^{\,a,}$\footnote{e-mail: malen.arias@ib.edu.ar},
	Marina Huerta$^{\,a,}$\footnote{e-mail: marina.huerta@ib.edu.ar},
    Andrei Rotaru$^{\,b,c,}$\footnote{e-mail: arotaru@sissa.it}      
	and Erik Tonni$^{\,b,c,}$\footnote{e-mail: etonni@sissa.it}
	}
	\vskip  .7cm

	\small
	{\em
		$^{a}\,$Centro At\'omico Bariloche, 8400 -S.C. de Bariloche, 
        R\'{\i}o Negro, Argentina 
		\vskip 0.07cm
		$^{b}\,$SISSA, via Bonomea 265, 34136 
        Trieste, Italy 
        \vskip 0.07cm
        $^{c}$\,INFN, Sezione di Trieste, via Valerio 2, 34127 
        Trieste, Italy
	}
	\normalsize

\end{center}

\vspace{0.3cm}
\begin{abstract} 

We study the entanglement negativity for the free, scalar chiral current in two spacetime dimensions,
which is a simple model violating the Haag duality in regions with nontrivial topology.
For the ground state of the system, 
both on the line and on the circle,
we consider the setups given by two intervals,
either adjacent or disjoint. 
We find analytic expressions for the moments of the partial transpose of the reduced density matrix and the logarithmic negativity. 
In the limit of small separation distance, this expression yields the same subleading topological contribution occurring in the mutual information.
In the limit of large separation 
distance between the two intervals,
the exponential decay of the logarithmic negativity
is obtained from its analytic expression. 
The analytic formulas are checked against exact 
numerical results from a bosonic lattice model, 
finding a perfect agreement.
We observe that, since the chiral current generates the neutral subalgebra of the full chiral Dirac fermion theory, this analysis highlights how symmetries produce nontrivial features in the entanglement structure that are analogue to those ones already observed in the mutual information for regions with nontrivial topology.

\end{abstract}

\newpage

	\tableofcontents

\newpage
\section{Introduction} 
\label{intro}

The algebraic formulation of quantum field theory provides a robust and conceptually precise framework for analyzing models with internal symmetries. In particular, as for the quantum information-theoretic measures, this approach is indispensable for characterizing the nontrivial structural modifications induced by the presence of global or local symmetries. A central manifestation of these effects is the breakdown of duality or additivity in the net of local operator algebras, which results in symmetry-induced contributions to entanglement measures.

This phenomenon becomes especially transparent in continuum field theories, where such non-additive contributions to the entanglement entropy arise in regions with nontrivial topology, typically disjoint unions subject to symmetry constraints. These contributions impact relative entropic quantities, such as the mutual information, as rigorously demonstrated in the context of global symmetries in \cite{Casini:2019kex} and extended to gauge (local) symmetries in \cite{Casini:2020rgj}.

The precise algebraic encoding of symmetries through local nets has broad and significant implications. It not only redefines our understanding of conventional order parameters relevant for confinement, but also sharpens the formulation of Noether charges, the treatment of anomalies, and the role of modular invariance, as discussed in \cite{Longo:2017mbg, Benedetti:2022zbb, Benedetti:2023owa, Benedetti:2024dku,Agon:2024xvs}.

In this work, we study the chiral 
$U(1)$ current in one spatial dimension. Within this perspective, this model can be defined as the 
neutral subalgebra of the chiral massless Dirac fermion and provides a simple yet nontrivial setting to investigate the role of symmetries in entanglement measures in nontrivial regions.

The algebra of observables of this model, analyzed in detail in \cite{Arias:2018tmw}, is generated either by the chiral current or, via bosonization, by the derivative of the massless scalar field
\begin{equation}
j=\psi^{\dagger}\psi=\partial\phi \,.
\end{equation}
Our goal is to explore how internal symmetries impact entanglement properties, particularly focusing on the manifestations of the breakdown of Haag duality in the local algebra, through the emergence of an extra topological contribution in the logarithmic negativity, a sensitive probe of mixed-state entanglement.

\begin{figure}[t!]
\begin{center}  
\includegraphics[width=0.3\textwidth]{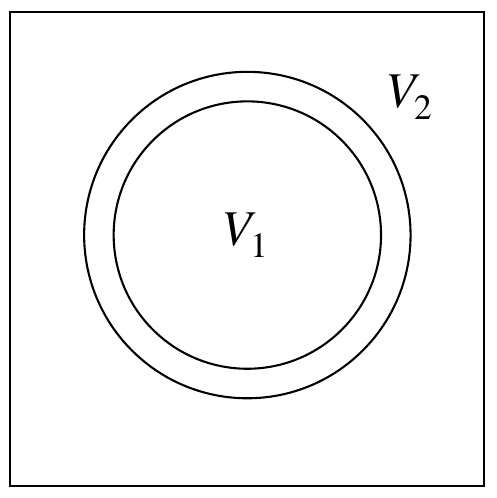}
\caption{A configuration in two spatial dimensions where $V_1$ and $V_2$ are nearly complementary.}
\label{set0}
\end{center}  
\end{figure}

More concretely, we consider a tripartition of the Hilbert space $\mathcal{H}=\mathcal{H}_V\otimes\mathcal{H}_{V'}$, with $V=V_1\cup V_2$, and define the reduced density matrix $\rho_V=\mathrm{Tr}_{V'}\rho$. In this setup, the logarithmic negativity is defined as the logarithm of the trace norm of the partially transposed reduced density matrix,
\begin{equation}
\mathcal{E}\equiv\log\lVert \rho_V^{\textrm{\tiny $\Gamma_2$}}\rVert=\log\mathrm{Tr}\,|\rho_V^{\textrm{\tiny $\Gamma_2$}}|
\label{defnegativity}
\end{equation}
where $\Gamma_2$ denotes the partial transpose with respect to subsystem $V_2$. Crucially, $\rho_V^{\textrm{\tiny $\Gamma_2$}}$ is not positive semidefinite and therefore has both positive and negative eigenvalues, in contrast with the reduced density matrix $\rho_V$. The logarithmic negativity serves as a standard entanglement witness, providing a quantitative measure of the entanglement between $V_1$ and $V_2$ \cite{Zyczkowski:1998yd, Eisert:1998pz, Zyczkowski:1999iw, Peres:1996dw, Vidal:2002zz, Plenio:2005cwa,Eisert:2006kue, Lee:2000}.
It is worth highlighting that, in contrast, the mutual information 
$I$ captures the total correlations, both quantum and classical, between the two subsystems. 
When $V_1$ and $V_2$ are nearly complementary regions,
as illustrated in Fig.\,\ref{set0} in a specific case, 
one finds $I=2S_{V_1}$. 
This relation provides a finite, geometrical definition of entanglement entropy in quantum field theory, at the cost of intrinsically involving a nontrivial geometry, unlike the original entanglement entropy for a single interval, which in turn makes it sensitive to the emergence of extra contributions in the case of incomplete models.
\\
The previous definitions and properties are discussed in 
Sec.\;\ref{sec:negativity}.

In one spatial dimension, considering the subsystem $V$ made by the union of two disjoint intervals 
$V_1 =(u_1, v_1)$ and $V_2 =(u_2, v_2)$, 
we exploit the connection between R\'enyi entropies 
and negativity R\'enyi entropies via the identity \cite{Calabrese:2012ew, Calabrese:2012nk, Calabrese:2014yza}
\begin{equation}
\label{eq:negativity_difinition}
\text{Tr} \big( \rho_V^{\textrm{\tiny $\Gamma_2$}} \big)^n 
= \big( \text{Tr}  \rho_V^n \big)
\big|_{u_2 \leftrightarrow v_2} 
\end{equation}
where the interchange $u_2 \leftrightarrow v_2$ accounts for the partial transposition over 
$V_2$ in this case.
Since our model is a chiral CFT, the relevant variable occurring in the setup provided by two disjoint intervals is the cross ratio $\eta \in (0,1)$ of their four endpoints
and the limits $\eta \to 0^+$ and $\eta \to 1^-$ 
correspond to the asymptotic regimes of  
large separation distance 
and adjacent intervals respectively.
In two spatial dimensions, the entanglement negativity has been studied e.g. in \cite{Eisler:2016Teo,DeNobili:2016nmj}.

In this manuscript,
 employing the analytical results for the chiral current spectrum of the modular Hamiltonian \cite{Arias:2018tmw}  
from which any functional entanglement measure of the density matrix can be computed, reviewed in Sec.\;\ref{sec:symmetries}, 
 we compute the moments of the partial transposes. 
By performing the analytic continuation 
characterising the replica limit  
\cite{Calabrese:2012ew, Calabrese:2012nk},
which involves the logarithm of the moments 
of the partial transposed  with even index, 
we also find an analytic expression 
for the logarithmic negativity $\mathcal{E}$
as a function of $\eta \in (0,1)$
(see Sec.\;\ref{sec:neg-cc-2int}).

At leading order in the limit of almost complementary regions, 
i.e. as $\eta\rightarrow 1^{-}$, 
one obtains half of the mutual Rényi entropy of order $1/2$, 
i.e. $\mathcal{E} = S^{(1/2)}_{V_2}=\tfrac{1}{2} I^{(1/2)} $,
as expected for pure states. 
More interestingly, in the same limit and at leading order, 
the topological contributions arising from the orbifold construction, already known to appear in the differences of mutual information 
between the full theory and the orbifold, 
also manifests in the logarithmic negativity, 
with the same coefficient $1/2$.
In particular, 
by introducing $\Delta \mathcal{E} \equiv
\widetilde{\mathcal{E}}^{\,\textrm{\tiny (W)}}  - \mathcal{E} $
and 
$\Delta I   \equiv I^{\textrm{\tiny (W)}} - I $,
where $\widetilde{\mathcal{E}}^{\,\textrm{\tiny (W)}} $ 
and $I^{\,\textrm{\tiny (W)}} $ are the negativity and the mutual information for a massless chiral fermion (Weyl) respectively according to the 
partial time-reversal prescription
introduced in \cite{Shapourian:2016cqu,Shapourian:2018lsz,Shapourian:2019xfi},
we find that
\begin{equation}
\label{eq:DeltaE_intro}
\Delta \mathcal{E} 
= 
\Delta I  
\approx
\frac{1}{2} 
\log \!\big( \!-\log (1-\eta) \big) 
\;\;\qquad \;\;
\eta\rightarrow 1^{-} \,.
\end{equation}
In order to derive the result \eqref{eq:DeltaE_intro}, we naturally had to consider the even-indexed fermionic R\'enyi negativities for the Dirac fermion in the prescription of \cite{Shapourian:2019xfi}, for two disjoint intervals in the infinite line. 
By employing the results of \cite{Herzog:2016ohd},
in Sec.\;\ref{sec-chiral-dirac-fermion}
we find an analytic expression for the moments of the partial transposed, 
whose replica limit leads to the following analytic result for the logarithmic negativity
\begin{equation}
    \widetilde{\mathcal{E}}
    = 
    2\, \widetilde{\mathcal{E}}^{\,\textrm{\tiny (W)}} 
    =
    - \frac{1}{4}\log(1-\eta)
\end{equation}
which is compatible with the 
the perturbative results obtained in \cite{Shapourian:2019xfi}.
The difference $\Delta\cal{E}$ is, in fact, what we generally expect for any incomplete model in the algebraic sense.
We test this idea in Sec.\;\ref{thermofield}, for a thermofield double state in the high temperature regime, constructed from the regular representation of a finite group $G$.
These are relatively simple but powerful toy models for exploring entanglement since in this regime, the 
thermofield double (TFD) state becomes maximally entangled 
where the mutual information and the negativity 
can be computed explicitly.

Remarkably, the logarithmic negativity also exhibits a distinctive behaviour in the limit of large separation distance,
which is explored in Sec.\;\ref{subsec:two-intervals-large-distance} for the chiral current. 
In contrast with the mutual information, whose falloff in this limit follows a power law, 
the logarithmic negativity displays an exponential decay. 
This characteristic feature of the logarithmic negativity has been observed numerically 
through lattice computations in various models 
\cite{Marcovitch:2008sxc, Wichterich:2008vfx, Calabrese:2012ew, Calabrese:2012nk, Calabrese:2013mi,Klco:2021cxq,Klco:2020rga, Parez:2022ind, Parez:2023xpj}.
For the chiral current, 
the analytic expression for the logarithmic negativity 
allows us to study its limit $\eta \to 0$,
finding the exponential decay given by  $\textrm{e}^{-2\pi\sqrt{\alpha/\eta}}$, 
where $\alpha \simeq 1.4458$ 
is obtained by solving numerically the following condition 
(see (\ref{eq-Re2F1-for-alpha}))
\begin{equation}
 {\text{Re}}\big[ \,
 _2F_1(1/2+\textrm{i} s,1/2+\textrm{i}s; 1;
 \alpha/s^2)
\,\big]=0 \,.
\end{equation}
Finally, in Sec.\,\ref{renyisvscorrelators},
we study the entanglement negativity in the lattice model 
considered in \cite{Arias:2018tmw}, whose continuum limit is described by
two copies of the chiral current that we are considering.
The numerical results obtained through exact numerical computations in this lattice model display an excellent agreement with the corresponding analytic expressions.

\section{The model and the entanglement entropies}
\label{sec:symmetries}

In this section we briefly review the chiral $U(1)$ current model, realized as the $U(1)$ orbifold of the chiral massless Dirac fermion. This framework highlights the role of symmetry in the structure of local algebras for disconnected regions and provides a natural setting to study violations of Haag duality. We summarize the current algebra, the appearance of non-local operators for unions of intervals, and the exact results for R\'enyi entropies and mutual information that will be used in the following analysis.

\subsection{Symmetries: the violation of the Haag duality or additivity}
\label{subsec:symmetries}

In models built as subsets of invariant operators under the action of a symmetry group, the local algebra net in the region $V$ is not uniquely defined.
Following \cite{Arias:2018tmw}, for a set $V=V_1\cup V_2$ as in Fig.\,\ref{set0}, we can define
\begin{equation}
{\cal{A}}_{\text{\tiny max}}(V)\equiv{\cal{A}}^{\prime}(V^{\prime})={\cal{A}}(V)\vee \{\text{a}\}
\label{max}
\end{equation}
and similarly for the complementary region $V^{\prime}$
\begin{equation}
{\cal{A}}_{\text{\tiny max}}(V^\prime)\equiv{\cal{A}}^{\prime}(V)={\cal{A}}(V^{\prime})\vee \{\text{b}\} \,.
\label{maxc}
\end{equation}
In (\ref{max}) and (\ref{maxc}), ${\cal{A}}_{\text{\tiny max}}$ is the maximal choice, $\cal{A}$ the additive minimum one, and  in between there is an infinite family of other choices. 
For global symmetries, the sets $\{a\}$ and $\{b\}$ are named as intertwiners and twist operators, which by definition, cannot be locally generated in $V$ and $V^{\prime}$ respectively.

The multiple algebra choices naturally results in new features for relative entropies. 
For the model $\cal{O}$ 
invariant under a symmetry group $G$ 
and contained in the full model $\cal{F}$,
i.e. $\cal{O}\subset \cal{F}$, 
it has been found \cite{Casini:2019kex} that 
the relative entropy 
\be
\label{rel-ent-Delta-I}
S_{\cal F}(\omega,\omega\circ E)
=
I_{\cal F} - I_{\cal O}
\ee
is the good entanglement measure, entanglement order parameter, for sensing the contributions of intertwiners in the algebra of the union.  
In (\ref{rel-ent-Delta-I}), $\omega$ is the vacuum state, $I_{\cal F}$ and $I_{\cal O}$ denote the mutual information of $\cal F$ and $\cal O$ respectively, 
and $E$ is the conditional  expectation
\be
E : {\cal{O}}(V) \vee {\cal{I}}_{V_1V_2}\rightarrow {\cal{O}}(V)
\ee
with ${\cal{I}}_{V_1V_2}$ the intertwiners that belong to ${\cal O^{\prime}}(V^{\prime})$, analogous to $\{a\}$ in (\ref{max}).
In other words, $S_{\cal F}(\omega,\omega\circ E)$ indirectly measures how the algebra choice for ${\cal{O}}$ in $V$, additive ${\cal{O}}(V)$  or dual ${\cal{O}}_{\text{\tiny max}}(V)$, manifests through the difference in the mutual information between the full model $\cal{F}$, where the intertwiners are locally generated and for the additive algebra in the orbifold without intertwiners.
Monotonicity of relative entropy implies that 
the order parameter $I_{\cal F} - I_{\cal O}$ is always 
positive. 
Although $\cal{F}$ occurs in the definition of this quantity, 
this property depends exclusively on $\cal{O}$ itself.

In $(1+1)$ dimensions, when $V_1$ and $V_2$ are almost complementary and $G= U(1)$, 
it has been found that \cite{Casini:2019kex}
\begin{equation}
\label{mutualdif}
\Delta I \, =I_{\cal F} - I_{\cal O}=\frac{1}{2}\log \! \big(\log(\ell/\delta) \big)
\end{equation}
where $\ell$ is the size of the interval and $\delta$ the separation distance with the complement.

\subsection{The chiral current model}

For the following discussion, consider the massless scalar in two dimensions. This can be described in terms of two identical copies of the chiral currents $j(x_{\pm})=\partial_{\pm} \phi$ with central charge $c_\pm = 1$. 
The Hamiltonian for this theory is given by
\be
H=\frac{1}{2}
\int j^2(x_{\pm})\, \textrm{d}x_\pm
=
\int (\partial_{\pm} \phi)^2
 \,\textrm{d}x_\pm
\ee
with null coordinates defined as $x_{\pm}=x_0\pm x_1$. The corresponding commutation relation is
\be
\label{eq:commutator_continuum}
\big[\,j(x)\, , j(y)\, \big]=
\ri \, \delta^{\prime}(x-y) \,.
\ee

The algebra of the current is generated by the operators of the fermion algebra that are invariant under the charge transformation $\psi(x)\rightarrow \textrm{e}^{\textrm{i}\alpha} \,\psi(x)$. In other words, the algebra of the current $\cal O$ corresponds to the orbifold that results by quotient of the fermionic model, denoted as the full algebra $\cal F$, by the $U(1)$ symmetry group.   

\begin{figure}[t]
\begin{center}  
\includegraphics[width=0.85\textwidth]{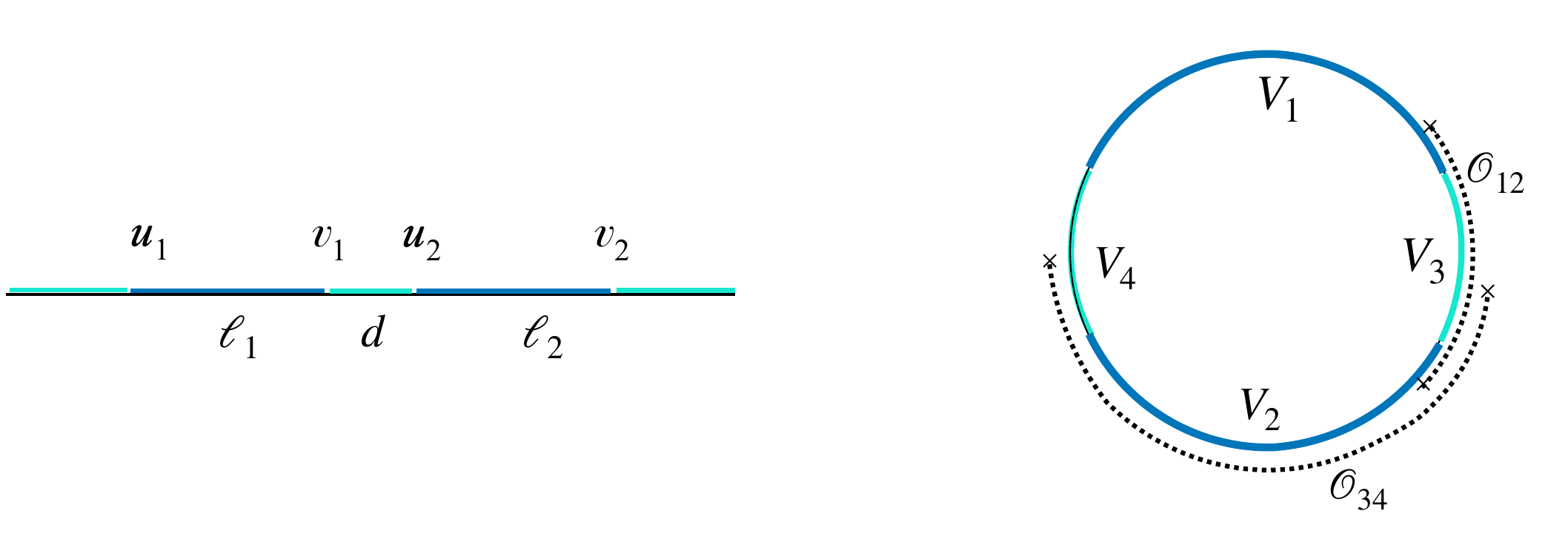}
\caption{The subsystem $V=V_1\cup V_2$ and its complement $V^{\prime} =V_3\cup V_4$, either on the line (left) or on the circle (right).}
\label{set2}
\end{center}  
\end{figure}

For clarity purposes, we place the theory in the circle, as shown in the right panel of Fig.\,\ref{set2}. In this setting, we can explicitly calculate intertwiner and twist operators \cite{Arias:2018tmw}.
Consider the operator ${\cal{O}}_{12}\in \cal{O}$ defined as
\be
{\cal{O}}_{12}=
\int  \alpha (x) \,\partial_{x} \phi \,\textrm{d}x
=
\phi_1-\phi_2
\ee
and similarly, for the complementary region
\be
{\cal{O}}_{34}=
\int  \alpha (x) \,\partial_{x} \phi \,\textrm{d}x
=
\phi_3-\phi_{4} \,.
\ee
It is clear that $\cal{O}_{\text{12}}$ does not commute with its dual counterpart $\cal{O}_{\text{34}}$.
The operators  
$\cal{O}_{\text{12}}$ and  $\cal{O}_{\text{34}}$ being derivatives of $\phi$ satisfy
$\cal{O}_{\text{12}}\in \cal O$ and 
$\cal{O}_{\text{34}}\in \cal O$ but  
$
 {\cal{O}}_{12}\notin {\cal O}_V
$ and 
$ {\cal{O}}_{34}\notin{\cal O}_{V^{\prime}}$,
since $\phi_1$ and $\phi_2$ by themselves do not exist in ${\cal O}(V)$ defined as ${\cal {O}}(V_1 \cup V_2)={\cal {O}}({V_1})\vee {\cal {O}}({V_2})$ and in turn, $\phi_3$ and $\phi_4$ by themselves do not exist in ${\cal O}(V^{\prime})={\cal O}(V_3 \cup V_4)$ defined as ${\cal {O}}(V^{\prime})={\cal {O}}({V_3})\vee {\cal {O}}({V_4})$.
Note that here, ${\cal{O}}_{12}$ and ${\cal{O}}_{34}$ are the analogues of the $\{a\}$ and $\{b\}$ non-local operators in (\ref{max}) and (\ref{maxc}) respectively.

\label{sec-renyi-entropies-two-intervals}

\begin{figure}[t!]
\begin{center}  
\includegraphics[width=0.55\textwidth]{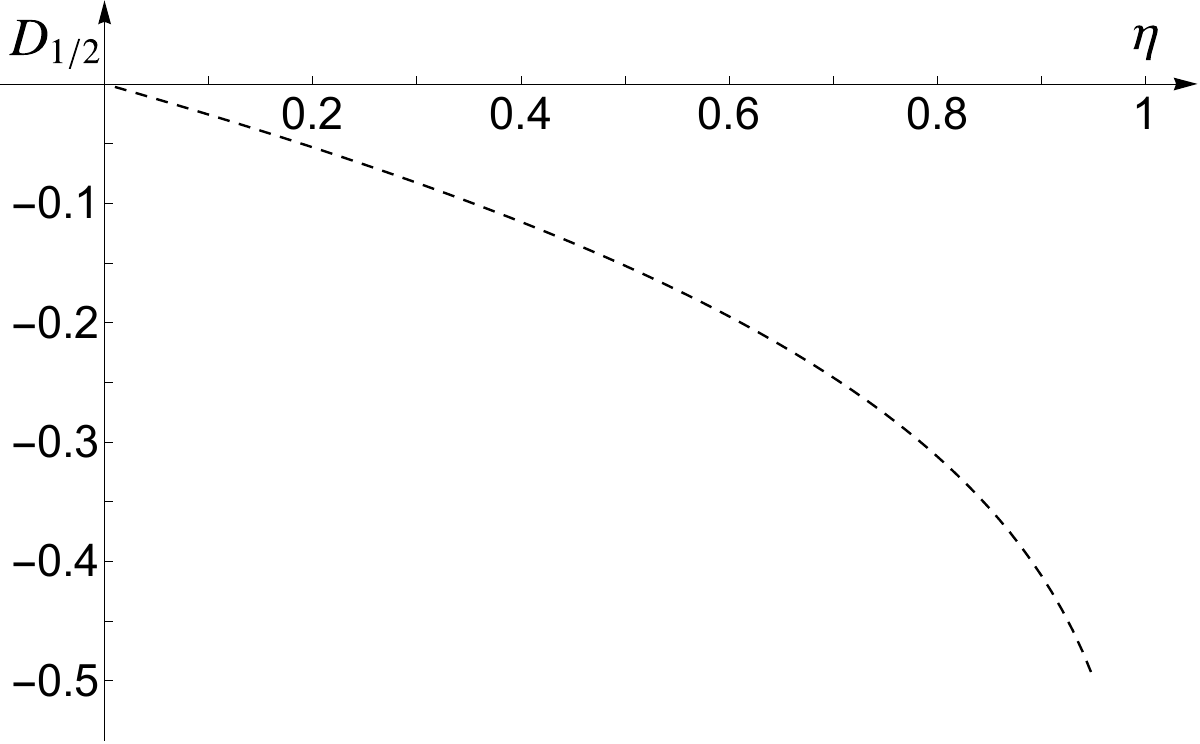}
\caption{The function $D_{1/2}$, 
from \eqref{Dn-integral-ACHP}.}
\label{urenyiplot}
\end{center}  
\end{figure}

The geometric modular action for conformal chiral models was previously studied in \cite{Longo:2009mn}, while the specific case of the chiral current was analysed in detail in \cite{Arias:2018tmw}.
From the solution of the spectral problem
for the modular Hamiltonian kernel in the union of multiple intervals \cite{Arias:2018tmw}, it is possible to compute in this nontrivial setup any entanglement measure related to the reduced density matrix or its powers. 
In particular,  
considering the domain $V \equiv V_1 \cup V_2$ 
made by the union of two disjoint intervals 
$V_1 =(u_1, v_1)$ and $V_2 =(u_2, v_2)$ with $u_1 < v_1 < u_2 < v_2$,
whose lengths are $\ell_1 \equiv v_1 - u_1$ and $\ell_2 \equiv v_2 - u_2$ respectively and their separation distance is $d\equiv u_2  - v_1$, 
the mutual R\'enyi entropies are
\be 
I^{(n)}
\equiv
S^{(n)}_{V_1}
+
S^{(n)}_{V_2}
-
S^{(n)}_V
\label{mutualrenyi}
\ee
with  $S^{(n)}_V$ defined as the R\'enyi entropy of $V$, namely
\be
\label{renyi-single-interval}
S^{(n)}_V 
\equiv 
\frac{1}{1-n}\,\log \! \big(\textrm{Tr} \, \rho_V^n \big)
\ee
while $S^{(n)}_{V_j}$ are the R\'enyi entropies of the single intervals
\be
\label{renyi-single-interval_Vj}
S^{(n)}_{V_j} 
= 
\frac{1}{1-n}\,\log \! \big(\textrm{Tr} \, \rho_{V_j}^n \big)
= \frac{1}{1-n} \, \log\!\left( \frac{c_n}{(\ell_j/\epsilon)^{\Delta_n}} \right)
\ee
where $\epsilon$ is the UV cutoff and 
\begin{equation}
\label{Delta-n-def}
    \Delta_n \equiv \frac{1}{12} \left( n - \frac{1}{n}\right)
\end{equation}
which corresponds to the scaling dimension of the branch-point twist fields for a generic CFT \cite{Calabrese:2004eu}
specialised to our case where the total central charge is $c=1/2$,
while $c_n$ is a normalization constant. 
Notice that, from (\ref{renyi-single-interval_Vj}), 
the entanglement entropy of an interval is 
$S_{V_j} = \tfrac{1}{6} \log(\ell_j /\epsilon) + \textrm{const}$,
as expected for a chiral CFT.
The moments of the reduced density matrix  
of $V$ for the chiral current read
\cite{Arias:2018tmw}
\begin{equation}
\label{moments-rhoA-current-explicit}
\tr  \rho_{V}^{n} 
\,=\,
c_n^2 
\left( \frac{\epsilon^2}{ \ell_1\, \ell_2\,(1-\eta)} \right)^{\Delta_n} 
\! \mathcal{F}_n(\eta)
\end{equation}
where $\eta$ is defined as the cross ratio of 
the four endpoints of $V_j$, with $j \in \{1,2\}$, 
which can be written as 
\begin{equation}
\label{cross-ratio-eta-def}
    \eta \equiv 
    \frac{(u_1 - v_1)\, (u_2 - v_2) }{ (u_1 - u_2)\, (v_1 - v_2)  }
    =\frac{\ell_1\, \ell_2}{ (\ell_1 + d)\,(\ell_2 + d) }
\;\;\;\;\qquad\;\;\;\;
\eta \in (0,1)
\end{equation}
where $d$ is the distance separating the two intervals. 
From (\ref{renyi-single-interval_Vj})-(\ref{moments-rhoA-current-explicit}),  the mutual R\'enyi entropies (\ref{mutualrenyi}) 
become
\begin{equation}
\label{eq:Renyi_MI_twodisjoint}
I^{(n)}
\,=\,
-
\frac{\Delta_n}{n-1} \,\log(1-\eta) 
-
\frac{D_n(\eta)}{2(n-1)}
\end{equation}
where we have introduced 
\begin{equation}
    \label{Dn-def}
    D_n(\eta) \equiv 
    -\, 2\log \!\big[ \mathcal{F}_n(\eta) \big] \,.
\end{equation}
In the chiral current model, 
the explicit expression for $D_n(\eta)$ 
has been found in \cite{Arias:2018tmw} and reads
\begin{equation}
\label{Dn-integral-ACHP}
D_n(\eta) \,=\,
\textrm{i} \,n 
\int_0^{\infty} \!
\big[\coth(\pi n\,s) -\coth(\pi s)\,\big]
\,\log \!\left( 
\frac{ F_{\textrm{i}s}(\eta) }{ F_{-\textrm{i}s}(\eta) } 
\right)
\textrm{d}s
\qquad
    F_{r}(\eta) \equiv \, 
{}_2 F_1(r,1-r;1;\eta)\,.
\end{equation}
In Fig.\,\ref{urenyiplot} 
we show this function for the special case of $n=1/2$.
For the integers $n \geqslant1$ 
we have that \cite{Gentile:2025koe, Abate:2025ywp}
\begin{equation}
\label{Dn-def-2F1}
    D_n(\eta) 
    =
\sum_{k=1}^{n-1}
\log\!\big[F_{k/n}(\eta) \big]\,.
\end{equation}
Notice that 
(\ref{Dn-integral-ACHP}) 
is related to a factor 
occurring in the moments $\tr  \rho_{V}^{n} $ 
for the massless scalar on the line 
and in its ground state, 
found in \cite{Calabrese:2009ez}.

Combining (\ref{Dn-def}) with (\ref{Dn-def-2F1}), we get
\begin{equation}
\label{mathcal-Fn-def}
\mathcal{F}_n(\eta)
=
\frac{1}{\sqrt{\prod_{k=1}^{n-1} F_{k/n}(\eta)}} \,.
\end{equation}

 The analytic continuation $n \to 1$ of 
(\ref{eq:Renyi_MI_twodisjoint}) 
provides the mutual information 
\cite{Calabrese:2009ez,Arias:2018tmw}
\begin{eqnarray}
\label{eq:MI_twodisjoint}
I
=
\lim_{n\,\to\, 1} 
I^{(n)}
&=&
-
\frac{1}{6} \log(1-\eta) 
-
\frac{1}{2} \big[ \partial_n D_n(\eta)\big]\big|_{n=1}
\nonumber
\\
\rule{0pt}{.8cm}
&=&
-
\frac{1}{6} \log(1-\eta) 
+
\frac{\textrm{i}}{2} 
\int_0^{\infty} \!
\frac{\pi s}{\big[ \sinh(\pi s) \big]^2}
\,\log \!\left( 
\frac{ F_{\textrm{i}s}(\eta) }{ F_{-\textrm{i}s}(\eta) } 
\right)
\end{eqnarray}
where (\ref{Dn-integral-ACHP}) has been employed to obtain the last expression. 
We conclude by reviewing the Rényi mutual information in its $\eta\rightarrow 0$ and $\eta\rightarrow 1$ limits.
These limits are particularly relevant and will be compared with those of the logarithmic negativity in the following section.

Considering the expression of $I^{(n)}$
obtained by combining (\ref{eq:Renyi_MI_twodisjoint}) 
and (\ref{Dn-def-2F1}), 
its expansion as $\eta \to 0^+$ reads 
\begin{equation}
\label{MI-renyi-expansion-eta-0}
    I^{(n)}
    \,=\,
    \sum_{r=2}^{+\infty} q_r(n)\, \eta^r
\end{equation}
where the coefficients of the first terms are given by 
\begin{equation}
    \label{q2-coeff-def}
    q_2(n)
    =
    \frac{(n^2 + 1)\, (n+1)}{240\, n^3}
\;\;\;\qquad\;\;\;
    q_3(n)
    =
    \frac{(4\,n^2 + 5)\, (4\,n^2 -1)\, (n+1)}{3780\, n^5}
\end{equation}
and
\begin{eqnarray}
\label{q-other-coeff-def}
    q_4(n)
    &=&
    \frac{
    \big[941\, n^4 (n^2+1)  - 529\, n^2 + 231\big]\, (n+1)
    }{241920\, n^7}
\\
\rule{0pt}{.8cm}
    q_5(n)
    &=&
    \frac{
    \big[1174\,n^6(n^2+1) -905\, n^4 +756\,n^2 -399\big]\, (n+1)
    }{332640\, n^9}
    \\
    \label{q6-coeff-def}
\rule{0pt}{.8cm}
    q_6(n)
    &=&
    \frac{
    \big[ 10000171\,n^8 (n^2+1)
    - 9507317\, n^6
    +11413583\,n^4
    - 11946754\, n^2
    +7190546\big]\, (n+1)
    }{3113510400\, n^{11}}
    \nonumber
    \\
    & &
\end{eqnarray}
and also $q_r(n)$ for $r \geqslant 7$ can be easily found.
The expansion (\ref{MI-renyi-expansion-eta-0}) up to $O(\eta^4)$ included
has been reported  also in \cite{Abate:2025ywp}.
Taking the limit $n\to 1$ in 
(\ref{MI-renyi-expansion-eta-0})-(\ref{q6-coeff-def}),
one finds the following expansion for the mutual information
\cite{Arias:2018tmw}
\begin{equation}
\label{expansion-MI-eta0}
I
=
\frac{1}{60}\; \eta^2 
+ \frac{1}{70}\; \eta^3
+ \frac{11}{840}\; \eta^4
+ \frac{5}{462}\; \eta^5
+ \frac{397}{36036}\; \eta^6
+ O\big(\eta^7\big) \,.
\end{equation}
This expansion is obtained also by expanding (\ref{eq:MI_twodisjoint}) for small $\eta$.
Hence, in the case of the mutual R\'enyi entropies, 
the two procedures given by 
the expansion for large separation distance 
and by the replica limit 
commute. 
Notice that the leading term of (\ref{expansion-MI-eta0})
is consistent with the result expected for a generic CFT, which is \cite{Calabrese:2010he,Cardy:2013nua,Agon:2015ftl, Chen:2016mya}
\begin{equation}
I\, \approx \frac{\sqrt{\pi}\;\Gamma(2\Delta+1)}{4^{2\Delta+1}\,\Gamma(2\Delta+3/2)}\;\eta^{2\Delta}
\end{equation}
that becomes the $O(\eta^2)$ term in 
(\ref{expansion-MI-eta0}) in the special case where  $\Delta=1$.

In the opposite limit of small separation distance, 
given by $d\ll \ell$, 
we have  $\eta\approx 1-2 \,d /\ell$.
From (\ref{eq:Renyi_MI_twodisjoint}), (\ref{Dn-def-2F1})
and the expansion $F_r(\eta) =  -\tfrac{\sin(\pi r)}{\pi} \log(1-\eta) +O(1)$ as $\eta \to 1^{-}$,
it is straightforward to find that 
the expansion of the mutual R\'enyi entropies 
in this asymptotic regime reads
\begin{equation}
\label{I-n-loglog-term}
I^{(n)}
=
-\frac{n+1}{12 \,n}\,\log(1-\eta)
- \frac{1}{2}
\log \!\big( \!-\log (1-\eta) \big)
+ \dots
\end{equation}
where the dots denote subleading terms. 
Notice the subleading divergence in (\ref{I-n-loglog-term}) is independent of $n$.
Since for chiral fermions the mutual information is given by
$I^{\textrm{\tiny (W)}} =-\frac{1}{6} \log(1-\eta) $, 
for $n = 1$ we find 
\begin{equation}
\Delta I
\approx 
\frac{1}{2} \log \!\big( \!-\log (1-\eta) \big)
\approx 
\frac{1}{2}\log\!\big( \log{(\ell/d)} \big)
\end{equation}
which is consistent with the behaviour expected from (\ref{mutualdif}).
This indicates the occurrence of a Haag duality violation,
which can be explicitly tested in this setup \cite{Arias:2018tmw}. 

\begin{figure}[t!]
\centering
\begin{subfigure}
    \centering
    \includegraphics[width=0.48\textwidth]{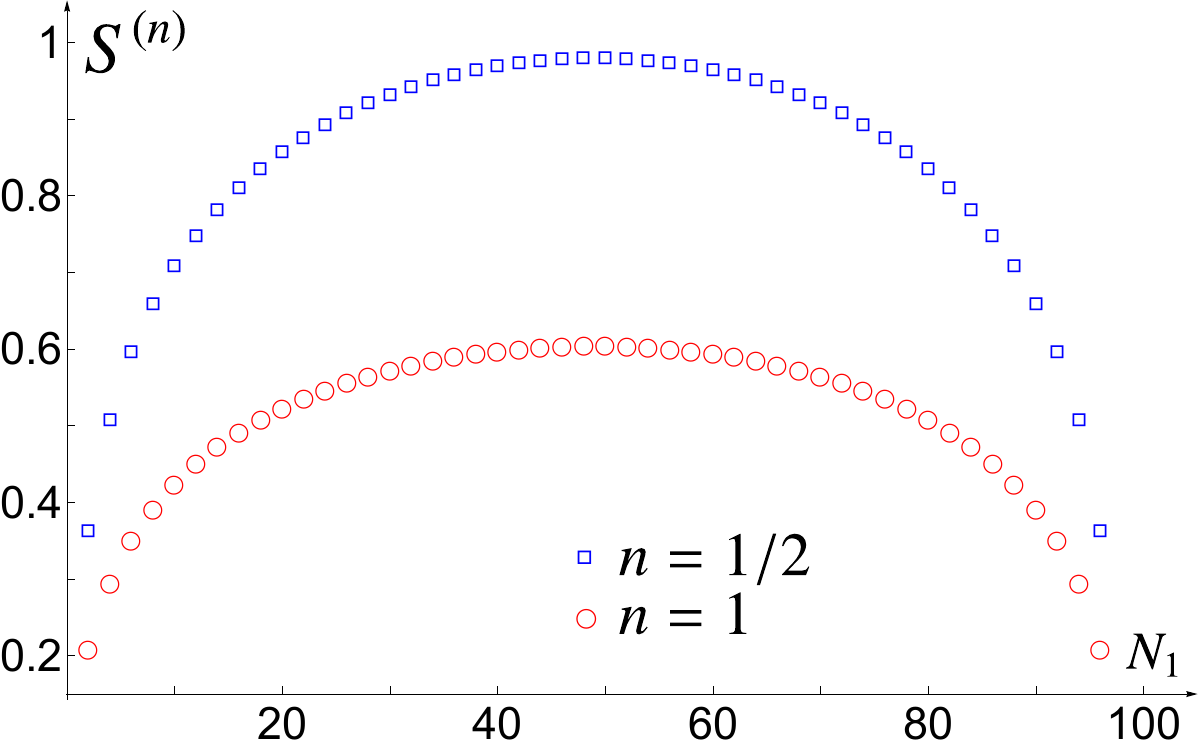}
\end{subfigure}
\hfill
\begin{subfigure}
    \centering
    \includegraphics[width=0.48\textwidth]{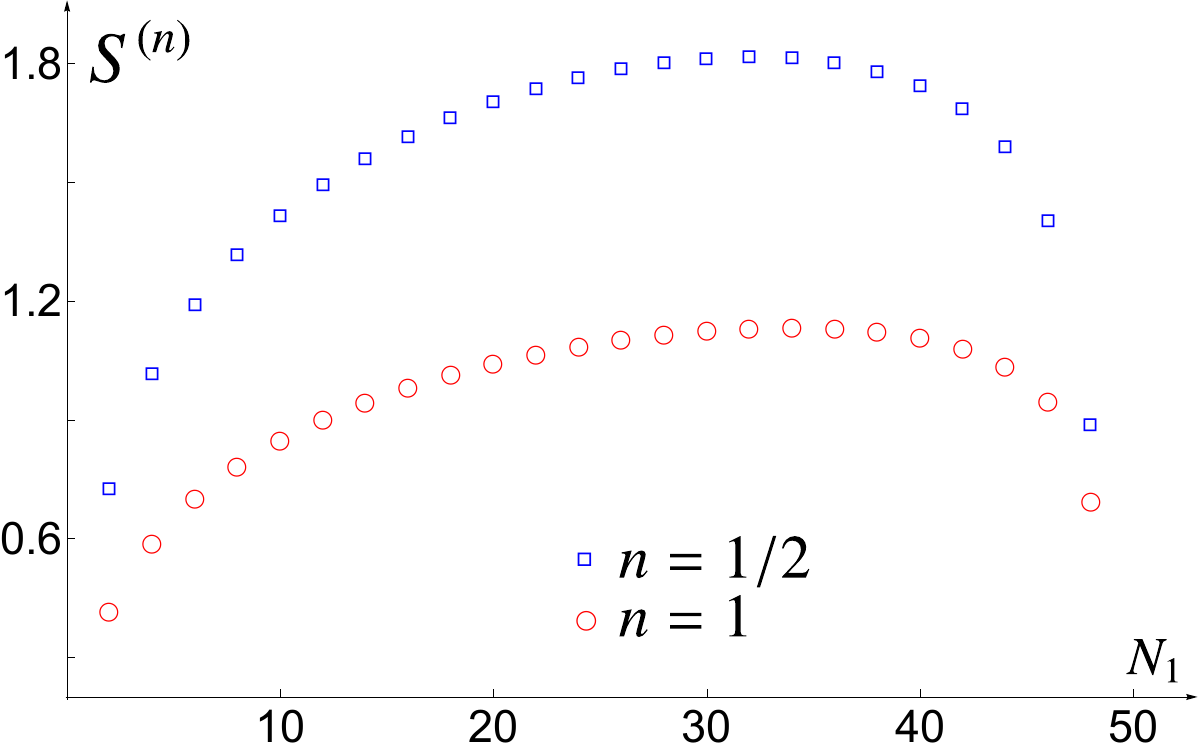}

\end{subfigure}
\caption{
Comparison of $S^{(1/2)}$ and $S$ for a subsystem given by either one block (left) 
or two equal blocks (right) 
in the circle made by $100$ sites.
}
\label{fig:renyi_compare}
\end{figure}

We begin by examining Haag duality on the circle for the configuration shown in Fig.\,\ref{set2}. Paying special attention to the $n=1/2$ case related to the negativity we will consider later, 
in Fig.\,\ref{fig:renyi_compare},
considering a circle made by $N=100$ sites, we show 
$S^{(1/2)}_{V_1}$ and $S_{V_1}$ for a block made by $N_1$ sites (left panel),
and $S^{(1/2)}_{V_1\cup V_2}$ and $S_{V_1\cup V_2}$
for two blocks of equal size $N_1=N_2$
(right panel).
The data points have been obtained from a specific lattice model in terms of the current operators in the null line that provides a natural discrete description of the model studied here, as discussed in Sec.\,\ref{renyisvscorrelators}.

From the symmetry of the curve in the R\'enyi entropy for one interval we immediately deduce that, in this case, there is no duality breaking.  In contrast, for the two interval case, with nontrivial topology, the R\'enyi entropy calculated with the local operator content, the additive algebra ${\cal {O}}({V_1})\vee {\cal {O}}({V_2})$, is not symmetric, that is, the algebra is not dual, so
\begin{equation}
{\cal{O}}(V_1\cup V_2)\neq {\cal{O}}(V_3 \cup V_4)\,.
\end{equation}

On the other hand, we can also test the Haag duality in the line by checking if the $D_n(\eta)$ function (\ref{Dn-def}) satisfies the dual property $D_n(\eta)=D_n(1-\eta)$. In Fig.\,\ref{urenyiplot} we plot $D_{1/2}$ calculated by numerically solving the integral (\ref{Dn-def}) and consistently find the Haag duality violation.
We note that the same is true for any other value of $n$.

\section{Entanglement negativity and symmetries}
\label{sec:negativity}

In this section, 
we first recall the definition of
the logarithmic negativity \cite{Peres:1996dw,Vidal:2002zz,Plenio:2005cwa}, 
a computable measure of the bipartite entanglement for mixed states, which serves both as an entanglement witness 
and as an upper bound to the distillable entanglement,
and the replica limit employed to explore it in quantum field theories \cite{Calabrese:2012ew,Calabrese:2012nk}
(see Sec.\,\ref{subsec-log-neg-def}).
Then, in Sec.\,\ref{thermofield},
we illustrate these ideas by employing 
a simple toy model. 
In particular, 
by considering a thermofield double state in the high-temperature regime, 
constructed from the regular representation of a finite group $G$, 
we find that both the negativity and the mutual information exhibit the same topological contribution. This behaviour is also observed in the case of the chiral current discussed in Sec.\,\ref{sec:neg-cc-2int}.

 \subsection{Logarithmic negativity and its replica limit}
\label{subsec-log-neg-def}

Consider a mixed state in the spatial region $V$ 
characterised by the density matrix $\rho_V$.
Let us introduce a spatial bipartition $V = V_1 \cup V_2$, 
by assuming that the Hilbert space factorises accordingly, 
namely $\mathcal{H} = \mathcal{H}_{V_1} \otimes \mathcal{H}_{V_2}$.
The partial transpose $\rho_V^{\textrm{\tiny $\Gamma_2$}}$ of $\rho_V$
e.g. with respect to $V_2$ can be introduced 
by defining its generic matrix element as follows
\be 
\label{partial-transposition-def}
\langle e_i^{(1)} e_j^{(2)}|\,
\rho_V^{\textrm{\tiny $\Gamma_2$}}
\,|e_k^{(1)} e_l^{(2)}\rangle
\,\equiv\,
\langle e_i^{(1)} e_l^{(2)}|\,
\rho_V\,
| e^{(1)}_k e^{(2)}_j\rangle
\ee
where the normalisation condition 
$\tr  \rho_V^{\textrm{\tiny $\Gamma_2$}} = \sum_i \lambda_i = 1$ is imposed. 
The crucial feature of the partial transpose $\rho_V^{\textrm{\tiny $\Gamma_2$}}$ is that also negative eigenvalues occur in its spectrum; hence it is not a proper density matrix.
However, the occurrence of negative eigenvalues allows to introduce the 
logarithmic negativity as \cite{Peres:1996dw,Vidal:2002zz,Plenio:2005cwa, Eisert:2006kue}
\begin{equation}
\label{log-neg-def}
    \mathcal{E} 
    \,\equiv \,
    \log \big|\!\big| \rho_V^{\textrm{\tiny $\Gamma_2$}} \big|\!\big|
    \,=\, 
    \log \!\big( 
    \tr \!
    \big|\rho_V^{\textrm{\tiny $\Gamma_2$}} \big| 
    \,\big)
\end{equation}
in terms of the trace norm of $\rho_V^{\textrm{\tiny $\Gamma_2$}} $, i.e. 
\be
\big|\!\big| \rho_V^{\textrm{\tiny $\Gamma_2$}} \big|\!\big|
\equiv
\tr \!
    \big|\rho_V^{\textrm{\tiny $\Gamma_2$}} \big| 
    \,\equiv\,
    \sum_i \big| \lambda_i \big|
    =
    1 -2 \sum_{\lambda_i < 0} \lambda_i 
\ee
where the last step is obtained by employing the normalization condition 
of $\rho_V^{\textrm{\tiny $\Gamma_2$}} $.

The moments of $\rho_V^{\textrm{\tiny $\Gamma_2$}} $ naturally lead to introduce 
\begin{equation}
\label{neg-moments-def}
    \mathcal{E}_{n} \equiv \,
    \log\!  
    \big[
    \tr \!\big( \rho_V^{\textrm{\tiny $\Gamma_2$}}\big)^{n} 
    \,\big] \,.
\end{equation}
Since the spectrum of $\rho_V^{\textrm{\tiny $\Gamma_2$}}$ contains also negative eigenvalues, 
a parity effect occurs in the moments of the partial transpose,
which allows to evaluate the logarithmic negativity (\ref{log-neg-def})
through a replica limit \cite{Calabrese:2012ew,Calabrese:2012nk}.
Indeed, for $n= n_{\textrm{\tiny e}}$ even 
and $n= n_{\textrm{\tiny o}}$ odd, 
they can be written respectively as 
\bea
\label{moments-rhoA-PT2-even}
\tr \!\big( \rho_V^{\textrm{\tiny $\Gamma_2$}}\big)^{n_{\textrm{\tiny e}}} 
&=&
\sum_i \lambda_i^{n_{\textrm{\tiny e}}}
\,= 
\sum_{\lambda_i>0} \big|\lambda_i\big|^{n_{\textrm{\tiny e}}}
+ 
\sum_{\lambda_i<0} \big|\lambda_i\big|^{n_{\textrm{\tiny e}}}
\\
\label{moments-rhoA-PT2-odd}
\rule{0pt}{.7cm}
\tr \!\big( \rho_V^{\textrm{\tiny $\Gamma_2$}}\big)^{n_{\textrm{\tiny o}}} 
&=&
\sum_i \lambda_i^{n_{\textrm{\tiny o}}}
\,= 
\sum_{\lambda_i>0} \big|\lambda_i\big|^{n_{\textrm{\tiny o}}}
-
\sum_{\lambda_i<0} \big|\lambda_i\big|^{n_{\textrm{\tiny o}}}\,.
\eea
The logarithmic negativity (\ref{log-neg-def}) 
can be obtained from (\ref{moments-rhoA-PT2-even}), 
by performing the following analytic continuation
\be
\label{neg-replica-limit}
\mathcal{E} = 
\lim_{n_{\textrm{\tiny e}} \to 1} 
\mathcal{E}_{n_{\textrm{\tiny e}}}\,.
\ee

It is worth considering the following ratio
\be
\label{Rn-ratio-def}
R_n \equiv 
\frac{
\tr \!\big( \rho_V^{\textrm{\tiny $\Gamma_2$}}\big)^{n}
}{
\tr  \rho_V^{n}
}\,.
\ee
Indeed, combining (\ref{neg-replica-limit}) with the normalization condition
$\tr  \rho_V = 1$, we have that 
\be
\mathcal{E} = \lim_{n_{\textrm{\tiny e}} \to 1}\log \big( R_{n_{\textrm{\tiny e}}}\big) \,.
\ee

When the bipartite system $V = V_1 \cup V_2$ is in a pure state, 
namely $\rho_V = |\Psi\rangle \langle \Psi|$,
the moments of the partial transpose w.r.t. $V_j$, 
with $j \in \{1,2\}$,
can be written in terms of the moments of $\rho_{V_j}$
as follows \cite{Calabrese:2012ew, Calabrese:2012nk}
\begin{equation}
\label{pure-states-moments-relation}
    \tr \! \big( \rho_V^{\textrm{\tiny $\Gamma_j$}} \big)^n = 
    \left\{
    \begin{array}{ll}
        \; \tr  \rho_{V_j}^{n_{\textrm{\tiny o}}} 
        \hspace{2cm} & 
        n=n_{\textrm{\tiny o}} \;\;\textrm{odd}
        \\
        \rule{0pt}{.6cm}
        \left( \tr \rho_{V_j}^{n_{\textrm{\tiny e}} / 2}\right)^2 
        & 
        n=n_{\textrm{\tiny e}} \;\;\textrm{even}
    \end{array}
    \right.
\end{equation}
where the parity of $n$ plays a crucial role. 
Combining (\ref{pure-states-moments-relation}) with (\ref{neg-replica-limit}), it is straightforward to observe that  
$\mathcal{E} = S_{V_2}^{(1/2)} = S_{V_1}^{(1/2)}$ 
in this case.

\subsection{Toy model for a finite group 
and the TFD state at infinite temperature
}
\label{thermofield}

We construct a minimal toy model designed to capture the difference in mutual information that arises when this quantity is computed in the full theory $\mathcal{F}$, endowed with a finite symmetry group $G$, and its corresponding orbifold $\mathcal{O}$, for a setup involving two almost complementary regions. This construction is analogous to that presented in~\cite{Casini:2019kex}, where the algebra of local operators is reduced to the subalgebra of intertwiners responsible for the emergence of the topological contribution to the mutual information difference.

Let $G$ be a finite group of order $|G|$ and $\mathbb{C}[G]$ the vector space over 
$\mathbb{C}$ with orthonormal basis 
$\{  |g \rangle  : g \in G \} $. To turn  $\mathbb{C}[G]$ 
into a Hilbert space
\be
\mathcal{H} = \mathbb{C}[G]
\ee
we equip it with the usual inner product
\be
\langle g|h\rangle =\delta_{g,h} \,.
\ee
Hence, $\mathbb{C}[G]$ becomes a complex Hilbert space of dimension $|G|$.
 This space carries the left regular representation
\be
U(h) |g \rangle =  |hg \rangle, \quad \forall \, h,g \in G \,.
\ee
Now consider two copies of $ \mathbb{C}[G]$, with Hilbert space
\be
\mathcal{H} = {\mathcal{H}}_1 \otimes {\mathcal{H}}_2= \mathbb{C}[G] \otimes \mathbb{C}[G]
\ee
with basis $ \{ |g_1,g_2 \rangle : g_1, g_2 \in G \} $,  and dimension $ |G|^2 $.

The strength of this simplified model lies in its tractability: it allows for an explicit computation of the entanglement between the algebras associated with each region, which are here mimicked by two identical copies labelled $1$ and $2$ of the same Hilbert space. In this context, we focus on the difference in logarithmic negativities, and compare it with the corresponding, previously known, difference in mutual information.

To this end, we consider the following pure maximally entangled state for the full algebra
\be
|\Psi\rangle = \frac{1}{\sqrt{|G|}} \sum_{g \in G} |g\rangle \otimes |g\rangle= \frac{1}{\sqrt{|G|}} \sum_{g \in G} |g,g\rangle 
\ee
which is invariant under the symmetry
\be
  U(h)|g_1, g_2\rangle = |h g_1, h g_2\rangle,\quad \forall h \in G \,.
\ee
 The global density matrix
 \be 
\rho = \frac{1}{|G|} 
\sum_{g,g'} |g, g \rangle \langle g', g'|
\ee
corresponds to a thermofield double in the high-temperature regime, where the two identical copies mimic the subalgebras associated with the complementary regions in the physical setup of interest.

The reduced density matrices for each subsystem, denoted here by the subscripts $1$ and $2$ respectively, are
\begin{equation}
\rho_1 = \rho_2 = \frac{1}{|G|} \sum_{g} |g \rangle \langle g|
\end{equation}
from which the mutual information results
\begin{equation}
I_{\mathcal{F}} = 2 \log |G|
\end{equation}
and the same holds for the R\'enyi mutual information for any index \( n \)
\begin{equation}
I^{(n)}_{\mathcal{F}} = \frac{2}{1-n} \log |G|^{1-n} 
= 2 \log |G| \,.
\end{equation}

For the logarithmic negativity, we use $I^{(n)} = 2 S^{(n)}_{i}$ and the relation valid for pure states given by $ \mathcal{E} = S_i^{(1/2)}$ on one of the subsystems ($i=1,2$), finding that
\begin{equation}
\mathcal{E}_{\cal{F}} = \log |G| \,.
\end{equation}

We now turn to the neutral algebra model. In this case, the density matrix is
\begin{equation}
\rho = \frac{1}{|G|} \sum_{g, g'} |g \rangle \langle g| \otimes |g' \rangle \langle g'|
\end{equation}
which describes the global mixed, separable state given by
\begin{equation}
\rho_{g_1 g_1', g_2 g_2'} = \frac{1}{|G|} 
\, \delta_{g_1 g_1'} \delta_{g_1' g_2} \delta_{g_2 g_2'} \,.
\end{equation}
As a result, the von Neumann entropies are
\begin{equation}
S_{\mathcal{O},12} = S_{\mathcal{O},1} = S_{\mathcal{O},2} = \log |G|
\end{equation}
and the mutual information reads
\begin{equation}
I_{\mathcal{O}} = \log |G| \,.
\end{equation}
In contrast to the previous case, the logarithmic negativity here vanishes. Since the global state is neither pure nor entangled, by construction, it follows that
\begin{equation}
\mathcal{E}_{\mathcal{O}} = 0\,.
\end{equation}

We can now compute the following differences
\begin{align}
\Delta I &= I_{\mathcal{F}} - I_{\mathcal{O}} = \log |G| 
\\
\Delta \mathcal{E} &= \mathcal{E}_{\mathcal{F}} - \mathcal{E}_{\mathcal{O}} = \log |G| \,.
\label{eq:topological_term_negativity}
\end{align}
Hence, both the negativity and the mutual information exhibit the same topological contribution, a result that we explicitly confirm for the current in the following sections.

\section{Entanglement negativity for the chiral current}
\label{sec:neg-cc-2int}
In this section, we explore the logarithmic negativity 
for the chiral current 
and the corresponding moments of the partial transpose,  
in connection with the Rényi mutual information. 
Our setup is designed to probe models that exhibit violations of Haag duality, such as the chiral current QFT. Building on previous results for the moments of the reduced density matrix $\rho_V$ of the current \cite{Arias:2018tmw}, we compute the moments of the partial transpose of $\rho_V$ for a generic integer power and then 
obtain the logarithmic negativity through the proper analytic continuation of the sequence made by the moments corresponding to even powers. 
We also quantitatively analyse both the short and long distance regimes of these quantities, for which the negativity displays distinct and characteristic behaviours.

\subsection{Two disjoint intervals}
\label{subsec:momentsPT2twointervals}

Considering the reduced density matrix 
of the union of two disjoint intervals $V_1 \cup V_2$
in a chiral CFT on the line and in its ground state,
we assume that the moments of its partial transpose w.r.t. $V_2$
take the following form
\begin{equation}
\label{moments-rhoA-T2-current-explicit}
\tr \!\big( \rho_V^{\textrm{\tiny $\Gamma_2$}}\big)^{n} 
=
c_n^2 \left( \frac{\epsilon^2}{\ell_1\, \ell_2\,(1-\eta) } \right)^{\Delta_n} 
\! \mathcal{G}_n(\eta)
\end{equation}
in terms of (\ref{Delta-n-def}) and of the cross ratio 
(\ref{cross-ratio-eta-def}), and where 
the function $\mathcal{G}_n(\eta)$ is model dependent. 
The r.h.s. of (\ref{moments-rhoA-T2-current-explicit}) corresponds to a four-point function of a primary field with scaling dimension $\Delta_n$ \cite{Calabrese:2012ew,Calabrese:2012nk}.
Plugging (\ref{moments-rhoA-T2-current-explicit}) 
into (\ref{neg-moments-def}) first, 
and then taking the replica limit (\ref{neg-replica-limit}),
we find that the logarithmic negativity 
is the UV finite function of the cross ratio given by 
\begin{equation}
    \label{neg-replica-limit-current-2intervals}
\mathcal{E} \,= 
\lim_{n_{\textrm{\tiny e}} \to 1} 
\log\!\big[ \mathcal{G}_{n_{\textrm{\tiny e}}}(\eta) \big] \,.
\end{equation}

Following \cite{Calabrese:2012ew,Calabrese:2012nk}, 
the partial transposition with respect to $V_2$ is defined
by exchanging the endpoints of $V_2$, 
namely through the swapping $u_2 \leftrightarrow v_2$.
In the cross ratio (\ref{cross-ratio-eta-def}),
this induces the transformation 
$\eta \mapsto \tilde{\eta} \equiv \eta/(\eta -1)$;
hence $\tilde{\eta} \in (-\infty, 0)$ as $\eta \in (0,1)$.
In particular, we impose that 
the moments of the partial transpose of the reduced density matrix 
are related to the moments of the reduced density matrix as follows
\begin{equation}
    \label{rhoAT2-rhoA-relation-2intervals}
    \tr \!\big( \rho_V^{\textrm{\tiny $\Gamma_2$}}\big)^{n} 
    =
    \big(
    \tr  \rho_{V}^{n} 
    \,\big)\big|_{u_2 \leftrightarrow v_2}
\end{equation}
where the exchange $u_2 \leftrightarrow v_2$ 
does not change the sign of the length of $V_2$.
Combining (\ref{moments-rhoA-current-explicit}) and (\ref{moments-rhoA-T2-current-explicit})
in (\ref{rhoAT2-rhoA-relation-2intervals}), 
one finds that 

\begin{equation}
\label{Gn-eta-explicit}
    \mathcal{G}_n(\eta)
    \,=\,
\left(\frac{1-\eta }{1 - \tilde{\eta}}\right)^{\Delta_n}
\!\mathcal{F}_n(\tilde{\eta})
\,=\,
(1-\eta)^{2\Delta_n}\,
\mathcal{F}_n\big(\eta/(\eta-1)\big) \,.
\end{equation}
From (\ref{moments-rhoA-current-explicit}), 
(\ref{moments-rhoA-T2-current-explicit}) and (\ref{Gn-eta-explicit}), 
the ratio (\ref{Rn-ratio-def}) in this setup reads
\begin{equation}
\label{Rn-ratio-Fn-explicit}
    R_n 
    =
    \frac{\mathcal{G}_n(\eta)}{\mathcal{F}_n(\eta)}
    = 
    (1-\eta)^{2\Delta_n}\,
    \frac{\mathcal{F}_n\big(\eta/(\eta-1)\big)}{\mathcal{F}_n(\eta)}
\end{equation}
which is a UV finite expression depending only on the cross ratio.

In the specific chiral CFT model given by the free chiral current, $\mathcal{F}_n(\eta)$ is given by (\ref{mathcal-Fn-def});
hence we can write explicit expressions for (\ref{Gn-eta-explicit}) 
and (\ref{Rn-ratio-Fn-explicit}).

We find it worth considering the special case of $n=2$ first,
where (\ref{mathcal-Fn-def}) can be expressed in terms of the 
complete elliptic integral of the first kind $K(z)$. 
In particular, we have that
\begin{equation}
\label{n=2-case-EllipticK}
F_{1/2}(\eta) = \frac{2}{\pi}\, K(\eta)
    \;\;\;\qquad\;\;\;
F_{1/2}(\tilde{\eta}) = \frac{2}{\pi}\, \sqrt{1-\eta}\;K(\eta) \,.
\end{equation}
By employing these expressions in (\ref{mathcal-Fn-def}), 
we find that (\ref{Rn-ratio-Fn-explicit}) drastically simplifies to 
\begin{equation}
\label{R2-identity}
R_2 = 1
\end{equation}
identically for $\eta\in(0,1)$. 
Within the approach to entanglement negativity based on the branch-point twist fields, this identity is a straightforward consequence of the characteristic property of these twist fields, as discussed in \cite{Calabrese:2012ew,Calabrese:2012nk}.
Thus, the validity of (\ref{R2-identity}) in the chiral current model indicates that the method based 
on the branch-point twist fields could be extended 
also to a chiral CFT.

For a generic integer value of the integer index $n>0$, 
by specializing (\ref{Gn-eta-explicit}) 
to the case of the chiral current, 
where (\ref{mathcal-Fn-def}) holds, 
we obtain an expression for $\mathcal{G}_n(\eta)$ 
that can be written in a form 
highlighting the role of the parity of $n$.
This can be done by using the following identity 
\begin{equation}
\label{tilde-F-def}
    F_{r}(\tilde{\eta}) 
    \, =\, 
    (1-\eta)^{r} \, \widetilde{F}_{r}(\eta)
\;\;\;\;\qquad\;\;\;
    \widetilde{F}_{r}(\eta) \equiv \, 
{}_2 F_1(r,r;1;\eta)
\end{equation}
which becomes the identity 
in the second expression of (\ref{n=2-case-EllipticK}) 
in the special case of $k/n = 1/2$
because $\widetilde{F}_{1/2}(\eta) = F_{1/2}(\eta)$.
In particular, 
for odd integers $n_{\textrm{\tiny o}} > 1$ we find
\begin{equation}
\label{G-n-odd-explicit-product}
    \mathcal{G}_{n_{\textrm{\tiny o}}}(\eta)   
\,=\,
\frac{(1-\eta)^{2\Delta_{n_{\textrm{\tiny o}}}}}{\sqrt{\prod_{k=1}^{n_{\textrm{\tiny o}}-1} F_{k/n_{\textrm{\tiny o}}}(\tilde{\eta})}}
\,=\,
    \frac{(1-\eta)^{2\Delta_{n_{\textrm{\tiny o}}}}}{\prod_{k=1}^{(n_{\textrm{\tiny o}}-1)/2} 
    F_{k/n_{\textrm{\tiny o}}}(\tilde{\eta})
    }
    \,=\,
    \frac{(1-\eta)^{\Delta_{n_{\textrm{\tiny o}}} /2}}{\prod_{k=1}^{(n_{\textrm{\tiny o}}-1)/2} 
    \widetilde{F}_{k/n_{\textrm{\tiny o}}}(\eta) }
\end{equation}
while for even integers $n_{\textrm{\tiny e}} > 2$ we get
\begin{equation}
\label{G-n-even-explicit-product}
    \mathcal{G}_{n_{\textrm{\tiny e}}}(\eta)
\,=\,
    \frac{(1-\eta)^{2\Delta_{n_{\textrm{\tiny e}}}}}{
    \sqrt{\prod_{k=1}^{n_{\textrm{\tiny e}}-1} 
    F_{k/n_{\textrm{\tiny e}}}(\tilde{\eta})}}
    \,=\,
    \frac{(1-\eta)^{2\Delta_{n_{\textrm{\tiny e}}}}}{
    \sqrt{F_{1/2}(\tilde{\eta})}\;
    \prod_{k=1}^{n_{\textrm{\tiny e}}/2-1} 
    \! F_{k/n_{\textrm{\tiny e}}}(\tilde{\eta})
    }
\,=\,
    \frac{\sqrt{\pi/2}\;
    (1-\eta)^{\Delta_{n_{\textrm{\tiny e}}/2}}
    }{
    \sqrt{K(\eta)}\;
    \prod_{k=1}^{n_{\textrm{\tiny e}}/2-1} 
    \!\widetilde{F}_{k/n_{\textrm{\tiny e}}}(\eta)
    }
\end{equation}
where we have first employed the identity 
${}_2 F_1(\alpha,\beta;\gamma;\eta) = {}_2 F_1(\beta,\alpha;\gamma;\eta)$,
then the fact that $F_{\theta}(\tilde{\eta}) > 0$ 
when $0< \theta < 1$ and $\eta \in (0,1)$,
and finally also the identity (\ref{tilde-F-def}).
In (\ref{G-n-even-explicit-product}) we used that 
the integer label $k$ in the product occurring in the definition of $\mathcal{F}_n$ in (\ref{mathcal-Fn-def}) 
takes also the value $n_{\textrm{\tiny e}}/2$.

As for the UV finite ratio (\ref{Rn-ratio-Fn-explicit}),
by employing (\ref{mathcal-Fn-def}), (\ref{n=2-case-EllipticK}), 
(\ref{G-n-odd-explicit-product}) and (\ref{G-n-even-explicit-product}),
we find 
\begin{equation}
\label{Rn-CFT-2intervals-explicit-2F1}
   R_n = 
   \left\{
\begin{array}{ll}
\displaystyle
(1-\eta)^{ \frac{\Delta_{n_{\textrm{\tiny o}}} }{2}}
\prod_{k=1}^{\frac{n_{\textrm{\tiny o}}-1}{2}}
    \frac{F_{k/n_{\textrm{\tiny o}}}(\eta)}{ 
    \widetilde{F}_{k/n_{\textrm{\tiny o}}}(\eta)}
    \hspace{1cm} 
    & 
    n=n_{\textrm{\tiny o}} \;\;\textrm{odd}
    \\
    \rule{0pt}{1.2cm}
\displaystyle
(1-\eta)^{\Delta_{n_{\textrm{\tiny e}}/2} }\,
    \prod_{k=1}^{ \frac{ n_{\textrm{\tiny e}} }{2}-1} 
    \frac{ F_{k/n_{\textrm{\tiny e}}}(\eta) }{
    \widetilde{F}_{k/n_{\textrm{\tiny e}}}(\eta) }
    \hspace{1cm} 
    & 
    n=n_{\textrm{\tiny e}} \;\;\textrm{even} \,.
   \end{array}
   \right.
\end{equation}

In order to obtain the logarithmic negativity through the replica limit 
(\ref{neg-replica-limit-current-2intervals}), 
let us focus on the 
moments where $n= n_{\textrm{\tiny e}}$ is even.
The logarithm of the last expression in (\ref{G-n-even-explicit-product}) reads
\be
\label{log-Gn-even-CFT}
    \log\!\big[\mathcal{G}_{n_{\textrm{\tiny e}}}(\eta)\big]
    \,=\,
    \Delta_{n_{\textrm{\tiny e}}/2} \, \log(1-\eta) 
    - 
    \frac{1}{2}\,
    \log\!\big[2 K(\eta)/\pi\big]
    - 
    \widetilde{D}_{n_{\textrm{\tiny e}}}(\eta)
\ee
where we have introduced the finite sum
\be
\label{tilde-Dn-def-even}
\widetilde{D}_{n_{\textrm{\tiny e}}}(\eta)
\equiv
\sum_{k=1}^{\frac{ n_{\textrm{\tiny e}} }{2}-1} \!
    \log\!\Big[ \widetilde{F}_{k/n_{\textrm{\tiny e}}}(\eta) \Big]
\ee
which can be written conveniently also as 
\be
\label{tilde-Dn-def-even-step-0}
\widetilde{D}_{n_{\textrm{\tiny e}}}(\eta)
=
\sum_{k=0}^{n_{\textrm{\tiny e}} /2} 
    \log\!\Big[ \widetilde{F}_{k/n_{\textrm{\tiny e}}}(\eta) \Big]
    -
    \log\!\Big[ \widetilde{F}_{1/2}(\eta) \Big] \,.
\ee
From  \eqref{moments-rhoA-T2-current-explicit} 
and \eqref{log-Gn-even-CFT}, we have
\begin{equation}
\label{neg-ren-even-cc-extended}
 \mathcal{E}_{ n_{\textrm{\tiny e}}} 
 = 
 -\frac{1}{24}
\left( n_{\textrm{\tiny e}} + \frac{2}{n_{\textrm{\tiny e}}} \right)
\log(1 - \eta)
 -\frac{1}{2}\log \!\big(2K(\eta)/\pi\big)
 -\widetilde{D}_{ n_{\textrm{\tiny e}}}(\eta)
 -\Delta_{ n_{\textrm{\tiny e}}}
 \log \!\big(\ell_1 \ell_2 / \epsilon^2 \big)
 +2\log (\tilde{c}_{ n_{\textrm{\tiny e}}} ) \,.
\end{equation}

In order to perform the analytic continuation 
$n_{\textrm{\tiny e}} \to 1$, 
for the finite sum in the r.h.s.
we can apply the following generalization of the Abel-Plana formula \cite{Saharian:2007ph,WhittakerWatson:ModernAnalysis} 
\begin{eqnarray}
\label{Abel-Plana-formula}
\sum_{k=a}^b f(k)
&= & 
\int_a^b  f(x)\, \textrm{d} x
+
\frac{f(a)+ f(b) }{2}
\nonumber
\\
& &
+\, \textrm{i} 
\int_0^{\infty} 
\frac{
f(a+ \textrm{i}\,z)
-f(a-\textrm{i}\,z)
-f(b+ \textrm{i}\,z)
+f(b-\textrm{i}\,z)}{\textrm{e}^{2 \pi z}-1}\;
\textrm{d} z
\end{eqnarray}
specialised to the case where $a=0$, $b= n_{\textrm{\tiny e}}/2$ and 
$f(k)= \log\!\big[ \widetilde{F}_{k/n_{\textrm{\tiny e}}}(\eta) \big]$. 
In the generalized Abel–Plana formula (\ref{Abel-Plana-formula}), the function $f(z)$ is assumed to be meromorphic in a horizontal strip of the complex plane
defined by $a \leqslant  \textrm{Re}(z) \leqslant b$ 
and to satisfy suitable analyticity and decay conditions that guarantee convergence of both the integrals and any residue contributions. In particular, $f(z)$ must be analytic except for isolated poles within the strip, with residues that decay rapidly enough to make their sum finite. 
Along the boundaries of the strip, $f(z)$ 
should be continuous and  it must vanish 
as $|\textrm{Im}(z)| \to \infty$ sufficiently fast 
to ensure the convergence of integrals 
involving $f(a \pm \ri z)$ and $f(b \pm \ri z)$.
In our case, the function $f(z)$ does not satisfy all the requirements of this formula because the function $f(z)$ has a countable set of branch points, whose density depends on the value of $\eta$. 
Since we were unable to prove that \eqref{Abel-Plana-formula} can be applied also in our case, 
we just assume that it still holds in our setup.

By employing (\ref{Abel-Plana-formula})
with $\widetilde{F}_{1/2}(\eta) = F_{1/2}(\eta)$ 
combined with the first formula in (\ref{n=2-case-EllipticK}),
we can write (\ref{tilde-Dn-def-even-step-0}) as follows
\bea
 \label{Dtilden-after-AP}
 \widetilde{D}_{n_{\textrm{\tiny e}}}(\eta)
    &=&
 -\frac{1}{2}\,
    \log\!\big[2 K(\eta)/\pi\big]
    +
    \int_0^{\frac{n_{\textrm{\tiny e}}}{2}}
    \log\!\Big[ \widetilde{F}_{s/n_{\textrm{\tiny e}}}(\eta) \Big]
    \,\textrm{d}s
    \nn
    \\
    \rule{0pt}{.7cm}
    & &
    \rule{0pt}{.8cm}
-\textrm{i} 
\int_0^{\infty} 
\frac{\coth(\pi s) - 1}{2}
\;\bigg\{
\log\!\Big[ \widetilde{F}_{1/2+\textrm{i}s/n_{\textrm{\tiny e}}}(\eta) \Big]
-
\log\!\Big[ \widetilde{F}_{1/2-\textrm{i}s/n_{\textrm{\tiny e}}}(\eta) \Big]
\nn
\\
    & &
    \rule{0pt}{.3cm}
    \hspace{5.cm}
-\,    
    \log\!\Big[ \widetilde{F}_{\textrm{i}s/n_{\textrm{\tiny e}}}(\eta) \Big]
+
\log\!\Big[ \widetilde{F}_{-\textrm{i}s/n_{\textrm{\tiny e}}}(\eta) \Big]
 \bigg\}\;
\textrm{d}s
\hspace{1cm} 
\eea
where the principal value of the logarithm is assumed, 
following the Mathematica software that we have employed 
throughout our analysis.
Changing the integration variable $s \mapsto n_{\textrm{\tiny e}}\, s$ 
in (\ref{Dtilden-after-AP}) first, and then plugging the result back into (\ref{log-Gn-even-CFT}), 
we find
\bea
\label{log-Gn-even-CFT-AbelPlana-original}
    \log\!\big[\mathcal{G}_{n_{\textrm{\tiny e}}}(\eta)\big]
    &=&
    \Delta_{n_{\textrm{\tiny e}}/2} \, \log(1-\eta) 
    - 
    \,n_{\textrm{\tiny e}}
    \int_0^{\frac{1}{2}}
    \log\!\Big[ \widetilde{F}_{s}(\eta) \Big]
    \,\textrm{d}s
    \\
    \rule{0pt}{.7cm}
    & &
    \rule{0pt}{.8cm}
+\textrm{i} \,n_{\textrm{\tiny e}}
\int_0^{\infty} 
\frac{\coth(\pi n_{\textrm{\tiny e}} s) - 1}{2}
\;\bigg\{
\log\!\Big[ \widetilde{F}_{1/2+\textrm{i}s}(\eta) \Big]
-
\log\!\Big[ \widetilde{F}_{1/2-\textrm{i}s}(\eta) \Big]
\nn
\\
    & &
    \rule{0pt}{.3cm}
    \hspace{5.5cm}
-\,    
    \log\!\Big[ \widetilde{F}_{\textrm{i}s}(\eta) \Big]
+
\log\!\Big[ \widetilde{F}_{-\textrm{i}s}(\eta) \Big]
 \bigg\}\;
\textrm{d}s
\nn
\eea
whose r.h.s. is real because
the combination of terms within the curly brackets
in the second integrand is purely imaginary. 

A slight manipulation of \eqref{log-Gn-even-CFT-AbelPlana-original} 
leads to a form that is simpler to evaluate numerically in Mathematica.
As for the choice of the branch for the logarithm of a complex number $z$, we consider the principal branch and define 
$\log z\equiv  \log |z|+\ri \,\text{Arg}(z)$,
where $\text{Arg}(z)$ denotes the principal value of the argument of $z$
and $\text{Arg}(z) \in (-\pi,\pi)$.
Then, for a complex valued function $g(z)$ we have that
$\log g(z)-  \log \overline{g(z)}=2\ri \,\theta(z)$, 
where $\theta(z)$ is provided by the decomposition of 
$g(z)=r(z) \,e^{\textrm{i}\theta(z)}$ in polar coordinates,
with $r(z) \geqslant 0$ and $\theta(z)\in [-\pi,\pi)$,
and it is independent of the choice of the principal branch. 
By using $\log g(z)-  \log \overline{g(z)}=2\ri \,\text{Arg}[g(z)]$ 
with $\text{Arg}(z)\in(-\pi,\pi)$ in the special case where
$g(z)=\widetilde{F}_{z}(\eta)$, we can write 
\eqref{log-Gn-even-CFT-AbelPlana-original} as follows
\bea
\label{log-Gn-even-CFT-AbelPlana-original-arg}
    \log\!\big[\mathcal{G}_{n_{\textrm{\tiny e}}}(\eta)\big]
    &=&
    \Delta_{n_{\textrm{\tiny e}}/2} \, \log(1-\eta) 
    - 
    \,n_{\textrm{\tiny e}}
    \int_0^{\frac{1}{2}}
    \log\!\Big[ \widetilde{F}_{s}(\eta) \Big]
    \,\textrm{d}s
    \\
    \rule{0pt}{.6cm}
    & &
-\,n_{\textrm{\tiny e}}
\int_0^{\infty} 
\big[ \coth(\pi n_{\textrm{\tiny e}} s) - 1 \,\big] 
\;\bigg\{
 \text{Arg}\Big[ \widetilde{F}_{1/2+\textrm{i}s}(\eta) \Big]
 - \text{Arg}\Big[ \widetilde{F}_{\textrm{i}s}(\eta) \Big]
 \bigg\}\;
\textrm{d}s \,.
\nn
\eea

Taking the replica limit 
(\ref{neg-replica-limit-current-2intervals}) 
in (\ref{log-Gn-even-CFT-AbelPlana-original}), 
we obtain the following real and analytic expression 
for the logarithmic negativity 
\bea
\label{log-neg-from-AbelPlana}
    \mathcal{E}
    &=&
    -\frac{1}{8} \, \log(1-\eta) 
    - 
    \int_0^{\frac{1}{2}}
    \log\!\Big[ \widetilde{F}_{s}(\eta) \Big]
    \,\textrm{d}s
    \\
    \rule{0pt}{.7cm}
    & &
    \rule{0pt}{.8cm}
+\textrm{i} 
\int_0^{\infty} 
\frac{\coth(\pi  s) - 1}{2}
\;\bigg\{
\log\!\Big[ \widetilde{F}_{1/2+\textrm{i}s}(\eta) \Big]
-
\log\!\Big[ \widetilde{F}_{1/2-\textrm{i}s}(\eta) \Big]
\nn
\\
    & &
    \rule{0pt}{.3cm}
    \hspace{5.5cm}
-\,    
    \log\!\Big[ \widetilde{F}_{\textrm{i}s}(\eta) \Big]
+
\log\!\Big[ \widetilde{F}_{-\textrm{i}s}(\eta) \Big]
 \bigg\}\;
\textrm{d}s
\nn
\eea
which is the main result of this manuscript.

\begin{figure}[t!]
    \centering
        \includegraphics[width=1\textwidth]{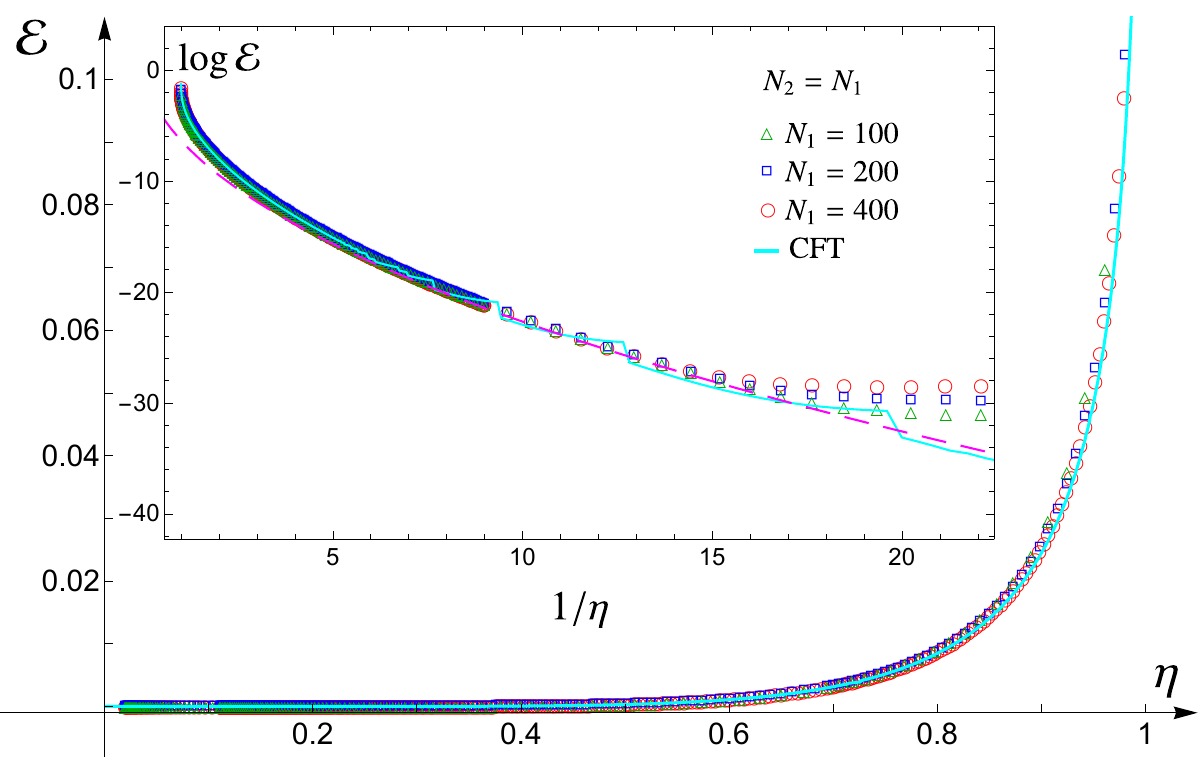}
    \caption{
    Logarithmic negativity for two disjoint intervals on the line.
    The solid cyan line corresponds to (\ref{log-neg-from-AbelPlana}). 
    The data points are obtained from the lattice model, 
    as discussed in Sec.\,\ref{renyisvscorrelators}, 
    for two equal blocks 
    (each of them made by $N_1$ consecutive sites) 
    with varying distance. 
    The inset highlights the large separation regime $\eta \to 0^+$, where an exponential decay \eqref{eq:EN_decay_exp} occurs (magenta dashed line). 
    } 
    \label{fig:EnegAlmostAdj_line}
\end{figure}

\begin{figure}[t!]
    \centering
      \subfigure{%
        \includegraphics[width=0.48\textwidth]{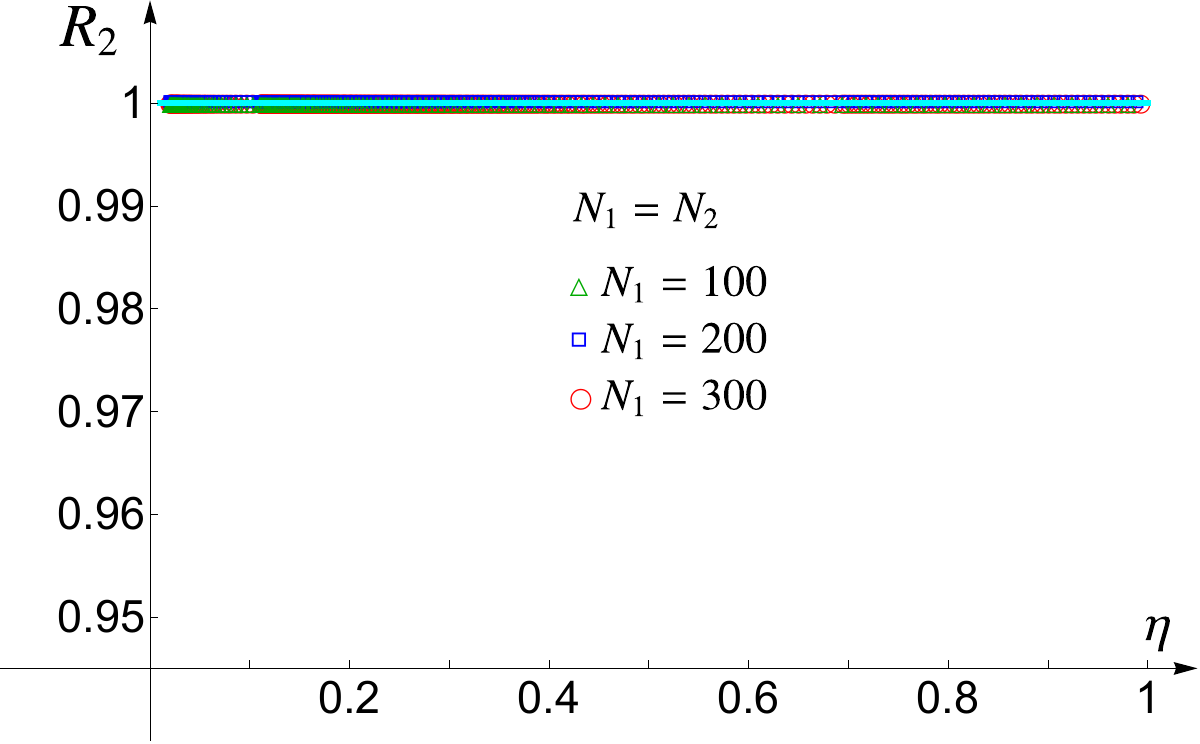}
        \label{fig:Rn2}
    }
    \hfill
          \subfigure{%
        \includegraphics[width=0.48\textwidth]{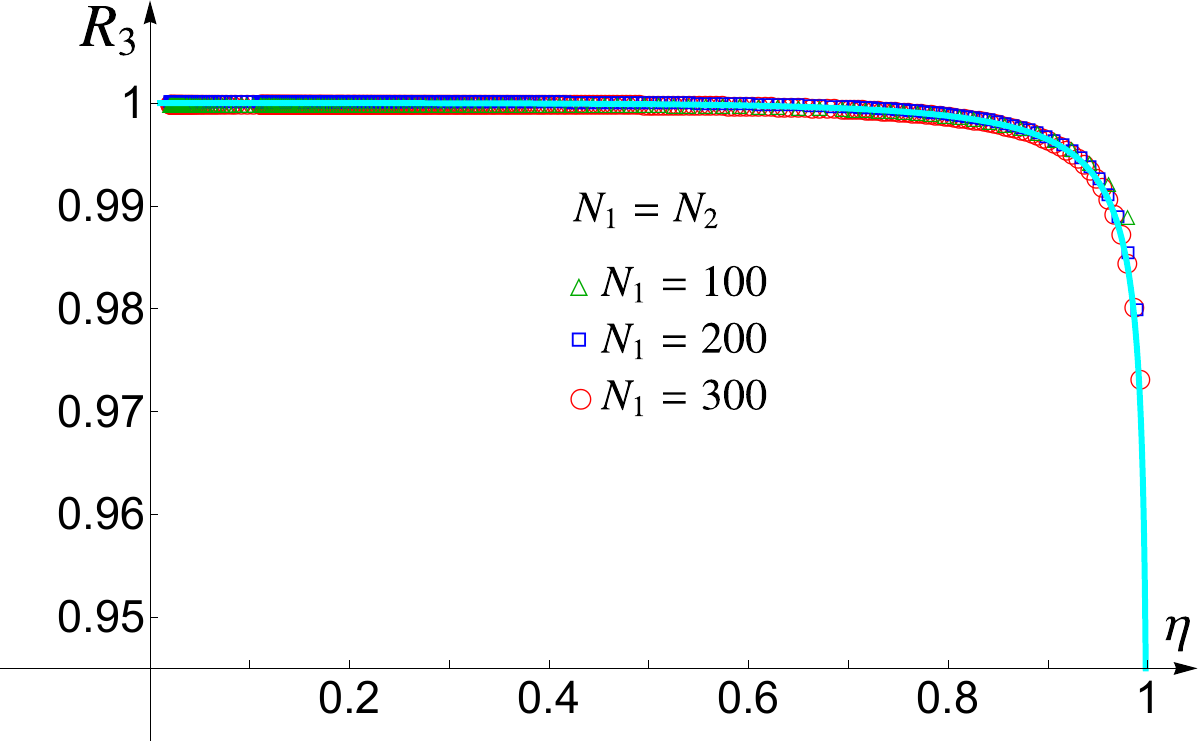}
        \label{fig:Rn3}
    }
    \hfill
    \subfigure{%
        \includegraphics[width=0.48\textwidth]{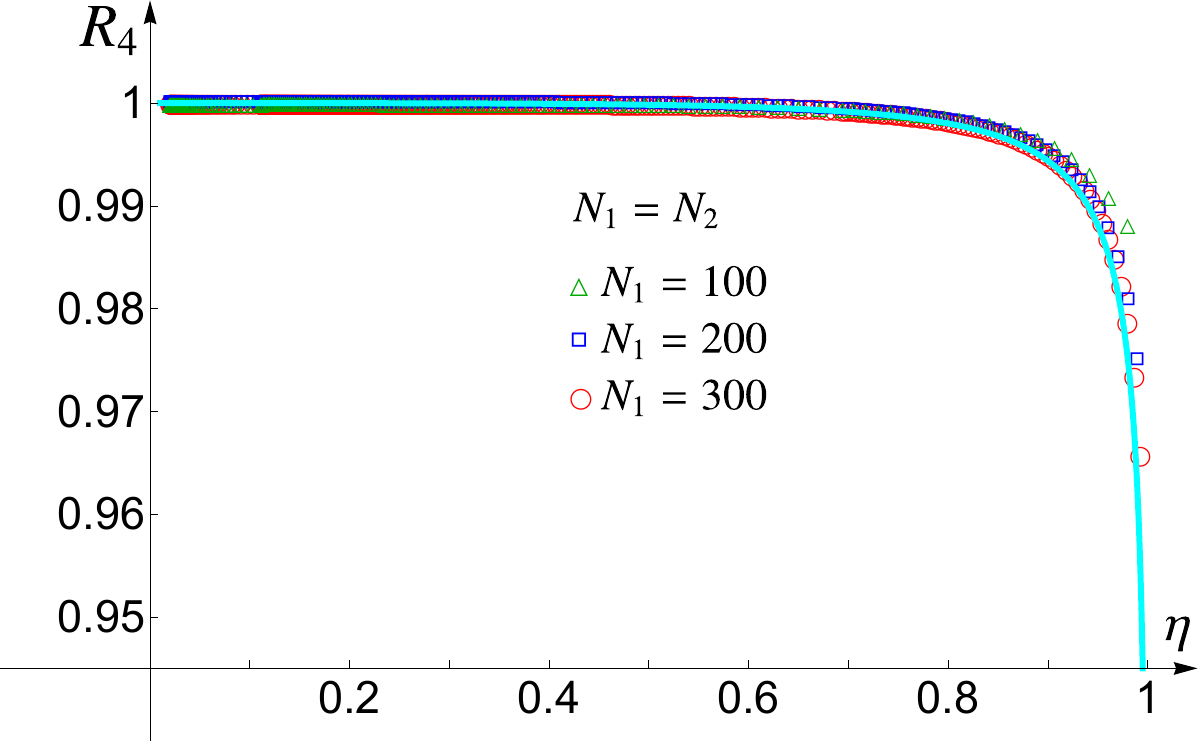}
        \label{fig:Rn4}
    }
    \hfill
    \subfigure{%
        \includegraphics[width=0.48\textwidth]{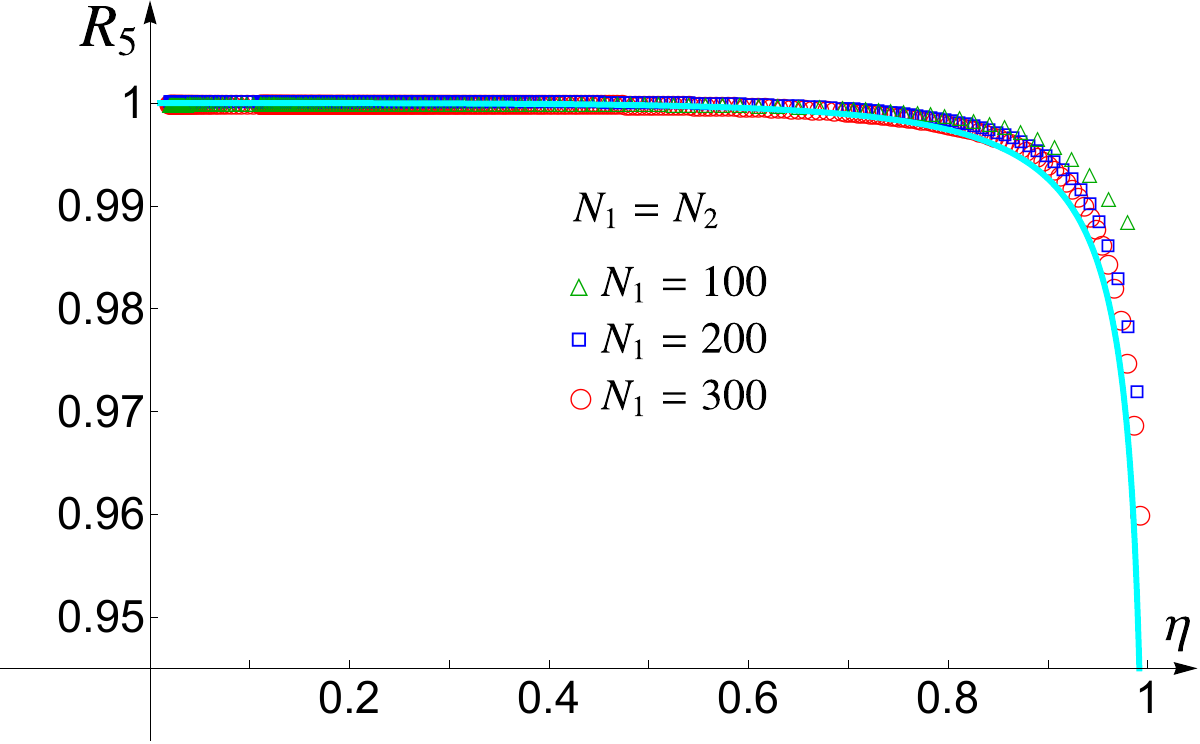}
        \label{fig:Rn5}
    }
\caption{Disjoint blocks in the line:
The ratio $R_n $ in (\ref{Rn-ratio-def}) for $n\in\{2,3,4,5\}$. 
The solid lines correspond to 
\eqref{Rn-CFT-2intervals-explicit-2F1}.
The data points are obtained from the lattice model for the chiral current (see Sec.\,\ref{renyisvscorrelators}), 
for two equal blocks and as described in the caption of 
Fig.\,\ref{fig:EnegAlmostAdj_line}. 
}
    \label{fig:LineRn}
\end{figure}

In Fig.\,\ref{fig:EnegAlmostAdj_line}, we compare the analytic expression (\ref{log-neg-from-AbelPlana}) for the logarithmic negativity against the numerical results obtained from the lattice model, as discussed in detail in Sec.\,\ref{sec:negativity_numerics}.
An excellent agreement is obtained in the whole domain of $\eta \in (0,1)$. We remark that also the exponential decay occurring in the large separation distance regime $\eta \to 0$ is nicely described,
as highlighted in the inset of Fig.\,\ref{fig:EnegAlmostAdj_line} and further discussed in Sec.\,\ref{subsec:two-intervals-large-distance}.

Notice that \eqref{log-Gn-even-CFT-AbelPlana-original} 
is the simplest expression that we have obtained
leading to the nice match with the lattice data points
shown in Fig.\,\ref{fig:EnegAlmostAdj_line}.
Trying to get simpler forms 
is a delicate procedure where the cuts of the hypergeometric functions play a crucial role
(indeed, e.g. the logarithm of the product of 
two complex numbers cannot be written 
as the sum of their logarithms).
Because of cuts in the hypergeometric functions, 
also the application of the generalised Abel-Plana formula
(\ref{Abel-Plana-formula}) needs some care.
In particular, its application to (\ref{log-Gn-even-CFT}),
which comes from the last expression in 
(\ref{G-n-even-explicit-product}), 
provides the result (\ref{log-neg-from-AbelPlana}) that matches with the numerical results from the lattice model, 
as already mentioned above; while, instead, 
taking the logarithm of the first expression in (\ref{G-n-even-explicit-product}) and then applying (\ref{Abel-Plana-formula}) 
leads to a curve that does not match with the lattice data points.

Notice that the analytic expression (\ref{log-neg-from-AbelPlana})
involves the numerical evaluation of integrals 
containing functions with singularities as $s \simeq 0$, 
coming from  $\tfrac{\coth(\pi  s) - 1}{2}=(e^{2\pi s}-1)^{-1}$. For $s \gg 1$, 
the same factor exponentially suppresses the integrand, rendering its contribution extremely small and therefore more  sensitive to numerical errors. Moreover, the occurrence of branch cuts in the integrands can introduce additional inaccuracies if the numerical integration routine does not correctly handle the associated phase discontinuities when crossing them. Our numerical checks confirm that, as $\eta \to 0$, the distribution of branch cuts in the arguments of the hypergeometric functions in  
\eqref{log-neg-from-AbelPlana} becomes increasingly dense, which likely accounts for the larger discrepancies between the analytical and lattice results in this regime.

In Fig.\,\ref{fig:LineRn}, we consider the ratio 
$R_n $ in (\ref{Rn-ratio-def}), for the same setup of  
Fig.\,\ref{fig:EnegAlmostAdj_line}.
The data points show the numerical results obtained in the lattice model for the chiral current discussed in 
Sec.\,\ref{sec:negativity_numerics}, 
while the solid lines are given by the analytic expression 
(\ref{Rn-CFT-2intervals-explicit-2F1}) in the continuum. 
An excellent agreement is obtained in the whole domain of $\eta \in (0,1)$, although the increasing of the finite size effects 
with the R\'enyi index $n$ is evident.

To the best of our knowledge, the expression 
(\ref{log-neg-from-AbelPlana}) is the first analytic result in CFT for the logarithmic negativity 
obtained from the partial transpose 
valid in the whole range of the cross ratio $\eta \in (0,1)$
(hence the fermionic partial time-reversal 
in fermionic models should not be considered).
Indeed, for the massless scalar field in the decompactification limit,
an analytic expression for the logarithmic negativity has been found only in the regime of small separation distance between the intervals \cite{Calabrese:2012ew, Calabrese:2012nk}.
For other simple CFT models like 
the compactified massless scalar \cite{Calabrese:2012ew, Calabrese:2012nk},
the free fermion \cite{Coser:2015eba}
and the Ising model \cite{Calabrese:2013mi,Alba:2013mg},
only the moments of the partial transpose are known in a closed form. 
Whenever the moments of the partial transpose are known analytically, the logarithmic negativity 
can be obtained numerically through an extrapolation method
\cite{Agon:2013iva,DeNobili:2015dla}.

\subsection{Large separation distance limit}
\label{subsec:two-intervals-large-distance}

It is worth studying the limiting regime of large separation distance 
(i.e. the limit $\eta \to 0^+$)
for the main analytic results obtained in Sec.\,\ref{subsec:momentsPT2twointervals}.

Considering (\ref{log-Gn-even-CFT}), 
one observes that its r.h.s. vanishes 
for $\eta = 0$, by using also that $K(0) = \pi/2$.
The Taylor expansion of (\ref{log-Gn-even-CFT})
as $\eta \to 0^+$ reads
\begin{equation}
\label{log-G-expansion-small-eta}
    \log\!\big[\mathcal{G}_{n_{\textrm{\tiny e}}}(\eta)\big]
    \,=\,
    \sum_{r=1}^{+\infty} p_r(n_{\textrm{\tiny e}})\, \eta^r
\end{equation}
where the coefficients of the first terms are given by 
\begin{equation}
\label{log-G-expansion-small-eta-p1}
    p_1(n_{\textrm{\tiny e}})
    =
    -\frac{n_{\textrm{\tiny e}}^2 - 1}{12\, n_{\textrm{\tiny e}}}
\;\;\;\;\;\qquad\;\;\;\;\;
        p_2(n_{\textrm{\tiny e}})
    =
    -\frac{
        9\,n_{\textrm{\tiny e}}^4 
        - 10\,n_{\textrm{\tiny e}}^2 
        + 1
        }{
        240\, n_{\textrm{\tiny e}}^3}
\end{equation}
and
\begin{eqnarray}
        p_3(n_{\textrm{\tiny e}})
    &=&
    -\frac{
        179\,n_{\textrm{\tiny e}}^6 
        - 210\,n_{\textrm{\tiny e}}^4
        + 21\,n_{\textrm{\tiny e}}^2
        + 10
        }{
        7560\, n_{\textrm{\tiny e}}^5}
\\
\rule{0pt}{.8cm}
        p_4(n_{\textrm{\tiny e}})
    &=&
    -\frac{
        4147\,n_{\textrm{\tiny e}}^8
        - 5040\,n_{\textrm{\tiny e}}^6
        + 462\,n_{\textrm{\tiny e}}^4
        + 200\,n_{\textrm{\tiny e}}^2
        + 231
        }{
        241920\, n_{\textrm{\tiny e}}^7}
    \\
\rule{0pt}{.8cm}
        p_5(n_{\textrm{\tiny e}})
    &=&
        -\frac{
        8893\,n_{\textrm{\tiny e}}^{10}
        - 11088\,n_{\textrm{\tiny e}}^8
        + 924\,n_{\textrm{\tiny e}}^6
        + 242\,n_{\textrm{\tiny e}}^4
        + 231\,n_{\textrm{\tiny e}}^2
        + 798
        }{
        665280\, n_{\textrm{\tiny e}}^9}
\end{eqnarray}
and the ones corresponding to $r \geqslant 6$ 
can be easily obtained.

We have computed $p_r(n_{\textrm{\tiny e}})$ 
for high values of $r$ (up to $r=20$), finding that its generic form 
is $p_r(n_{\textrm{\tiny e}}) \propto 
\tilde{p}_{2r}(n_{\textrm{\tiny e}}) / n_{\textrm{\tiny e}}^{2r-1}$, where  $\tilde{p}_{2r}(n_{\textrm{\tiny e}})$ 
is a polynomial of degree $2r$ 
having integer coefficients 
such that $\tilde{p}_{2r}(1) =0 $.
It would be useful to have a closed formula for $p_r(n_{\textrm{\tiny e}})$ valid for a generic $r \geqslant 1$, in order to check this property at any order in the expansion (\ref{log-G-expansion-small-eta}).
Assuming that this property holds, 
since the logarithmic negativity is obtained through the replica limit (\ref{neg-replica-limit-current-2intervals}) of (\ref{log-Gn-even-CFT}), we can conclude that $\mathcal{E}$ vanishes faster than any power law.
This feature distinguishes the logarithmic negativity 
from the mutual information, 
which exhibits the power law decay 
in (\ref{expansion-MI-eta0}).
This discussion leads us to observe that, 
as for the moments of the partial transpose 
and the logarithmic negativity, 
the two procedures of taking the large separation distance expansion 
and the replica limit do not commute. 
This is in contrast with the case of the R\'enyi entropies for two intervals and of the mutual information, 
which has been discussed in 
Sec.\,\ref{sec-renyi-entropies-two-intervals}.

Within the approach based on branch-point twist fields, 
the decay of $\mathcal{E}$ faster than any power law at large separation distance 
has been highlighted for any CFT in \cite{Calabrese:2012nk},
by exploiting the generalised operator product expansion introduced in \cite{Calabrese:2010he} 
and observing that, after the replica limit, 
only one-point functions remain,
which  vanish identically on the complex plane.
However, an analytic expression for the decay 
of the logarithmic negativity has not been found
within this approach. 
The exponential nature of this decay is known only 
through numerical computations in few lattice models
\cite{Marcovitch:2008sxc, Wichterich:2008vfx, Calabrese:2012nk, Calabrese:2013mi,Klco:2021cxq,Klco:2020rga, Parez:2022ind, Parez:2023xpj}.

In the case of the chiral current that we are investigating, 
the exponential decay of the logarithmic negativity 
in (\ref{log-neg-from-AbelPlana}) can be obtained as follows. 
Since all the terms in (\ref{log-neg-from-AbelPlana})
are analytic in $\eta$, 
the only possible source of a non analytical contribution 
are the cuts occurring in the logarithm of the functions 
$\widetilde{F}_{1/2+\textrm{i}s}(\eta) $ 
and  $\widetilde{F}_{-\textrm{i}s}(\eta) $.
In order to find this non analytical contribution, 
we isolate in (\ref{log-neg-from-AbelPlana}) 
the following integral
\begin{eqnarray}
 \label{expansion-cut-2F1}
& &
  \textrm{i} \int_0^{\infty}
  \frac{\coth(\pi s)-1}{2}\,
  \bigg\{
\log\!\Big[ \widetilde{F}_{1/2+\textrm{i}s}(\eta) \Big]
-
\log\!\Big[ \widetilde{F}_{1/2-\textrm{i}s}(\eta) \Big]
\bigg\}
\nonumber
 \\
 \rule{0pt}{.7cm}
 & &
  \approx 
  \int_0^{\infty}
  \frac{\pi\,\Theta\big(s-s_0(\eta)\big)}{e^{2\pi s}-1}
  \, \textrm{d}s
  \,\approx \, 
  \pi \int_{s_0(\eta)}^{\infty} 
  e^{-2\pi s} \, \textrm{d}s
  \,=\,
  \frac{e^{-2\pi s_0(\eta)}}{2}
  \end{eqnarray}
where the first approximation in the second line corresponds to the contribution from the first cut
in the limit of $s \gg 1$ and $\eta\rightarrow 0$. 
In order to find $s_0(\eta)$, let us first consider the expansion 
$\widetilde{F}_{r}(\eta) = 1 + r^2 \eta +O(\eta^2)$ as $\eta \to 0^+$ specialised to the case where $r =1/2 + \textrm{i}s$. 
This gives an expansion whose real and imaginary parts read respectively
\begin{equation}
\label{Re-Im-part-tildeF}
  \textrm{Re}\big[\widetilde{F}_{1/2 + \textrm{i}s}(\eta) \big]
  = 1 + \big(1/4 - s^2\big)\eta +O(\eta^2)
  \;\qquad\;
  \textrm{Im}\big[\widetilde{F}_{1/2 + \textrm{i}s}(\eta) \big]
  = s\,\eta +O(\eta^2) \,.
\end{equation}
From the first expansion, one observes that,
for $s \gg 1$, the $O(\eta)$ term becomes comparable with the $O(1)$ term when
\begin{equation}
\eta \approx  \frac{\alpha}{s^2}
\label{alpha}
\end{equation}
for some constant value of $\alpha$, 
where the higher order terms have been discarded. 
Combining the second expansion in (\ref{Re-Im-part-tildeF}) with (\ref{alpha}), we have that 
$\textrm{Im}\big[\widetilde{F}_{1/2 + \textrm{i}s}(\eta) \big]$ vanishes as $s \gg 1$.
The expression (\ref{alpha}) provides an approximation 
for the value of $s$ corresponding to the first cut,
which can be used to find 
the explicit value of $\alpha$ 
by solving the following equation
\begin{equation}
\label{eq-Re2F1-for-alpha}
 \textrm{Re}\big[\widetilde{F}_{1/2 + \textrm{i}s}(\alpha/s^2) \big]
=
 \textrm{Im}\big[
 \widetilde{F}_{1/2 + \textrm{i}s}(\alpha/s^2) 
 \big]
 =0
\;\;\;\qquad\;\;\;\;
s \gg 1
\end{equation}
where the first equality is observed for the case that we are considering, but it does not occur for 
$\widetilde{F}_{a + \textrm{i}s}(\alpha/s^2)$ with $a \neq 1/2$.
Solving (\ref{eq-Re2F1-for-alpha}) numerically, 
for the lowest root we find $\alpha \simeq 1.4458$.
From (\ref{alpha}) and this numerical value for $\alpha$, 
we have that $s_0(\eta) \approx \sqrt{\alpha/\eta}$ 
in (\ref{expansion-cut-2F1}).
Assuming that (\ref{expansion-cut-2F1}) provides the leading contribution in the large separation distance regime of 
(\ref{log-neg-from-AbelPlana}), 
for the logarithmic negativity in the $\eta\rightarrow 0$ limit 
we find 
\begin{equation}
\label{eq:EN_decay_exp}
 \mathcal{E} \approx e^{-2\pi\sqrt\alpha/\sqrt\eta}
 \;\;\;\qquad\;\;\;
 2\pi\sqrt\alpha \simeq 7.555 \,.
 \end{equation}
This result provides the dashed magenta curve in the inset of Fig.\,\ref{fig:EnegAlmostAdj_line}, which nicely describes 
the exponential decay in the regime of large separation distance
of the numerical results 
obtained from the lattice model
as discussed in Sec.\,\ref{sec:negativity_numerics}. 

The procedure described above
(from (\ref{expansion-cut-2F1}) to (\ref{eq:EN_decay_exp}))
for the logarithm of $\widetilde{F}_{1/2+\textrm{i}s}(\eta) $ 
can be adapted to the logarithm of $\widetilde{F}_{-\textrm{i}s}(\eta) $,
by considering the integral coming from the last two terms in (\ref{log-neg-from-AbelPlana}) and applying to it the same approximations employed in (\ref{expansion-cut-2F1}).
This gives $e^{-2\pi \tilde{s}_0(\eta)}/2$, where $\tilde{s}_0(\eta)$ can be obtained as explained above by considering the expansion 
$\widetilde{F}_{r}(\eta)$ as $\eta \to 0^+$ up to $O(\eta^2)$ included, specialised to the case where $r =- \textrm{i}s$.
The same argument reported above adapted to this case leads to 
$\eta\approx \tilde{\alpha}/s^2$, 
with some constant values $\tilde{\alpha}$, 
for the approximate value of $s$ corresponding to the first cut.
The numerical value of $\tilde{\alpha}$ is obtained by solving numerically the condition for the cut, namely  
$ \textrm{Im}\big[ \widetilde{F}_{- \textrm{i}s}(\alpha/s^2) 
 \big] =0$, which gives $\tilde{\alpha} \simeq 6.5936$ 
 for the lowest root.
 This leads to a term $e^{-2\pi\sqrt{\tilde{\alpha}} /\sqrt\eta}$ 
 with $2\pi\sqrt{\tilde{\alpha}} \simeq 16.134$, 
 which is subleading with respect to (\ref{eq:EN_decay_exp}).

Notice that, by adapting the analysis described above to
 (\ref{log-Gn-even-CFT-AbelPlana-original}), 
 for a generic value of the even integer $n_{\textrm{\tiny e}} \geqslant 2$,
 one finds the power law decay given by 
(\ref{log-G-expansion-small-eta}) 
and (\ref{log-G-expansion-small-eta-p1}).

\subsection{Small separation distance limit}
\label{subsec:two-intervals-small-distance}

The limiting regime where the two intervals of given lengths 
become very close to each other corresponds 
to $\eta \to 1^{-}$ in (\ref{cross-ratio-eta-def})
and its analysis is similar to the one reported 
for the massless scalar in \cite{Calabrese:2012nk}. 
In particular, we need that 
\begin{equation}
\label{eta-to-one-asymptotics}
    K(\eta) =-\frac{1}{2}\log(1-\eta) + \log 4+ o(1)
    \;\;\;\qquad\;\;\;
    \widetilde{F}_{r}(\eta) 
    =
\frac{\pi}{\sin(2\pi r)\, \Gamma(1-r)^2\, \Gamma(2r)} + o(1)
\end{equation}
as $\eta \to 1^{-}$, where $r<1/2$.
Since $K(\eta)$ occurs in (\ref{Gn-eta-explicit}) 
only for even values of $n$, a parity effect occurs 
in the asymptotic behaviour of (\ref{moments-rhoA-T2-current-explicit}) 
for small separation distance.
Indeed, combining (\ref{eta-to-one-asymptotics}) with 
(\ref{G-n-odd-explicit-product}) and (\ref{log-Gn-even-CFT}), 
for odd and even $n$ we find respectively
\begin{eqnarray}
\label{log-G-eta-1-expansion-odd}
    \log\!\big[\mathcal{G}_{n_{\textrm{\tiny o}}}(\eta)\big]
    &=&
    \frac{ \Delta_{n_{\textrm{\tiny o}}} }{2} \, \log(1-\eta)  +O(1)
\\
\label{log-G-eta-1-expansion-even}
    \log\!\big[\mathcal{G}_{n_{\textrm{\tiny e}}}(\eta)\big]
    &=&
    \Delta_{n_{\textrm{\tiny e}}/2} \, \log(1-\eta) 
    - 
    \frac{1}{2}
    \log\!\big( \! - \log(1-\eta)\big)
+O(1)
\end{eqnarray}
where the $O(1)$ terms depend on $n$ in a highly nontrivial way. 
In particular, notice that the term 
$\widetilde{D}_{n_{\textrm{\tiny e}}}(\eta)$, 
defined in  (\ref{tilde-Dn-def-even}),
gives a contribution to $O(1)$ in the expansion 
(\ref{log-G-eta-1-expansion-even}). 
We remark that, while the leading divergence is always logarithmic, 
the subleading log-log divergence in (\ref{log-G-eta-1-expansion-even}) 
is independent of $n_{\textrm{\tiny e}}$ 
and does not occur in (\ref{log-G-eta-1-expansion-odd}). 
This feature has been observed also for the massless scalar 
\cite{Calabrese:2012nk}.
Thus, taking the replica limit (\ref{neg-replica-limit-current-2intervals}) in (\ref{log-G-eta-1-expansion-even}) we find the following asymptotic behaviour for the logarithmic negativity in the regime of the small separation distance
\begin{equation}
\label{log-neg-2int-expansion-eta-1}
    \mathcal{E} = 
    - \frac{1}{8} \log(1-\eta) 
    - 
    \frac{1}{2}
    \log\!\big( \! - \log(1-\eta)\big)
+O(1)
\end{equation}
where the specific value of the constant term $O(1)$ has been obtained in \cite{Calabrese:2012nk} and it is not relevant for our discussion. 
Comparing the expansion (\ref{log-neg-2int-expansion-eta-1}) for the chiral current against the corresponding expansion for the real massless scalar, given in Eq.\,(106) of \cite{Calabrese:2012nk}, 
we observe that while the leading logarithmic divergence differs by a factor $1/2$, the subleading log-log divergence is the same in the two models. 

In the following, we provide the explicit expressions for the moments of the partial transpose when  
$V_1$ and $V_2$ are adjacent intervals. 
Although these results are based only on the leading term 
in the expansion as $\eta \to 1^{-}$
of the corresponding results for disjoint intervals,
it is important for us to report them explicitly 
both because they provide a connection 
with the corresponding results valid in a generic CFT,
and obtained through the branch-point twist fields method \cite{Calabrese:2012ew, Calabrese:2012nk}.
Hence, this is a crucial consistency check of the analytic expressions reported in Sec.\,\ref{subsec:momentsPT2twointervals}.
As for the cross ratio $\eta$ 
defined in (\ref{cross-ratio-eta-def}), 
we have that 
$1 -\eta = \tfrac{ \ell_1 + \ell_2}{\ell_1 \ell_2} \,d +O(d^2)$ 
as $d \to 0$, while $\ell_1$ and $\ell_2$ are kept fixed.
In the following, we consider a naive approach to this limit
where we first take $d\to 0$ with $\ell_1$ and $\ell_2$ fixed,
ignoring the constraint given by  $d \gg \epsilon$.
Then, the powers of $\epsilon$ and $d$ in the resulting expression are simply discarded 
and the dependence of the UV cutoff 
is finally reintroduced by replacing each length 
$\ell$ with the corresponding dimensionless ratio $\ell/\epsilon$.

By applying this procedure to the moments 
(\ref{moments-rhoA-T2-current-explicit}),
for odd values of $n=n_{\textrm{\tiny o}}$, 
from (\ref{G-n-odd-explicit-product}) 
(see also \eqref{log-G-eta-1-expansion-odd}),
we find 
\begin{equation}
\label{moments-adj-odd}
    \tr \!\big( \rho_V^{\textrm{\tiny $\Gamma_2$}}\big)^{n_{\textrm{\tiny o}}} 
    \propto
    \left(\frac{ \ell_1 \ell_2 \,(\ell_1 + \ell_2 ) }{\epsilon^3}\right)^{-\Delta_{n_{\textrm{\tiny o}}}/2}
\end{equation}
while for even values of $n=n_{\textrm{\tiny e}}$,
the expression (\ref{moments-rhoA-T2-current-explicit}) 
combined with (\ref{G-n-even-explicit-product}) 
(see also \eqref{log-G-eta-1-expansion-even})
leads to 
\begin{equation}
\label{moments-adj-even}
    \tr \!\big( \rho_V^{\textrm{\tiny $\Gamma_2$}}\big)^{n_{\textrm{\tiny e}}} 
    \propto
    \left( \frac{\ell_1 \ell_2}{\epsilon^2} \right)^{-\Delta_{n_{\textrm{\tiny e}}/2}} 
    \left( \frac{\ell_1 + \ell_2}{\epsilon} \right)^{-\Delta_{n_{\textrm{\tiny e}}}+\Delta_{n_{\textrm{\tiny e}}/2}} \,.
\end{equation}
The logarithmic negativity of adjacent intervals 
can be found by taking the replica limit (\ref{neg-replica-limit})
with the expression in (\ref{moments-adj-even}),
and the result is 
\begin{equation}
\label{log-neg-adj}
    \mathcal{E} = \frac{1}{8} \, \log\!\left( \frac{\ell_1 \, \ell_2}{(\ell_1 + \ell_2)\, \epsilon} \right) + \textrm{const}
\end{equation}
which is the expected result for a chiral CFT with central charge $c=1$.

The case of adjacent intervals on a circle of length $L$ can be studied by mapping the line into the circle through a conformal transformation. 
The resulting expressions for the moments and the logarithmic negativity are given by  (\ref{moments-adj-odd}), (\ref{moments-adj-even}) and (\ref{log-neg-adj}) 
where each dimensionless ratio $\ell/\epsilon$,
with $\ell \in \{\ell_1, \ell_2, \ell_1 + \ell_2\}$,
is replaced by $\tfrac{L}{\pi \epsilon} \sin ( \pi \ell/L)$,
that becomes $\ell/\eps$ when $\ell/L \ll 1$.
These replacements lead respectively to
\begin{eqnarray}
\label{moments-adj-odd-temp}
    \tr \!\big( 
    \rho_V^{\textrm{\tiny $\Gamma_2$}}
    \big)^{n_{\textrm{\tiny o}}} 
    &\propto&
    \left(\frac{ L^3\,
    \sin(\pi\ell_1/L)\,\sin(\pi\ell_2/L)\, 
    \sin\!\big((\pi(\ell_1+\ell_2)/L\big)\,
    }{\pi^3\, \epsilon^3}\right)^{-\Delta_{n_{\textrm{\tiny o}}}/2}
    \\
    \label{moments-adj-even-temp}
    \rule{0pt}{.9cm}
        \tr \!\big( \rho_V^{\textrm{\tiny $\Gamma_2$}}\big)^{n_{\textrm{\tiny e}}} 
    &\propto&
    \left( \frac{L^2\,\sin(\pi\ell_1/L)\,\sin(\pi\ell_2/L)}{\pi^2\,\epsilon^2} \right)^{-\Delta_{n_{\textrm{\tiny e}}/2}} 
    \left( \frac{L\, \sin\!\big((\pi(\ell_1+\ell_2)/L\big)}{\pi \,\epsilon} \right)^{-\Delta_{n_{\textrm{\tiny e}}}+\Delta_{n_{\textrm{\tiny e}}/2}}
    \hspace{1.2cm}
\end{eqnarray}
and
\begin{equation}
    \label{log-neg-adj-temp}
    \mathcal{E} = \frac{1}{8} \, 
    \log\!\left( \frac{L\,\sin(\pi\ell_1/L)\,\sin(\pi\ell_2/L)
    }{\pi \,\epsilon\,\sin\!\big((\pi(\ell_1+\ell_2)/L\big)} \right) + \textrm{const}
\end{equation}
which are compared against the numerical results from the lattice in Sec.\,\ref{sec-numerics-adj-blocks} 
(see Fig.\,\ref{fig:Substract_Vdjacent_PBC}).

We remark that this well-known procedure to find the analytic expressions on the circle strongly indicates that the moments 
$\tr \!\big( \rho_V^{\textrm{\tiny $\Gamma_2$}}\big)^{n} $
behave like the three-point functions 
of the branch-point twist fields,
as discussed in \cite{Calabrese:2012ew, Calabrese:2012nk},
also for the specific chiral CFT model that we are exploring.

\section{Comparisons with the massless chiral Dirac field}
\label{sec-chiral-dirac-fermion}

In this section, considering the subsystem 
$V = V_1 \cup V_2$ 
made by the union of two disjoint intervals on the line, 
we compare  the logarithmic negativity  
and the moments of the partial transpose 
for the chiral current 
with the corresponding quantities
for the complex Weyl fermion, 
which is chiral and has central charge
$c = 1$. 
Both these models are in their ground state. 
For a free fermion, the reduced density matrix $\rho_V$ is Gaussian but $\rho_V^{\textrm{\tiny $\Gamma_2$}}$ is not Gaussian anymore \cite{Eisler:2015tgq}
and the absence of this crucial feature makes the computation of its logarithmic negativity difficult to explore \cite{Coser:2015eba,Herzog:2016ohd}.
Thus, the notion of fermionic partial time-reversal 
has been studied in \cite{Shapourian:2016cqu, Shapourian:2018ozl}, 
whose action on $\rho_V$ gives the Gaussian operator
$\rho_V^{\textrm{\tiny R$_2$}}$, 
where we are considering the partial time-reversal
with respect to $V_2$ without loss of generality.

\subsection{Fermionic partial time-reversal for the massless Dirac field}
\label{sec:Dirac_fermion}

Considering the reduced density matrix $\rho_V$ of the massless Dirac field, its partial time reversal with respect to $V_2$
is defined as
\cite{Shapourian:2016kvr,Shiozaki:2017ive,Shapourian:2018ozl,Shapourian:2016cqu}
\begin{equation}
\rho_V^{\textrm{\tiny R$_2$}}
\equiv
\big(\mathbb{I}_{V_1} \otimes \mathsf{T}_{V_2}\big)\,
\rho_V\,
\big(\mathbb{I}_{V_1} \otimes \mathsf{T}_{V_2}^{-1}\big)
\end{equation}
where $\mathsf{T}_{V_2}$ acts as a composition of complex conjugation in the fermionic coherent-state basis of $V_2$, and a unitary transformation that depends on the fermion parity. 
In order to make contact with 
\cite{Shapourian:2016cqu} and  \cite{Eisler:2015tgq}, 
we find it worth introducing the following notation 
\begin{equation}
\label{eq:moments_Ryu_Dirac}
O_+ \equiv \rho_V^{\textrm{\tiny R$_2$}}
\;\;\;\qquad\;\;\;
O_- \equiv \big(\rho_V^{\textrm{\tiny R$_2$}}\big)^{\dagger}
\end{equation}
which can be formally written through 
Grassmannian path integrals, as explained in \cite{Shapourian:2016cqu}. 
The operators in the lattice whose continuum limit provides 
\eqref{eq:moments_Ryu_Dirac} are discussed in the Appendix\;\ref{app:fermion_numerics}
(see \eqref{eq:GaussianOperators_majorana}).

The fermionic logarithmic negativity 
based on this partial time-reversal 
is defined from the trace norm of $\rho_V^{\textrm{\tiny R$_2$}}$ as follows
\begin{equation}
\label{eq:fermionic_negativity_ryu}
\widetilde{\mathcal{E}}
\equiv
\log\big|\rho_V^{\textrm{\tiny R$_2$}}\big|
=
\log\!\left[
\operatorname{Tr}\!
\big(\sqrt{O_+\, O_{-}} \,\big)
\right]
\end{equation}
which can be found by considering the following sequence
\begin{equation}
\label{eq:even_moments_Ryu_Dirac}
\widetilde{\mathcal{E}}_{n_{\textrm{\tiny e}}}
\equiv
\log\!\big[
\operatorname{Tr}\!\big((O_+ O_-)^{n_{\textrm{\tiny e}}/2}\big)
\big]
\end{equation}
where $n_{\textrm{\tiny e}}$ is a generic even integer and taking the analytic continuation $n_{\textrm{\tiny e}} \to 1$. 
The sequence corresponding to odd integers $n_{\textrm{\tiny o}}$ has been studied through 
the supersymmetric trace construction of \cite{Shapourian:2016cqu} or 
the twisted partial-transpose approach of \cite{Shapourian:2019xfi}.
We focus only on (\ref{eq:even_moments_Ryu_Dirac})
in our analysis.

For the setup that we are considering and the operators (\ref{eq:moments_Ryu_Dirac}), 
it has been found that \cite{Coser:2015eba}
\begin{equation}
\label{eq:CFT_equation}
\operatorname{Tr}\!\left(\,
\prod_{i=1}^n O_{s_i} \!\right)
= \,
\tilde{c}_{n}^2
\left(\frac{1 - \eta}{\ell_1 \ell_2/\epsilon^2}\right)^{2 \Delta_n}
\left|
\frac{\Theta[\mathbf{e}](\tilde{\tau})}{\Theta(\tilde{\tau})}
\right|^2
\;\;\qquad \;\;\;
\mathbf{e} = 
\bigg(
\begin{array}{c}
\boldsymbol{0} 
\\
\boldsymbol{\delta}
\end{array}
\bigg)
\end{equation}
in terms of the lengths of the intervals and the cross ratio (\ref{cross-ratio-eta-def}),
where $\tilde{\tau}$ is given by the following 
$(n-1)\times (n-1)$ symmetric matrix 
\cite{Calabrese:2012ew,Calabrese:2012nk}
\begin{equation}
    \tilde{\tau}_{i, j} 
    \,\equiv\,
    \mathrm{i} \,\frac{2}{n} 
    \sum_{k=1}^{n-1} \sin (\pi k / n) \,
    \frac{F_{k/n}\big(1/(1-\eta)\big) 
    }{F_{k/n}\big(\eta/(\eta-1)\big)} \,
    \cos \!\big[2 \pi k (i-j)/n\big] 
\end{equation}
in terms of the hypergeometric function 
introduced in (\ref{Dn-integral-ACHP}),
which is a purely imaginary matrix 
whose imaginary part is positive definite 
when $\eta \in (0,1)$.
The normalisation constant $\tilde{c}_{n}$ in (\ref{eq:CFT_equation})
plays the same role as $c_n$ in (\ref{moments-rhoA-current-explicit}) and (\ref{moments-rhoA-T2-current-explicit}) within this model. 
In (\ref{eq:CFT_equation}), the characteristic $\mathbf{e}$ of the Riemann-Siegel theta function is the special one where $\mathbf{0} $ is the $(n-1)$-dimensional vector made by zeros,
while $\boldsymbol{\delta} $ is the $(n-1)$-dimensional vector whose elements are $\delta_i \in \big\{0\,,1/2 \big\}$.
The values of $\delta_i$ in the r.h.s. of (\ref{eq:CFT_equation}) are determined by the sequence of $s_i \in \{+,-\}$
in the l.h.s. of the same equation, according to the correspondence discussed in 
\cite{Coser:2015eba,Coser:2015dvp}.
For instance, 
$\tr O_{+}^{n}$ corresponds to 
$\boldsymbol{\delta} = \boldsymbol{0}$
for any integer value of $n \geqslant 1$; 
hence the Riemann-Siegel theta does not occur in this case.
For the special case given by the operator occurring 
within the trace in (\ref{eq:even_moments_Ryu_Dirac}),
we have that $\delta_i = 1/2$ for $1 \leqslant i \leqslant n-1$; hence we denote by
$\mathbf{e}_{\textrm{\tiny $1/2$}}$
the corresponding characteristic in (\ref{eq:CFT_equation}).
By employing the Thomae formulae 
\cite{Nakayashiki:1997Thomae,FarkasZemel:2011Thomae},
it has been observed in \cite{Herzog:2016ohd} that, 
for even integers $n_{\textrm{\tiny e}}$, the following identity holds
\begin{equation}
\label{eq:Thomae_result}
\left|
\frac{\Theta[\mathbf{e}_{\textrm{\tiny $1/2$}}](\tilde{\tau})}{\Theta(\tilde{\tau})}
\right|^2
= (1 - \eta)^{-n_{\textrm{\tiny e}}/4} \,.
\end{equation}

From (\ref{eq:CFT_equation}) 
and (\ref{eq:Thomae_result}),
for the moments occurring in \eqref{eq:even_moments_Ryu_Dirac} 
we obtain
\begin{equation}
\label{eq:evenTr_string}
\textrm{Tr}
\big(
(O_+ O_- )^{ n_{\textrm{\tiny e}} / 2}
\big)
= \,
\tilde{c}_{ n_{\textrm{\tiny e}}}^2 
\left(\frac{1 - \eta}{\ell_1 \ell_2/\epsilon^2}\right)^{2 \Delta_{ n_{\textrm{\tiny e}}}}
(1 - \eta)^{- n_{\textrm{\tiny e}} / 4}
\end{equation}
hence \eqref{eq:even_moments_Ryu_Dirac} becomes 
\begin{equation}
\label{eq:Ryu_CFT_LogNegMoments}
\widetilde{\mathcal{E}}_{ n_{\textrm{\tiny e}}} 
=
-\frac{1}{12}
\left( n_{\textrm{\tiny e}} + \frac{2}{n_{\textrm{\tiny e}}} \right)
\log(1 - \eta)
-2\Delta_{ n_{\textrm{\tiny e}}}
\log \!\big( \ell_1 \ell_2 / \epsilon^2 \big)
+ 
2 \log(\tilde{c}_{ n_{\textrm{\tiny e}}})
\end{equation}
The analytic continuation 
$ n_{\textrm{\tiny e}} \to 1$ of this expression 
provides the fermionic logarithmic negativity 
for two disjoint intervals on the line and,
by using that $\tilde{c}_1 = 1$, it reads
\begin{equation}
\label{eq:negativity_Dirac}
\widetilde{\mathcal{E}}  
= -\frac{1}{4}
\log(1 - \eta)\,.
\end{equation}
By considering the short-distance limit of  \eqref{eq:negativity_Dirac} through the procedure outlined in Sec.\,\ref{subsec:two-intervals-small-distance}, 
one recovers the results of \cite{Calabrese:2012nk} for the logarithmic negativity of adjacent intervals in the line.

Remarkably, our result \eqref{eq:Ryu_CFT_LogNegMoments} coincides
with the expression in Eq.\,(129) of \cite{Shapourian:2019xfi}
specialised to even values of the R\'enyi index, 
which provides the leading terms in the expansion as $\eta \to 1^{-}$.
Thus, from the comparison with \eqref{eq:Ryu_CFT_LogNegMoments}, which  holds for any $\eta \in (0,1)$, 
we conclude that the subleading terms that have not been computed in \cite{Shapourian:2019xfi} do not occur.

\subsection{Comparison with the chiral component of 
the Dirac field}
\label{sec-comparison-chiral-fermion}

A chiral component of the massless Dirac field,
i.e. a complex Weyl fermion, is a chiral CFT with $c=1$.
The R\'enyi negativities 
$\widetilde{\mathcal{E}}^{\,\textrm{\tiny (W)}}_{n_{\textrm{\tiny e}}}$
in this model can be obtained by taking one half of 
the corresponding quantities for the massless Dirac field;
hence, from \eqref{eq:Ryu_CFT_LogNegMoments} we find 
\begin{equation}
\label{eq:Ryu_CFT_LogNegMoments_Weyl}
  \widetilde{\mathcal{E}}^{\,\textrm{\tiny (W)}}_{n_{\textrm{\tiny e}}} 
=
-\frac{1}{24}
\left( n_{\textrm{\tiny e}} + \frac{2}{n_{\textrm{\tiny e}}} \right)
\log(1 - \eta)
-\Delta_{ n_{\textrm{\tiny e}}}
\log \!\big( \ell_1 \ell_2 / \epsilon^2 \big)
+ 
 \log(\tilde{c}_{ n_{\textrm{\tiny e}}})
\end{equation}
whose analytic continuation 
$ n_{\textrm{\tiny e}} \to 1$ gives the 
corresponding fermionic logarithmic negativity
\begin{equation}
\label{eq:negativity_Majorana}
 \widetilde{\mathcal{E}}^{\,\textrm{\tiny (W)}}  
= -\frac{1}{8}\log(1 - \eta) \,.
\end{equation}

Since R\'enyi negativities for the Weyl fermion in \eqref{eq:Ryu_CFT_LogNegMoments_Weyl} and 
the corresponding quantity for the chiral current, 
given in \eqref{neg-ren-even-cc-extended}, 
have the same leading divergence as $\eta \to 1^-$, 
it is natural to examine their difference, namely
\begin{equation}
\label{Delta-E-ne-loglog}
\Delta \mathcal{E}_{ n_{\textrm{\tiny e}}} 
\equiv 
  \widetilde{\mathcal{E}}^{\,\textrm{\tiny (W)}}_{n_{\textrm{\tiny e}}}
  -
  \mathcal{E}_{ n_{\textrm{\tiny e}}}
= \frac{1}{2}\log \!\big(
2K(\eta)/\pi\big)
+ \widetilde{D}_{ n_{\textrm{\tiny e}}}(\eta)
+ \log \!
\big( \tilde{c}_{ n_{\textrm{\tiny e}}}/ c^2_{ n_{\textrm{\tiny e}}}\big) \,.
\end{equation}
This analytic expression  
provides the black dashed curves in 
Fig.\,\ref{fig:diff_neg}.
As $\eta\rightarrow 1^-$, 
from (\ref{tilde-Dn-def-even}) and the second expression in 
\eqref{eta-to-one-asymptotics},
it is straightforward to realise that 
$\widetilde{D}_{n_{\textrm{\tiny e}}}(\eta)\approx\mathcal{O}(1)$. 
Hence, by using the first expression in 
\eqref{eta-to-one-asymptotics}, we find that 
\begin{equation}
\label{eq:topo_term_even_n}
\Delta \mathcal{E}_{ n_{\textrm{\tiny e}}} 
=
\frac{1}{2}\log \!\big|\log(1-\eta)\big| 
+ \mathcal{O}(1) 
\;\;\;\qquad\;\;\; 
\eta\rightarrow 1^-
\end{equation}
whose leading term is divergent and 
independent of $ n_{\textrm{\tiny e}}$;
hence also the logarithmic negativity has the same leading divergence
occurring in the r.h.s. of (\ref{eq:topo_term_even_n}).
This result is in analogy with the $\log|G|$ contribution 
in  \eqref{eq:topological_term_negativity}, 
for the toy model discussed in Sec.\,\ref{thermofield}.

Remarkably, the same leading divergence in (\ref{eq:topo_term_even_n}) 
occurs also in Eq.\,(3.58) of \cite{Benedetti:2024dku} 
for the difference between the R\'enyi mutual information 
of these models, for any value of the R\'enyi index. 
Comparing the special case of the mutual information in 
this result, 
which contains both classical and quantum correlations,
with the special case given by the logarithmic negativity in (\ref{eq:topo_term_even_n}), 
which contains only quantum correlations,
we infer that the leading divergence in the r.h.s. of (\ref{eq:topo_term_even_n}) originates 
only from the quantum correlations.

We find it instructive to perform the analysis 
described above for the fermionic partial time reversal 
also for the partial transposition of the free fermion 
discussed in 
\cite{Eisler:2015tgq,Coser:2015eba,Coser:2015mta,Coser:2015dvp}, 
which provides the operator 
$\rho_{V;{\textrm{\tiny W}}}^{\textrm{\tiny $\Gamma_2$}}$
in the case of the complex Weyl fermion.
Thus, we introduce 
\begin{equation}
\label{eq:DeltaEprime_def}
\Delta \mathcal{E}^{\textrm{\tiny ($\Gamma_2$)}}_{ n_{\textrm{\tiny e}}}
\equiv \,
\log \!\Big[
\tr 
\big(
\rho_{V;{\textrm{\tiny W}}}^{\textrm{\tiny $\Gamma_2$}}
\big)^{ n_{\textrm{\tiny e}}}
\Big]
- 
\mathcal{E}_{ n_{\textrm{\tiny e}}}
\end{equation}
where $\mathcal{E}_{ n_{\textrm{\tiny e}}}$ 
is (\ref{neg-moments-def}) 
for the chiral current (see (\ref{neg-ren-even-cc-extended})).
In Appendix\;\ref{app:negativity-difference-Eisler}, 
by employing some inequalities 
conjectured in \cite{Herzog:2016ohd},
we argue that
for any value of the even integer 
$n_{\textrm{\tiny e}}$,
\eqref{eq:DeltaEprime_def},
displays  the following asymptotic behaviour
\begin{equation}
\label{eq:DeltaEprime_asymptotics}
    \Delta\mathcal{E}_{ n_{\textrm{\tiny e}}}^{\textrm{\tiny ($\Gamma_2$)}}
    =
    \frac{1}{2} \log \!\big|\log(1-\eta) \big| 
    + \mathcal{O}(1) 
    \;\;\;\qquad \;\;\;
    \eta\rightarrow 1^-  \,.
\end{equation}
Hence, this difference specialised to the logarithmic negativity, 
which corresponds to $n_{\textrm{\tiny e}} \to 1$,
displays the same divergence.
It would be interesting to obtain (\ref{eq:DeltaEprime_asymptotics}) 
in a rigorous way.

It would be interesting to consider 
$\widetilde{\mathcal{E}}_{n_{\textrm{\tiny o}}}$
for an odd integer $n_{\textrm{\tiny o}}$.
Indeed, finding an analytic expression for this quantity 
(as done in (\ref{eq:Ryu_CFT_LogNegMoments}) for an even integer $n_{\textrm{\tiny e}}$)
would lead to a quantitative comparison 
with the short-distance expansion reported in  \cite{Shapourian:2016cqu}
and also to further comparisons with the corresponding results 
for the chiral current.

\section{Lattice model for the chiral current}
\label{renyisvscorrelators}

In this section, we discuss a method to evaluate the moments of the partial transpose and the logarithmic negativity in a lattice model for the chiral current. 
The explicit expressions for these quantities 
are reported in Sec.\,\ref{sec:negativity_numerics}, 
while in Sec.\,\ref{sec:negativity_numerics}, 
for the sake of completeness,
we describe  the computation of the entanglement entropies
(i.e. the R\'enyi entropies and of the entanglement entropy)
in the same lattice model. 
The numerical results for the moments of the partial transpose 
and the logarithmic negativity are compared 
with the corresponding analytic expressions 
reported in Sec.\,\ref{subsec:momentsPT2twointervals} 
and Sec.\,\ref{sec-comparison-chiral-fermion}.

\subsection{Entanglement entropies}
\label{subsec-renyi-lattice}

The procedure to compute the entanglement entropies 
in a Gaussian state  for a free bosonic lattice model 
(e.g. the harmonic lattice) made by $N$ sites, 
whose operators 
$\hat{q}_i $ and $\hat{p}_i $ associated to the $i$-th site 
satisfy the canonical commutation relations
\begin{equation}
\label{CCR-relations}
    \big[\,\hat{q}_i\,, \,\hat{p}_j\,\big]=\ri \delta_{i, j} 
    \;\;\;\qquad\;\;\;
    \big[\,\hat{q}_i \,,\, \hat{q}_j\,\big]
    =
    \big[\,\hat{p}_i \,,\, \hat{p}_j\,\big]
    =0 
    \;\;\; \qquad  \;\;\;
    1\leqslant i,j\leqslant N
\end{equation}
is well known
\cite{Botero:2004vpl,Audenaert:2002xfl,peschel_reduced_2009,Casini:2009sr,Weedbrook:2011wxo}. 
In particular, considering the ground state of the system for simplicity, from the correlation matrices $Q$ and $P$
whose generic element is given by the two-point correlator
$Q_{i,j} = \langle \,\hat{q}_i \, \hat{q}_j \rangle$ and $P_{i,j} = \langle \,\hat{p}_i \, \hat{p}_j \rangle$ respectively, the covariance matrix is defined as 
$\gamma = Q \oplus P$.

The spatial bipartition is provided by 
the region $V$ and its complement $\overline{V}$,
which induces the factorisation of the Hilbert space 
given by $\mathcal{H} = \mathcal{H}_V \otimes \mathcal{H}_{\overline{V}}$.
The reduced density matrix $\rho_V \equiv \textrm{Tr}_{\mathcal{H}_{\overline{V}}}
\big(|0\rangle\langle 0|\big)$ is also a Gaussian state,
and therefore it can be described by its 
reduced covariance matrix  
$\gamma_V \equiv Q_V \oplus P_V$,  
where $(Q_V)_{i,j} = Q_{i,j}$ 
and $(P_V)_{i,j} = P_{i,j}$, with $i,j \in V$.
Denoting by $N_V$ the number of sites in $V$,
the $(2N_V)\!\times\!(2N_V)$ matrix $\gamma_V$ is real, symmetric and positive definite;
hence its Williamson decomposition can be considered. 
This decomposition provides the symplectic spectrum
$\big\{\sigma_k \big\}$ of $\gamma_V$, 
where $1 \leqslant k \leqslant N_V$,
and it can be shown that it coincides with the positive eigenvalues of $\sqrt{Q_V P_V}$ in the case that we are considering. 
The Rényi entropies and the entanglement entropy can be obtained from the symplectic spectrum respectively as 
\begin{equation}
\label{renyi-from-ss}
S_V^{(n)} =
\frac{1}{n-1}\sum_{k=1}^{N_V}
\log\!\Big[(\sigma_k+1/2)^n-(\sigma_k-1/2)^n\Big]
\end{equation}
and \begin{equation}
\label{EE-from-ss}
S_V =
\sum_{k=1}^{N_V}\!\Big[
(\sigma_k+1/2)\log(\sigma_k+1/2)
-(\sigma_k-1/2)\log(\sigma_k-1/2)
\Big] \,.
\end{equation}

\subsubsection{Algebra in the null line}
\label{sec:staggered_boson_review}

We consider the class of quadratic models 
defined by the following generalized commutation relation
\be
\label{eq:commutator_generic_f}
\big[\, \hat{b}_i,\hat{b}_j \big]
= 
\textrm{i}\, Y_{i,j}
\ee
where $\hat{b}_i$ is the Hermitian operator associated to the $i$-th site and $Y$ is a real  
and antisymmetric $N_0\times N_0$ matrix 
\cite{Arias:2018tmw}.
Considering the Gaussian ground state of the generic model within this class, we assume that one can construct 
the correlation matrix $B$, whose generic element is 
$B_{i,j} \equiv \big\langle \,\hat{b}_i \,\hat{b}_j \big\rangle$.
From \eqref{eq:commutator_generic_f}, 
it is straightforward to realise that a consistency condition of this correlation matrix is 
$Y_{i,j}=2 \, \textrm{Im}
\big( \big\langle \,\hat{b}_i \,\hat{b}_j \big \rangle \big)$.

The lattice model for the chiral current 
belongs to this class. 
Indeed, its Hamiltonian on the circle made by $N_0$ sites,
where periodic boundary conditions are imposed, 
is defined as follows \cite{Arias:2018tmw,Berenstein:2023tru} 
\be
\label{eq:Hamiltonian_bmodes}
H=\frac{1}{2}\sum_{i=1}^{N_0} \hat{b}_i^{\,2}\,
\ee
where the Hermitian operators $\hat{b}_i$ satisfy 
the non-canonical commutation relation (\ref{eq:commutator_generic_f}), 
with
\be
Y_{i,j}\equiv \delta_{j,i+1}-\delta_{j,i-1}+\delta_{j,1}\delta_{i,N_0}-\delta_{j,N_0}\delta_{i,1}
\label{eq:commutator-chiral}
\ee
which is obtained from the discretization of \eqref{eq:commutator_continuum}. We will refer to this choice of algebra as the null line algebra.
Since the dispersion relation of the model $E(k)=2\sin k$ 
with $k\in[0,2\pi)$ has two zeros (at $k=0$ and $k=\pi$)
for even values of $N_0$,
a doubling of the degrees of freedom occurs in this case
\cite{Arias:2018tmw, Berenstein:2023tru}. 
This feature, 
that has been extensively studied in fermionic lattice models \cite{Kaplan:1992bt,Ginsparg:1981bj,Nielsen:1981hk},
has been explored also for bosonic lattice models
\cite{Arias:2018tmw,Berenstein:2023tru,Berenstein:2023ric}.
These doubled degrees of freedom 
are related by the parity operation 
$\hat{b}_i\rightarrow (-1)^i \,\hat{b}_i$, 
which is also a symmetry of the Hamiltonian \eqref{eq:Hamiltonian_bmodes}. 
The continuum limit of the lattice model defined by \eqref{eq:Hamiltonian_bmodes} with $N_0$ even is given 
by two decoupled copies of the chiral current.
When $N_0$ is odd, the dispersion relation has only one zero, 
as shown in \cite{Berenstein:2023ric}. 
In this case, the doubling does not occur and the continuum limit is given by a single copy of the chiral current 
with antiperiodic boundary conditions.
In consequence, the parity of $N_0$ determines not only the presence of the doubling but also the specific boundary conditions in the continuum limit.  
This resembles the fact that the system size parity in the Majorana chain determines the spin structure in the Majorana fermion in the continuum limit, as discussed in \cite{Seiberg:2023cdc}.
Hereafter, we consider an even number $N_0$ of sites in our analysis.

In the setup given by (\ref{eq:Hamiltonian_bmodes}) 
and (\ref{eq:commutator-chiral}),
the algebra of the model has a global center generated by the following operators
\begin{equation}
\label{eq:zero_modes}
\hat{\Psi}_{\text{\tiny even}}=\sum_{i=1}^{N_0} \hat{b}_i 
\;\;\;\;\qquad\;\;\;
\hat{\Psi}_{\text{\tiny odd}}=\sum_{i=1}^{N_0}  (-1)^{i}\, \hat{b}_i \,.
\end{equation}
The algebra of the model is then obtained from the union of $N_0/2-1$ copies of the Heisenberg algebra and the center \cite{Berenstein:2023tru}.
 As a consequence, 
 the different representations of this algebra are naturally labelled by the values $(\Psi_{\text{\tiny even}},\Psi_{\text{\tiny odd}})$ of the two operators in \eqref{eq:zero_modes}. 
 
Some subtleties occur about the translational invariance 
of the model.
Indeed, the translational symmetry for the model given by the shift $\mathcal{S}_1$ defined by $\hat{b}_j\rightarrow \hat{b}_{j+1}$ 
applied to all the sites of the chain
is a symmetry of the Hamiltonian and of the commutation relations. 
However, its action on \eqref{eq:zero_modes} is nontrivial 
because it leads to $\Psi_{\text{\tiny odd}}\rightarrow -\Psi_{\text{\tiny odd}}$ and therefore it
induces a change $(\Psi_{\text{\tiny even}},\Psi_{\text{\tiny odd}})\rightarrow(\Psi_{\text{\tiny even}},-\Psi_{\text{\tiny odd}})$ at the level of representations. 
Instead, the symmetry $\mathcal{S}_2$ defined 
by shift $\hat{b}_i\rightarrow \hat{b}_{i+2}$ 
for all the sites leaves \eqref{eq:zero_modes},
\eqref{eq:Hamiltonian_bmodes} 
and \eqref{eq:commutator-chiral} invariant.
As for the shifts $\mathcal{S}_{p}$ 
given by $\hat{b}_i\rightarrow \hat{b}_{i+p}$ 
with integer $p \geqslant 3$, 
they can be decomposed into a sequence made by 
$\mathcal{S}_2$ and a single $\mathcal{S}_1$, that occurs only when $p$ is odd.
Hence, every $\mathcal{S}_{p}$ with even $p$ 
leaves the system invariant, 
while all the $\mathcal{S}_{p}$ with odd $p$ 
act like $\mathcal{S}_{1}$.

It is worth considering a general deformation of (\ref{eq:Hamiltonian_bmodes}) through the following term
\cite{Berenstein:2023tru}
\begin{equation}
H_{\text{def}}=\sum_{i=1}^{N_0}\sum_{j=1-i }^{N_0-i} m_{i,j} \,\hat{b}_i \, \hat{b}_{i+j}
\;\;\;\;\qquad\;\;\;
m_{i,j}\in\mathbb{R} \,.
\end{equation}
However, this term does not preserve both the parity and the translation symmetry of the system in general 
and this feature influences the continuum limit 
in a crucial way \cite{Berenstein:2023tru}. 
This makes it difficult to remove the zero mode for the staggered boson
by standard methods \cite{Botero:2004vpl}.

The translational invariance of the model given by (\ref{eq:Hamiltonian_bmodes})-(\ref{eq:commutator-chiral}) implies that 
$B_{i,j}=B(i-j)$ for the generic element of the correlation matrix.
In terms of $x=i-j$, it reads \cite{Arias:2018tmw}
\be
\label{B-function-def}
B(x)=
\left\{
\begin{array}{ll}
\displaystyle
\;\frac{\sin (2 \pi/N_0)}{N_0}\; 
\frac{
\big[ \cos(\pi x/2)\, \big]^2
}{
\sin\!\big(\pi (x+1)/N_0\big)\, 
\cos\!\big(\pi(x-1+N_0/2)/N_0\big)} 
\hspace{1.cm}
&
|x|\neq 1
\\
\rule{0pt}{.6cm}
\displaystyle
\;\frac{\ri}{2} \;Y(x) 
&
|x|=1 \,.
\end{array}
\right.
\ee
In the infinite line limit, given by $N_0\rightarrow \infty$,
this expression becomes 
\be
\label{B-matrix-line}
B(x)=
\left\{
\begin{array}{ll}
\displaystyle
\;\frac{1+(-1)^x}{\pi (1-x^2)}\hspace{1.5cm}
&
|x|\neq 1
\\
\rule{0pt}{.7cm}
\displaystyle
\;\frac{\ri}{2} \;Y(x) \hspace{1.5cm}
&
|x|=1 \,.
\end{array}
\right.
\ee

Consider a subsystem $V$ made by $N_V$ sites. 
Since a zero mode occurs when $N_V$ is odd
\cite{Berenstein:2023tru}, 
in our analysis we restrict to the cases where $N_V$ is even. 
Let us introduce the $N_V\times N_V$ matrices $Y_V$ and $B_V$ 
obtained by restricting to $V$ the matrices $Y$ and $B$, 
defined by 
(\ref{eq:commutator-chiral}) and 
(\ref{B-function-def}) respectively. 
These matrices allow to construct the following matrix 
\be
\label{eq:Vmatrixdef}
W_V
\equiv 
- \,\ri \,Y_V^{-1}\, B_V-\frac{1}{2}\,\mathbb{I}_{N_V}
=
-\,\textrm{i} \,Y_V^{-1}\,\text{Re}(B_V)
\ee 
where $\mathbb{I}_{N_V}$ is the $N_V\times N_V$ identity matrix. 
The spectrum of (\ref{eq:Vmatrixdef})
provides the entanglement entropies of the lattice model of the chiral current that we are exploring, as discussed in 
\cite{Arias:2018tmw, Gentile:2025koe}.
In particular, the R\'enyi entropies $S_V^{(n)}$ are given by 
\be
\label{eq:renyi_V}
S_V^{(n)}=\frac{1}{n-1}\,
\textrm{Tr} \left\{
\Theta(W_V)\, \log \!\left[
\left(W_V+\frac{1}{2}\right)^n \! - \left(W_V-\frac{1}{2}\right)^n
\,\right]
\,\right\}\,
\ee
where $\Theta(W_V)$ is the projector 
on the subspace of the eigenstates of $W_V$ 
associated to the positive eigenvalues;
hence only the positive eigenvalues of $W_V$ occur in the sum. Then,  for the entanglement entropy one finds \cite{Sorkin:2012sn}
\begin{eqnarray}
\label{eq:EE_V}
S_V
&=&
\operatorname{Tr}\!
\left[
\left(W_V + \frac12\right)
\log \left|W_V + \frac12 \right| \;
\right]
\\
\rule{0pt}{.8cm}
&=&
\operatorname{Tr}
\left\{
\Theta(W_V)
\left[ (W_V + \tfrac12)\log \left(W_V + \frac12 \right)
       + \left(\frac12 - W_V \right)
       \log\left(W_V - \frac12\right) \right]
       \,\right\}
       \nonumber
\end{eqnarray}
which is another sum where only the positive eigenvalues of $W_V$ occur.

\subsubsection{Generalized real time approach}
\label{sec:realtimeapproach}

While the algebra in the null line 
discussed in Sec.\,\ref{sec:staggered_boson_review}
provides the entanglement entropies,
in order to study 
the logarithmic negativity and the moments of the partial transpose,
the loss of spatial localization of operators
induces us to consider a different method,
similar to the real time approach \cite{Casini:2009sr}.

Consider a bosonic lattice system on a circle made by $N$ sites, whose Hamiltonian is 
\begin{equation}
\widetilde{H}=
\frac{1}{2}
\Bigg( 
\sum_{i=1}^{N} {\hat{p}_i} ^2
+\sum_{i,j=1}^{N} 
\hat{\tilde{q}}_i \,\widetilde{M}_{i,j} \,\hat{\tilde{q}}_j
\Bigg)
\label{eq:hamiltonian_basic_noncan}
\end{equation}
where $\hat{\tilde{q}}_i $ and $\hat{p}_j$ 
are Hermitian operators satisfying 
the following non-canonical commutation relations
 \begin{equation}
 \label{eq:corr_noncan}
\big[\, \hat{\tilde{q}}_i \, ,\, \hat{p}_j \,\big]
=\,\textrm{i} \,T_{i,j}
\;\;\;\qquad\;\;\; 
\big[\,\hat{\tilde{q}}_i \,,\, \hat{\tilde{q}}_j\, \big]
=
\big[\, \hat{p}_i \, ,\, \hat{p}_j \,]
=0
\end{equation}
where $T$ is a real matrix. 
The commutation relations (\ref{eq:corr_noncan})
define the algebra in the $t=0$ slice. 
The standard real time approach
\cite{Casini:2009sr} corresponds to $T_{i,j} = \delta_{i,j}$.

This lattice model can be studied by relating it to a standard quadratic model based on 
the Hermitian operators 
$\hat{q}_i $ and $\hat{p}_i$ satisfying the canonical commutation relations (\ref{CCR-relations}).
By introducing the vectors 
$ \hat{\tilde{\mathbf{r}}} \equiv
\big(\, \hat{\tilde{q}}_1,\dots,\hat{\tilde{q}}_N,\hat{p}_1,\dots,\hat{p}_N \big)^{\text{t}}$,
and 
$\hat{\mathbf{r}}= \big( \hat{q}_1,\dots,\hat{q}_N,\hat{p}_1,\dots,\hat{p}_N \big)^{\text{t}}$, 
one finds that they are related through the following linear transformation
\begin{equation}
\label{eq:canonical_t=0_variables}
    \hat{\tilde{\mathbf{r}}}=\Gamma \, \hat{\mathbf{r}}
    \;\;\;\qquad \;\;\;\; 
    \Gamma=T\oplus \mathbb{I}_N
\end{equation}
where $\mathbb{I}_N$ is the $N \times N$ identity matrix and the matrix $T$ characterises the algebra (\ref{eq:corr_noncan}).
For the canonical variables $\hat{r}_i$ we have that 
$ \big[\, \hat{r}_i\, ,\, \hat{r}_j\, \big]=\ri \,J_{i,j}$,
where $J$ is the $(2N) \times (2N)$ real and antisymmetric matrix 
having the $N \times N$ matrix made by zeros on the diagonal blocks and the $N \times N$ matrices $\mathbb{I}_N$ and $-\mathbb{I}_N$
in the lower and upper block respectively. 
Combining this characteristic feature of the canonical variables and 
(\ref{eq:canonical_t=0_variables}), it is straightforward to obtain that
$ \big[\, \hat{\tilde{r}}_i\, ,\, \hat{\tilde{r}}_j\, \big]=
\ri ( \Gamma J \,\Gamma^\textrm{t} )_{i,j}$,
which tells us that $\Gamma J \,\Gamma^\textrm{t} $ 
plays the role of the matrix $Y$ in (\ref{eq:commutator_generic_f}).

In terms of the canonical variables $\hat{q}_i $ and $\hat{p}_i$, the Hamiltonian (\ref{eq:hamiltonian_basic_noncan}) 
reads 
\begin{equation}
\label{Ham-noncan-M-version}
\widetilde{H}
=\frac{1}{2}
\Bigg( \sum_{i=1}^N {\hat{p}_i}^2+
\sum_{i,j=1}^N \hat{q}_i \,M_{i,j} \, \hat{q}_j
\Bigg)
\;\;\;\;\qquad\;\;\;
M \equiv T^\textrm{t}\,\widetilde{M} \,T \,.
\end{equation}
Since the two-point correlators associated to the ground state of this Hamiltonian are \cite{Botero:2004vpl,Audenaert:2002xfl}
\begin{equation}
\label{two-point-corr-M}
\big\langle 
\hat{q}_i \, \hat{q}_j
\big\rangle 
=
\frac{1}{2}\big(\sqrt{M^{-1}}\, \big)_{i,j} 
\;\qquad\;
\big\langle 
\hat{p}_i \, \hat{p}_j
\big\rangle 
=
\frac{1}{2}\big(\sqrt{M}\, \big)_{i,j} 
\;\qquad\;
\big\langle 
\hat{q}_i \, \hat{p}_j
\big\rangle 
=
\frac{\textrm{i}}{2}\,\delta_{i,j}
\end{equation}
the two-point correlators for the operators satisfying (\ref{eq:corr_noncan}) 
can be obtained straightforwardly by combining 
(\ref{Ham-noncan-M-version}) 
and (\ref{two-point-corr-M}). 
This gives 
\begin{equation}
\label{eq:correlators_genericM}
\big\langle 
\hat{\tilde{q}}_i \, \hat{\tilde{q}}_j
\big\rangle 
=
\frac{1}{2}\left(\sqrt{T \widetilde{M}^{-1} T^{\text{t}}}\,\right)_{i,j} 
\;\qquad\;
\big\langle 
\hat{p}_i \, \hat{p}_j
\big\rangle 
=
\frac{1}{2}\left( \sqrt{T^\text{t}\widetilde{M}\,T} \,\right)_{i,j} 
\;\qquad\;
\big\langle 
\hat{\tilde{q}}_i \, \hat{p}_j
\big\rangle 
=
\frac{\textrm{i}}{2}\,T_{i,j} \,.
\end{equation}

Given a subsystem $V$ made by $N_V$ sites, 
consider the restriction to $V$ 
of the correlation matrices in (\ref{eq:correlators_genericM}), namely
\begin{equation}
\label{eq:reducedA_correlators_genericM_rtilde}
Q_ {V;\tilde{\mathbf{r}}}
\equiv
\frac{1}{2}\,
\sqrt{T\,\widetilde{M}^{-1}\,T^\text{t}}\; \Big|_V
\;\;\qquad \;\;
P_ {V;\tilde{\mathbf{r}}}
\equiv
\frac{1}{2}\,\sqrt{T^\text{t}\,\widetilde{M}\,T}\;\Big|_V
\;\;\qquad \;\;
E_ {V;\tilde{\mathbf{r}}}
\equiv 
\frac{\ri}{2}\,T_V
\end{equation}
where $T_V$ is the restriction to $V$ of the matrix $T$ in (\ref{eq:corr_noncan}).
The restriction to $V$ of (\ref{eq:canonical_t=0_variables}) 
gives the relation between the variables 
$\hat{\tilde{q}}_i $ and $\hat{p}_i$ in $V$
to the canonical variables 
$\hat{q}_i $ and $\hat{p}_i$ in $V$,
through the matrix $\Gamma_V\equiv T_V\oplus \mathbb{I}_{N_V}$.
Combining this relation with (\ref{eq:reducedA_correlators_genericM_rtilde})
leads to the two-point correlation matrices of the canonical variables reduced to $V$, that read
\begin{equation}
\label{eq:reducedA_correlators_genericM_r}
Q_ {V;\mathbf{r}}
=
T_V^{-1} \, Q_ {V;\tilde{\mathbf{r}}} \,T_V^{-\text{t}} 
\;\;\;\qquad \;\;\;\;
P_ {V;\mathbf{r}}
=
P_ {V;\tilde{\mathbf{r}}}
\;\;\;\qquad\;\;\; 
 E_ {V;\mathbf{r}}
 =
T_V^{-1} \,E_ {V;\tilde{\mathbf{r}}}
=
\frac{\ri }{2} \, \mathbb{I}_{N_V} \,.
\end{equation}
The entanglement entropies can be obtained 
from these correlation matrices 
in the standard way mentioned at the beginning of Sec.\,\ref{subsec-renyi-lattice}, 
i.e. by considering the spectrum of 
\begin{equation}
\label{CVr-def}
C_ {V;\mathbf{r}} \equiv 
\sqrt{Q_ {V;\mathbf{r}} \,P_ {V;\mathbf{r}}}
\end{equation}
and applying \eqref{renyi-from-ss} and
\eqref{EE-from-ss} with the positive eigenvalues of 
(\ref{CVr-def}).

\subsubsection{Lattice model for the chiral current}
\label{subsubsec:lattice current model}

The lattice model for the chiral current 
is the special case of 
\eqref{eq:hamiltonian_basic_noncan} and \eqref{eq:corr_noncan} 
given by  
\begin{equation}
\label{eq:chiralcurrent_M_T}
\widetilde{M}_{i,j}=\delta_{i,j} 
\;\;\;\; \qquad \;\;\;\;
T_{i,j}=\delta_{i,j}-\delta_{i,j-1}-\delta_{i,N}\delta_{j,1} 
\end{equation}
where the condition $T_{i+N,j}=T_{i,j}$ is also required, 
when periodic boundary conditions are imposed.
From the canonical position and momentum operators on a circle made by $N$ sites,
which satisfy \eqref{CCR-relations},
the relevant variables $\hat{\tilde{r}}_i$ are obtained as follows
\begin{equation}
\label{eq:operatoralgebra_circle}
\hat{\tilde{q}}_i=\hat{q}_{i}-\hat{q}_{i+1} 
\;\;\;\qquad\;\;\;
\hat{p}_i=\dot{\hat{q}}_i  
\;\;\;\qquad\;\;\;
1\leqslant i\leqslant N
\end{equation}
where the $\hat{\tilde{q}}_i$ are realised in the lattice as link operators while the $\hat{p}_i$ are attached to vertices, as explained in detail in Appendix\;\ref{app:purestatea_links}.

The Hamiltonian  (\ref{eq:Hamiltonian_bmodes}) 
and the commutation relations given by \eqref{eq:commutator_generic_f}
and (\ref{eq:commutator-chiral})
are recovered 
through the following identifications
\begin{equation}
\label{eq:from_b_to_r}
\hat{b}_{2j-1}\equiv \hat{p}_j 
\;\;\;\; \qquad \;\;\;\;
\hat{b}_{2j}\equiv \hat{\tilde{q}}_{j}
\end{equation}
which provide an equivalence 
between the lattice model in this section, made by $N$ sites, 
and the lattice model considered in Sec.\,\ref{sec:staggered_boson_review}, 
made by $N_0=2N$ sites.
This observation tells us that
the bosonic degrees of freedom 
corresponding to the $j$-th site 
in the formulation based on (\ref{eq:hamiltonian_basic_noncan})-(\ref{eq:corr_noncan})
are delocalised on two consecutive sites 
in the null line formulation 
of the lattice model for the chiral current
 \cite{Berenstein:2023tru}. 
From \eqref{eq:from_b_to_r}, one infers that 
the shift $\hat{b}_i\rightarrow \hat{b}_{i+2}$ 
 is indeed a proper lattice translation, while the shift $\hat{b}_i\rightarrow \hat{b}_{i+1}$ acts as a duality between non-canonical position and momentum operators \cite{Berenstein:2023ric}.

Notice that $T$ in (\ref{eq:chiralcurrent_M_T}) is not invertible; 
hence the correlator 
 $\big\langle \hat{q}_i \, \hat{q}_j \big\rangle$ in 
 (\ref{two-point-corr-M}) is divergent. 
However, the correlators in \eqref{eq:correlators_genericM} 
 are well defined because $T^{-1}$ does not occur in their definition. 
Furthermore, the lack of $T^{-1}$ could prevent 
the evaluation of the logarithmic negativity for the bipartition of the whole system in a pure state. 
This potential obstacle is overcome through a
proper choice of the algebra, as discussed in the Appendix\;\ref{app:purestatea_links}.

The generic correlation matrix element
\eqref{eq:correlators_genericM} simplifies to
\begin{equation}
\label{qq-pp-from-TTtranspose}
\big\langle 
\hat{\tilde{q}}_i\,\hat{\tilde{q}}_j
\big\rangle 
=\big\langle 
\hat{p}_i\,\hat{p}_j
\big\rangle =
\frac{1}{2}\big(\sqrt{T\,T^{\text{t}}} \,\big)_{i,j}
\end{equation}
where  we have used that
\begin{equation}
\label{TTtranpose-element}
\big( \,T\, T^{\text{t}} \big)_{i,j}
=2\,\delta_{i,j}-\delta_{i,j+1}-\delta_{i,j-1}-\delta_{i,N}\delta_{j,1} -\delta_{i,1}\delta_{j,N}  \,.
\end{equation}

\begin{figure}[t!]
\begin{center}  
\includegraphics[width=0.65\textwidth]{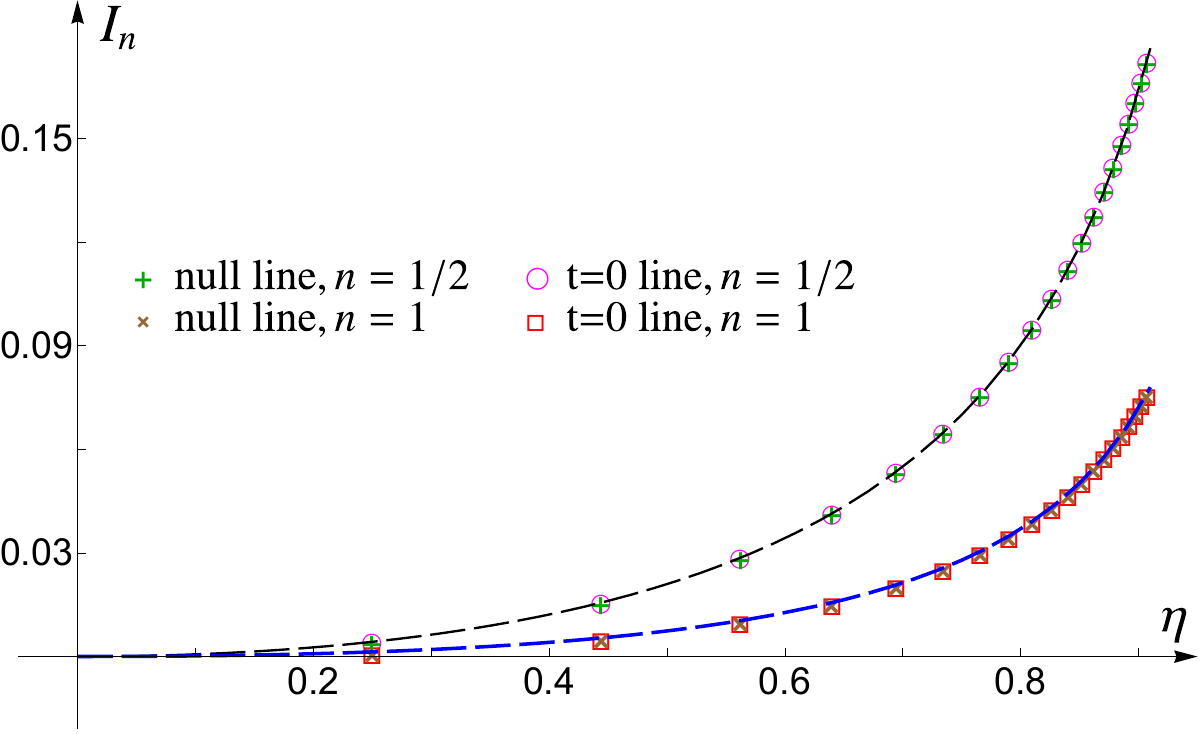}
\caption{The quantities 
$I_{1/2} $ and $I$ (top and bottom curve respectively),
evaluated through the algebra in the null line or in the $t=0$ line, 
discussed in  Sec.\,\ref{sec:staggered_boson_review} 
and  Sec.\,\ref{sec:realtimeapproach} respectively.  
The dashed black and dashed blue curves are given by \eqref{eq:Renyi_MI_twodisjoint} specialised to $n=1/2$
and \eqref{eq:MI_twodisjoint} respectively.
}
\label{fig:renyi_mutual_info}
\end{center}  
\end{figure}

The eigenproblem associated to the matrix $T\, T^{\text{t}}$ can be solved analytically.
Its eigenvalues and eigenvectors read respectively
\begin{equation}
\lambda_p=2 \big[1-\cos (2\pi p/N)\big] 
\;\;\qquad\;\; 
\Psi_j^{(p)}=\frac{e^{2\pi\ri \,p\, j/N}}{\sqrt{N}}
\;\;\qquad\;\; 
 1 \leqslant p \leqslant N \,.
\end{equation}
This leads to
\begin{equation}
\big(\sqrt{T \,T^{\text{t}}}\, \big)_{i,j}
=
\sum_{p=1}^{N} 
\sqrt{\lambda_p} \;
\Psi^{(p)}_i \,\overline{\Psi^{(p)}_j}
=
\frac{2 \sin (\pi/N)}{
N \big[\cos \! \big(2 \pi  (i-j)/N \big)- \cos (\pi /N) \big]} \,.
\end{equation}
Hence, (\ref{qq-pp-from-TTtranspose}) becomes
\begin{equation}
\label{eq:correlator_circle_sym}
\big\langle
\hat{\tilde{q}}_i \, \hat{\tilde{q}}_j
\big\rangle
\,=\,
\big\langle\hat{p}_i\,\hat{p}_j
\big\rangle
\,=\,
\frac{\sin (\pi/N)}{
N \big[\cos \! \big(2 \pi  (i-j)/N \big)- \cos (\pi /N) \big]}
\end{equation}
whose limit $N\rightarrow\infty$
provides the corresponding correlator on the line 
\begin{equation}
\label{qq-pp-large-N}
\lim_{N\to \infty}
\big\langle \hat{\tilde{q}}_i \,\hat{\tilde{q}}_j
\big\rangle\,
=
\lim_{N\to \infty}
\big\langle\hat{p}_i\,\hat{p}_j
\big\rangle
\,=\,
\frac{2}{ \pi \big[ 1-4(i-j)^2 \big]} \,.
\end{equation}

The expressions in \eqref{eq:correlator_circle_sym} 
and \eqref{qq-pp-large-N} restricted to the subsystem $V$
provide the correlation matrices 
$Q_ {V;\tilde{\mathbf{r}}}=P_ {V;\tilde{\mathbf{r}}}$ 
occurring in \eqref{eq:reducedA_correlators_genericM_rtilde} 
for this case, where $\widetilde{M}$ is the identity matrix.
Thus, \eqref{CVr-def} becomes
\begin{equation}
C_ {V;\mathbf{r}}
=
\sqrt{
T_V^{-1}\,Q_ {V;\tilde{\mathbf{r}}}\, 
T_V^{-\text{t}}\,Q_ {V;\tilde{\mathbf{r}}}
}  \;.
\end{equation}
We remark that $T_V$ is invertible when $N_V < N$, 
while this property does not hold for $N_V = N$.

It is worth checking that the two choices of algebra 
(i.e. the null line algebra of size $2N$  
and the $t=0$ slice algebra with $N$ bosonic degrees of freedom,
discussed in Sec.\,\ref{sec:staggered_boson_review}
and Sec.\,\ref{sec:realtimeapproach} respectively)
provide the same results and they are compatible with the corresponding field theory computations.  

The analytic expressions in the continuum to compare with 
have been obtained in \cite{Arias:2018tmw} 
and correspond to \eqref{eq:Renyi_MI_twodisjoint} and  \eqref{eq:MI_twodisjoint}. 
The results of this analysis are shown in Fig.\,\ref{fig:renyi_mutual_info}. For the points, for each fixed value of $\eta$, we compute the mutual information $I$ and the R\'enyi mutual $I^{(1/2)}$ across ten equivalent configurations. The system consists of two intervals of equal size $N_1 = N_2$ separated by a distance $D$. 
We then scale the entire geometry by a factor $k = 2, 4, \dots, 20$. The data points in the figure correspond to the continuum limit,
extrapolated for $k \to \infty$.
A perfect agreement is observed between the two choices of algebra
in the lattice and also with the analytic formulas in the continuum. 
In the case of the algebra in the null line, 
numerical analyses for the mutual information 
and the R\'enyi mutual information 
have been previously performed in \cite{Arias:2018tmw,Gentile:2025koe}.

\subsection{Entanglement negativity}
\label{sec:negativity_numerics}

The approach based on the algebra on the $t=0$ slice,
described in Sec.\,\ref{sec:realtimeapproach},
can be employed to compute the logarithmic negativity 
and the moments of the partial transpose 
in the lattice model for the chiral current. 
While both the algebra on the null line 
and the algebra on the $t=0$ slice 
can be used to evaluate the entanglement entropies, 
only in the latter one
it is understood how to evaluate these quantities
because a notion of momentum explicitly occurs.

It is worth reviewing first
the standard procedure to compute the moments of the partial transpose and the logarithmic negativity for a free bosonic system in a Gaussian state described by canonical variables
\cite{Simon:1999lfr,Audenaert:2002xfl, Marcovitch:2008sxc}
as done for the entanglement entropies
in  Sec.\,\ref{subsec-renyi-lattice}.
Given a reduced density matrix $\rho_V$, 
its partial transpose 
$\rho_V^{\textrm{\tiny $\Gamma_2$}}$ is also a Gaussian operator. 
Hence, 
from the reduced covariance matrix $\gamma_V$ characterising $\rho_V$, 
a real, symmetric and positive definite matrix $\gamma_V^{\textrm{\tiny $\Gamma_2$}}$ can be introduced 
to characterise $\rho_V^{\textrm{\tiny $\Gamma_2$}}$.

Since in this setup the partial transposition with respect to 
a certain region is implemented by the time reversal of the momenta 
in that region only, the matrix 
$\gamma_V^{\textrm{\tiny $\Gamma_2$}}$ 
can be constructed from $\gamma_V$ as follows 
\begin{equation}
\label{gamma-V-gamma2-def-Lambda}
\gamma_V^{\textrm{\tiny $\Gamma_2$}} 
= \Lambda \,\gamma_V \, \Lambda
\;\;\; \qquad \;\;\;
\Lambda \equiv \mathbb{I}_{N_V} \oplus \mathsf{R}_2
\;\;\; \qquad \;\;\;
\mathsf{R}_2 \equiv \mathbb{I}_{V_1} \oplus \big(\! -\!\mathbb{I}_{V_2} \big)
\end{equation} 
where $\mathbb{I}_{V_j}$ is the $N_j \times N_j$ identity matrix,
being $N_j$ the number of sites in $V_j$, for $j \in \{1,2\}$.

The Williamson decomposition of $\gamma_V^{\textrm{\tiny $\Gamma_2$}}$
provides its symplectic spectrum 
$\big\{\tilde{\nu}_i\, ; 1 \leqslant i \leqslant N_V \big\}$.
A key feature distinguishing this symplectic spectrum 
from the one of $\gamma_V$ is that 
$\tilde{\nu}_i\leqslant 1/2$ is allowed for certain symplectic eigenvalues of $\gamma_V^{\textrm{\tiny $\Gamma_2$}}$.
The symplectic spectrum of $\gamma_V^{\textrm{\tiny $\Gamma_2$}}$ 
provides  the moments of $\rho_V^{\textrm{\tiny $\Gamma_2$}}$ 
and the corresponding logarithmic negativity 
(defined in \eqref{neg-moments-def} 
and \eqref{log-neg-def} respectively)
as
\begin{equation}
\label{eq:logarithmic_moments_symplectic_spectrum}
\log \!\big[ 
\tr \! \big(\rho_V^{\textrm{\tiny $\Gamma_2$}}\big)^n
\,\big]
=
-\sum_{i=1}^{N_V}  
\log\left[
\big(\tilde{\nu}_i+ 1/2\big)^n
-
\big(\tilde{\nu}_i - 1/2\big)^n
\right]
\end{equation}
and
\begin{equation}
\label{eq:negativity_symplectic_spectrum}
\mathcal{E}= 
-\sum_{i=1}^{N_V} 
\log \! \Big[ \,
\big|\tilde{\nu}_i+ 1/2\big|
-
\big|\tilde{\nu}_i- 1/2\big|  
\,\Big]
\end{equation}
respectively.
When $\gamma_V$ is block diagonal, 
(\ref{gamma-V-gamma2-def-Lambda}) becomes 
 $\gamma_V^{\textrm{\tiny $\Gamma_2$}} = 
Q_V \oplus \widetilde{P}_V$,
with $\widetilde{P}_V \equiv \mathsf{R}_2 \,P_V \,\mathsf{R}_2$,
and the symplectic spectrum is given by the positive eigenvalues of $\widetilde{C}_V = \sqrt{Q_V \widetilde{P}_V}$.

In order to find a proper implementation of the partial transposition in the lattice model for the chiral current, it is worth considering 
the algebra on the $t=0$ slice.
It would be instructive to find an implementation 
also in the algebra on the null line.  
Considering the setup discussed in Sec.\,\ref{sec:realtimeapproach}
and the algebra on the $t=0$ line, 
we implement the partial transposition 
by acting on the reduced correlation matrices
in \eqref{eq:reducedA_correlators_genericM_rtilde}.
In particular, while the reduced correlation matrix
$Q_ {V;\mathbf{r}}$ is left unchanged 
under a partial transposition w.r.t. $V_2$,
the reduced correlation matrix $P_ {V;\mathbf{r}}$ 
is modified as follows
\begin{equation}
\widetilde{P}_ {V;\mathbf{r}} \equiv
\mathsf{R}_2 \,P_ {V;\mathbf{r}} \,\mathsf{R}_2 \,.
\end{equation}
As a consequence, the matrix  
$\widetilde{C}_ {V;\mathbf{r}} \equiv \sqrt{Q_ {V;\mathbf{r}} \, \widetilde{P}_ {V;\mathbf{r}}}$
specialised to the lattice model for the chiral current becomes
\begin{equation}
\label{eq:Ctilde_matrix}
\widetilde{C}_ {V;\mathbf{r}} 
= 
\sqrt{T_V^{-1} \,Q_ {V;\widetilde{\mathbf{r}}} \,
T_V^{-\text{t}} \,\mathsf{R}_2 \,
Q_ {V;\tilde{\mathbf{r}}} \,\mathsf{R}_2
}
\end{equation}
whose positive eigenvalues provide 
the moments of the partial transpose and the logarithmic negativity through 
\eqref{eq:logarithmic_moments_symplectic_spectrum}
and
\eqref{eq:negativity_symplectic_spectrum} 
respectively. The elements of $Q_ {V;\tilde{\mathbf{r}}}$ are obtained from  \eqref{eq:correlator_circle_sym} and \eqref{qq-pp-large-N}, 
for the circle and the line, respectively.

The discussion above allows us to outline the implementation of the partial transposition 
through the null line algebra.  
In particular, 
from \eqref{eq:from_b_to_r} 
and the associated discussion, 
it becomes clear that, considering  
a subsystem $\widetilde{V} = \widetilde{V}_1\cup \widetilde{V}_2$ made by $2N_V$ sites,
where $\widetilde{V}_1$ and $\widetilde{V}_2$ are two blocks made by 
$2N_1$ and $2N_2$ consecutive sites respectively, 
the partial transpose in $\widetilde{V}_2$
affects only the elements 
of the restricted correlation matrix $B_{\widetilde{V}}$
(from either (\ref{B-function-def}) or (\ref{B-matrix-line}))
that involve the modes $\hat{b}_{2j+1}$ with $j \in V_2$, 
by flipping their sign.
The rest of the procedure follows, paralleling the one 
described in Sec.\,\ref{sec:staggered_boson_review}. 
We will not pursue further this procedure, because it is
equivalent to the method discussed above, 
to get \eqref{eq:Ctilde_matrix}.

In Sec.\,\ref{sec-numerics-adj-blocks} 
and Sec.\,\ref{sec-numerics-disjoint-blocks}
we discuss the results of some numerical computations in the lattice model for the chiral current and compare them 
 with the corresponding analytic expressions found in Sec.\,\ref{sec:neg-cc-2int}.
In particular, 
\eqref{qq-pp-large-N} and \eqref{eq:correlator_circle_sym} 
are first employed to construct the matrix $\widetilde{C}_ {V;\mathbf{r}} $ in \eqref{eq:Ctilde_matrix}, 
whose spectrum is found numerically.  
Then, these eigenvalues determine 
the moments of the partial transpose 
and the logarithmic negativity 
through \eqref{eq:logarithmic_moments_symplectic_spectrum}
and \eqref{eq:negativity_symplectic_spectrum} 
respectively. 
We consider the lattice model for the chiral current 
either on the infinite line or on the circle made by $N$ sites.
The subsystem $V$, which contains $N_V$ sites, 
is the union of two blocks $V_1$ and $V_2$ made by 
$N_1$ and $N_2$ consecutive sites respectively
(hence $N_V = N_1 + N_2$), separated by $D$ 
consecutive sites.
In Sec.\,\ref{sec-numerics-adj-blocks}, 
the special case of adjacent blocks is explored, 
while in Sec.\,\ref{sec-numerics-disjoint-blocks}
we consider configurations where the blocks are disjoint
(i.e. $D=0$ and $D\neq 0$ respectively). 
Denoting by $a$ the lattice spacing, the continuum limit is defined in the standard way, namely by taking $a \to 0$ 
and $N_1$, $N_2$, $D$ and $N$ to $+\infty$ while the products 
$N_1 a = \ell_1$, $N_2 a = \ell_2$, $D a = d$ and $N a = L$ 
are kept fixed.
The setup on the infinite line corresponds to $L \to +\infty$.

An important consistency check is provided by the bipartition of a pure state, where the identity 
$\mathcal{E}  = S^{(1/2)}_{V_i} $ 
and its generalisation 
(\ref{pure-states-moments-relation}),
involving the moments of the partial transpose and the R\'enyi entropies, hold
\cite{Vidal:2002zz, Calabrese:2012ew, Calabrese:2012nk}.
This analysis, which requires a proper identification of the site and link operators through the choice of the algebra
\cite{Arias:2018tmw,Casini:2013rba,Casini:2014aia,Huerta:2022cqw}, is reported 
in the Appendix\;\ref{app:purestatea_links}.

\subsubsection{Numerical results for adjacent blocks}
\label{sec-numerics-adj-blocks}

In the following, we consider subregions 
$V= V_1 \cup V_2$ made by adjacent blocks either 
on the infinite line 
or on the circle made by $N$ sites. 

On the infinite line, our numerical analysis has been performed by 
keeping $N_V$ fixed and changing the value of $N_1$ such that 
$1\leqslant N_1\leqslant N_V$.
To compare the numerical results obtained from the lattice 
with the corresponding analytic formulas given by
(\ref{moments-adj-odd}), (\ref{moments-adj-even}) 
and (\ref{log-neg-adj}) in the continuum,
let us rewrite the logarithm of 
(\ref{moments-adj-odd}) and  (\ref{moments-adj-even})
in terms of the ratio $\chi_1 \equiv N_1/N_V$ and $N_V$
respectively as 
\begin{equation}
\label{eq:log_moments_odd_z}
\mathcal{M}_V^{( n_{\textrm{\tiny o}})}(\chi_1)
\equiv
\frac{1}{24}\left(\frac{1}{ n_{\textrm{\tiny o}}}- n_{\textrm{\tiny o}} \right)\log \!\left[
\chi_1\! \left(1-\chi_1\right)
\right]  
+
\frac{3}{24}\left(\frac{1}{ n_{\textrm{\tiny o}}}- n_{\textrm{\tiny o}} \right)\log (N_V) 
+
\log ( c_{ n_{\textrm{\tiny o}}} )
\end{equation}
and 
\begin{equation}
\label{eq:log_moments_even_z}
\mathcal{M}_V^{( n_{\textrm{\tiny e}} )}(\chi_1)
\equiv
\frac{1}{12}\left(\frac{2}{ n_{\textrm{\tiny e}}}
-
\frac{ n_{\textrm{\tiny e}}}{2}\right)
\log \! \big[\chi_1 \big(1-\chi_1 \big)\big] 
+
\frac{2- n_{\textrm{\tiny e}}^2}{8  n_{\textrm{\tiny e}}}
\log (N_V) 
+
\log (c_{ n_{\textrm{\tiny e}}}) \,.
\end{equation}
As for the logarithmic negativity,
taking the replica limit $ n_{\textrm{\tiny e}}\rightarrow 1$
of (\ref{eq:log_moments_even_z}), we find 
\begin{equation}
\label{eq:logneg_z}
\mathcal{E}(\chi_1)
=
\frac{1}{8}\log [
\chi_1\!
\left(1-\chi_1\right)] 
+
\frac{1}{8}\log (N_V)
\end{equation}
where $c_{1} = 1$ has been used. 

\begin{figure}[t!]
    \centering
      \subfigure{%
       \includegraphics[width=0.48\textwidth]{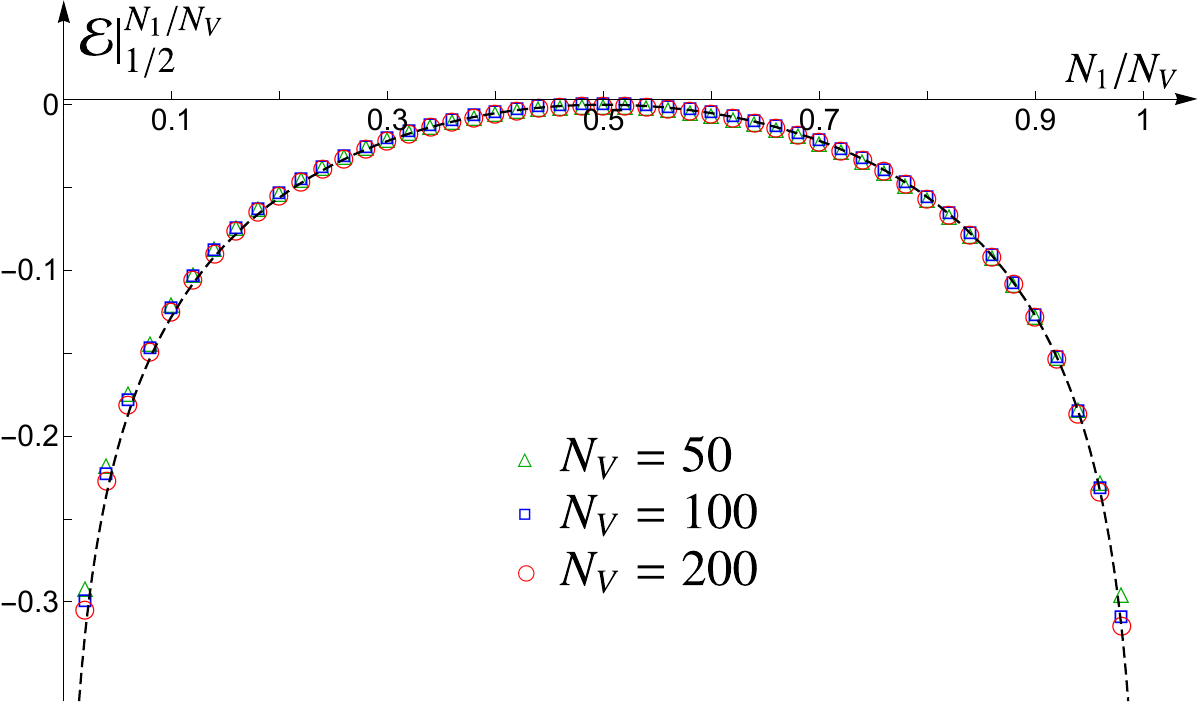}
        \label{fig:SubstractedLogNeg}
    }
    \hfill
\subfigure{%
        \includegraphics[width=0.49\textwidth]{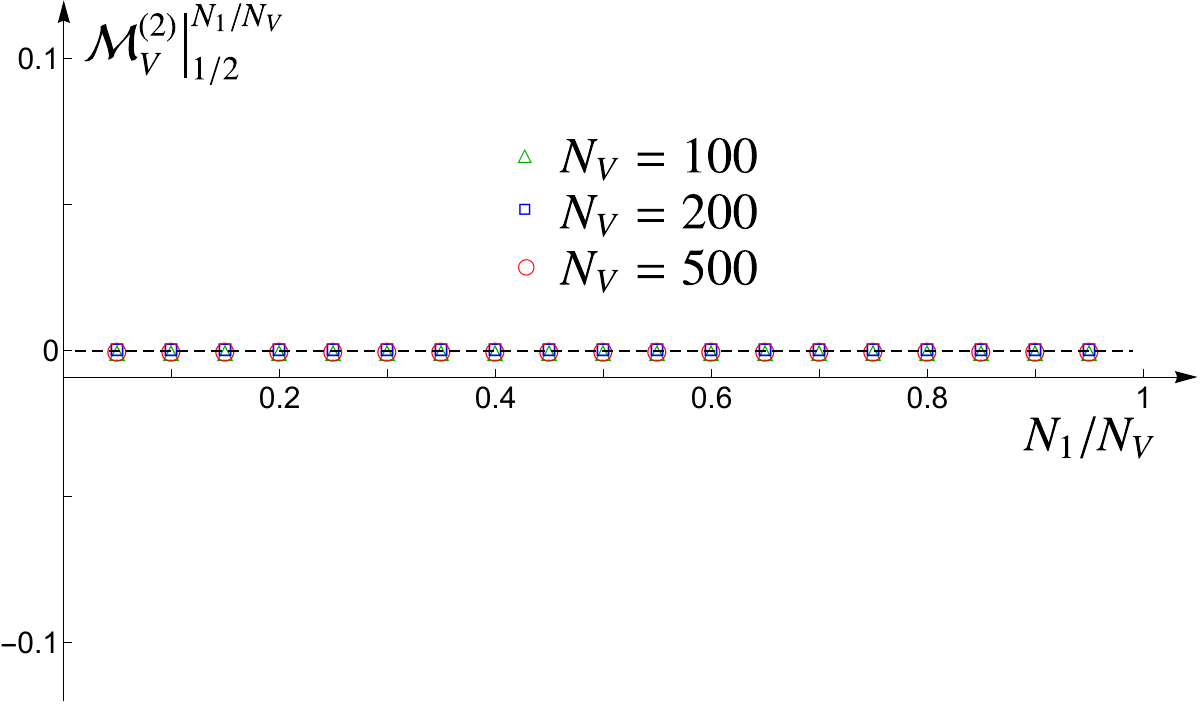}
        \label{fig:SubstractedRenyi2}
      
    }
    \hfill
   \subfigure{%
      \includegraphics[width=0.48\textwidth]{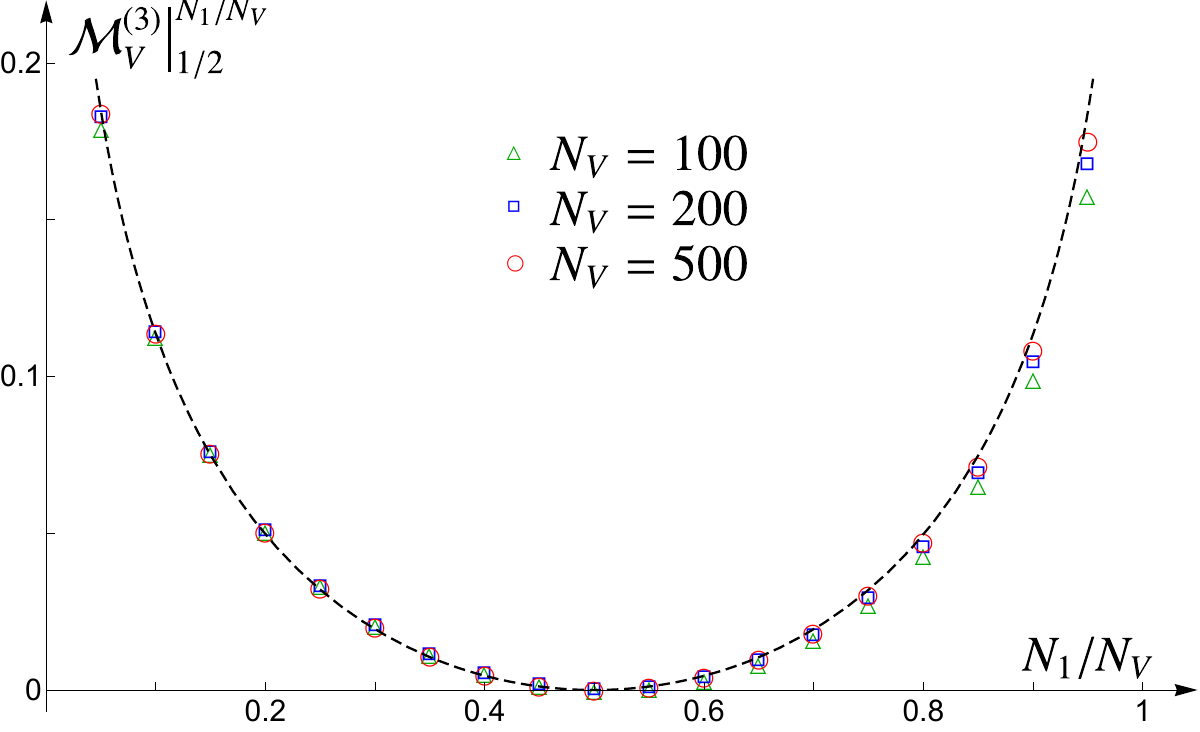}
        \label{fig:SubstractedRenyi3}
     
    }
    \subfigure{%
       \includegraphics[width=0.48\textwidth]{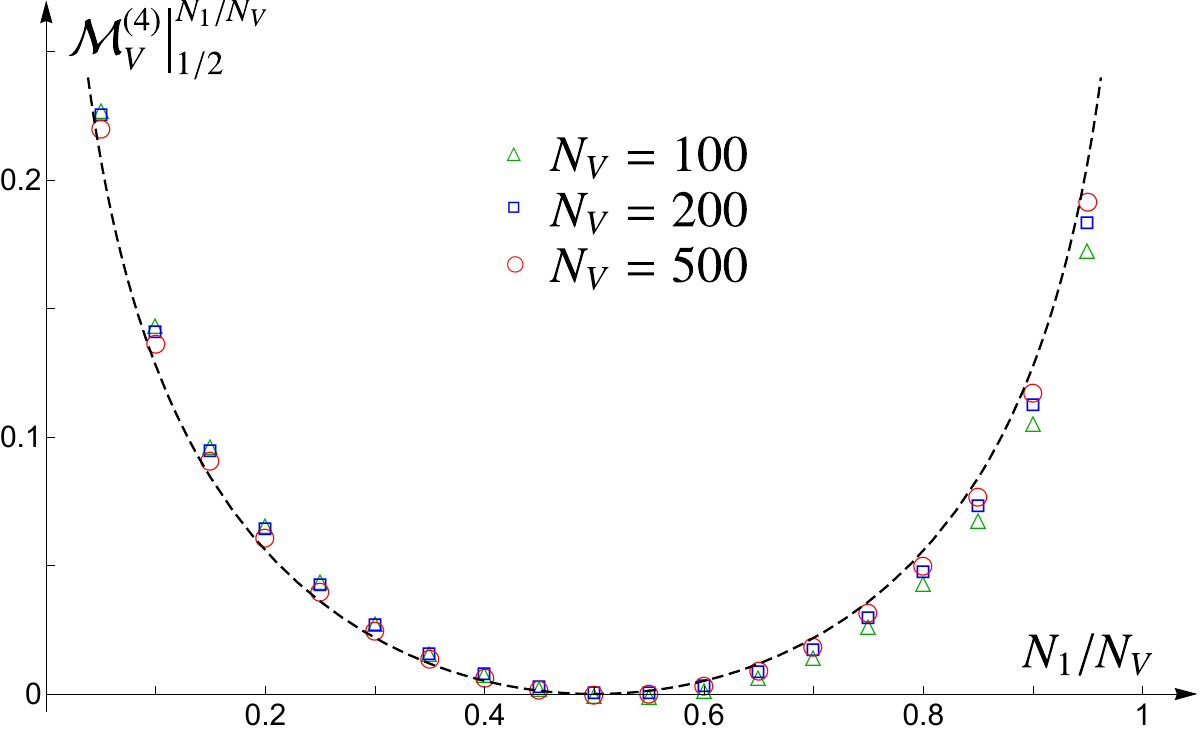}
        \label{fig:SubstractedRenyi4}
    
    }
    \hfill
    \subfigure{%
        \includegraphics[width=0.48\textwidth]{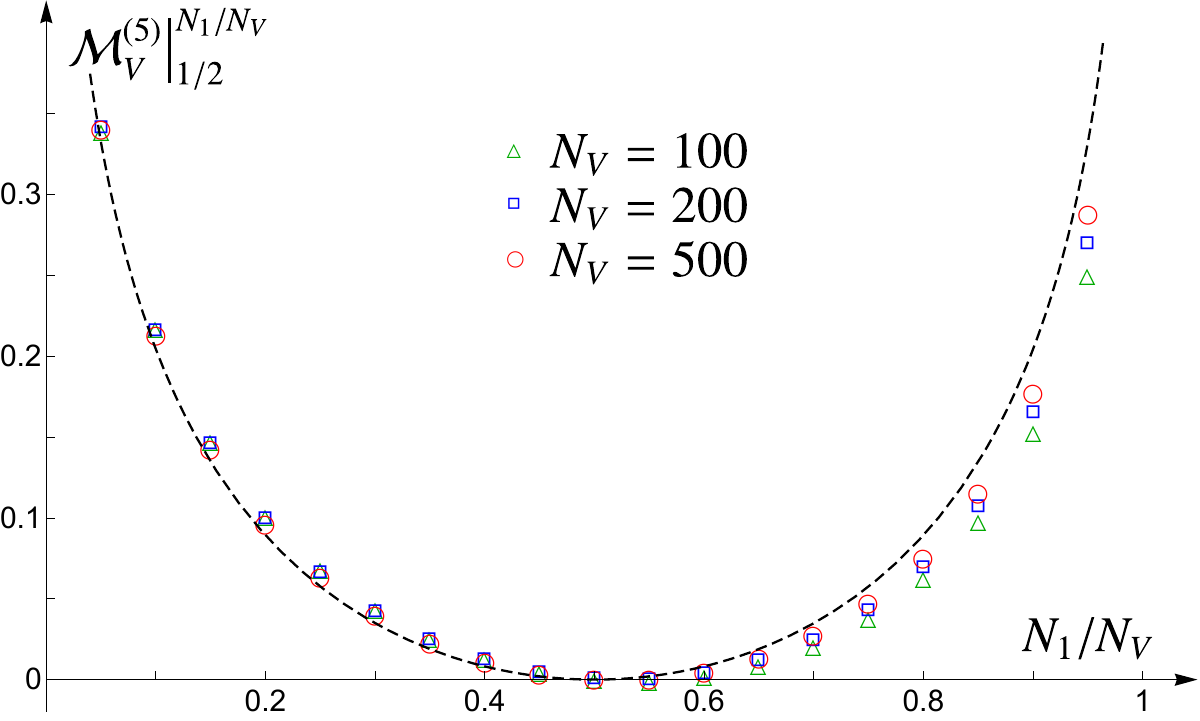}
        \label{fig:SubstractedRenyi5}
       
    }    
    \subfigure{%
     \includegraphics[width=0.48\textwidth]{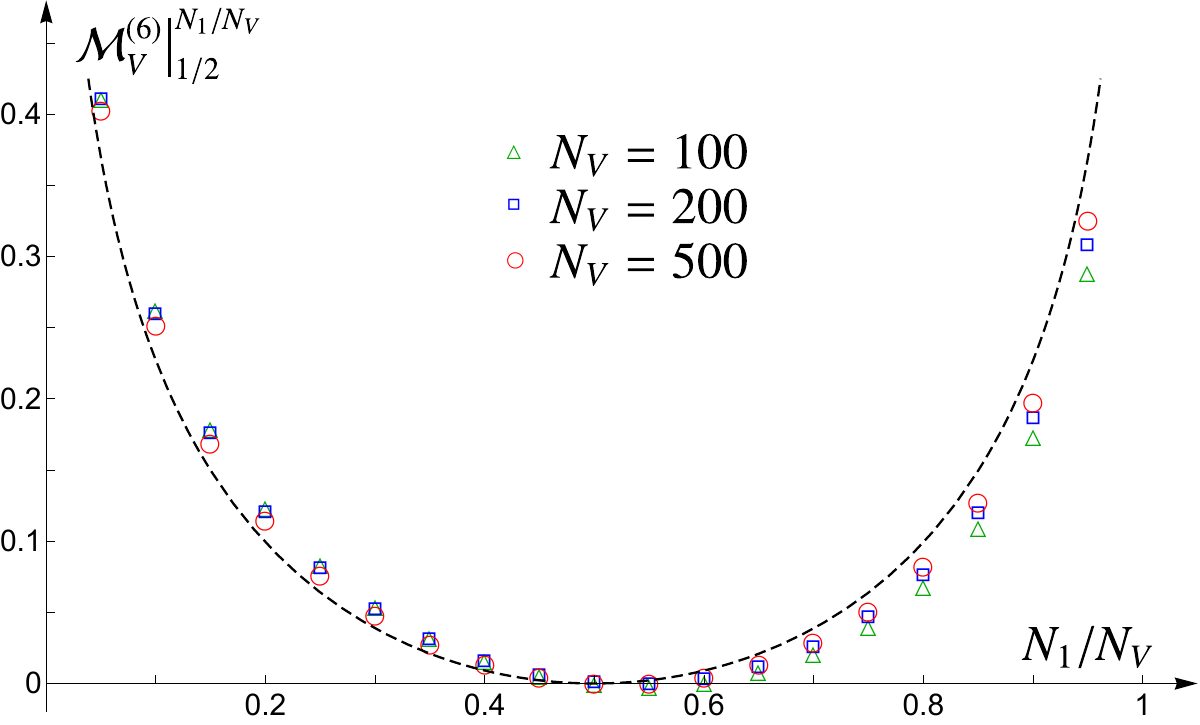}
        \label{fig:SubstractedRenyi6}
      
    }    
    \caption{Adjacent blocks in the line:
    The quantities 
    (\ref{eq:logneg_lattice_adjacent}) (top left panel)
    and (\ref{eq:moments_lattice_adjacent}) 
    for various values of $N_V$ and 
     $2 \leqslant n \leqslant 6$.
    The black dashed curves 
    are obtained from the analytic expressions
    \eqref{eq:log_moments_odd_z}-\eqref{eq:logneg_z}.
    }
    \label{fig:Substract_Vdjacent_line}
\end{figure}

In order to capture the dependence on $\chi_1$
and get rid of $c_n$ and $N_V$, let us consider 
the  differences given by 
\begin{equation}
\label{eq:moments_lattice_adjacent}
\mathcal{M}_V^{(n)}
\Big|^{N_1/N_V}_{1/2}
\! \equiv\,
\mathcal{M}_V^{(n)}(N_1/N_V)-\mathcal{M}_V^{(n)}(1/2)
\end{equation}
and
\begin{equation}
\label{eq:logneg_lattice_adjacent}
\mathcal{E}
\big|^{N_1/N_V}_{1/2}
\equiv\,
\mathcal{E}(N_1/N_V)-\mathcal{E}(1/2) \,.
\end{equation}

In Fig.\,\ref{fig:Substract_Vdjacent_line},
the numerical results obtained for these quantities in the lattice model for the chiral currents are compared 
with the corresponding analytic expressions
in the continuum limit (dashed curves),
for $2 \leqslant n \leqslant 6$,
observing a nice agreement. 
However, while perfect agreement occurs
for the logarithmic negativity (see the top left panel), 
the finite size effects increase with $n$,
as expected (see e.g. \cite{Calabrese:2012ew}).
Let us highlight the 
vanishing result obtained for $n_{\textrm{\tiny e}} = 2$
(see the top right panel of Fig.\,\ref{fig:Substract_Vdjacent_line}), 
as expected from the identity 
$\tr \!\big(\rho_V^{\textrm{\tiny $\Gamma_2$}} \big)^2
=\tr \rho_V^2$ in the continuum. 
It is remarkable that the numerical data points from the lattice also follow this identity. 
This suggests that a nontrivial relation involving the underlying correlation matrices might occur.

\begin{figure}[t!]
    \centering
      \subfigure{%
        \includegraphics[width=0.48\textwidth]{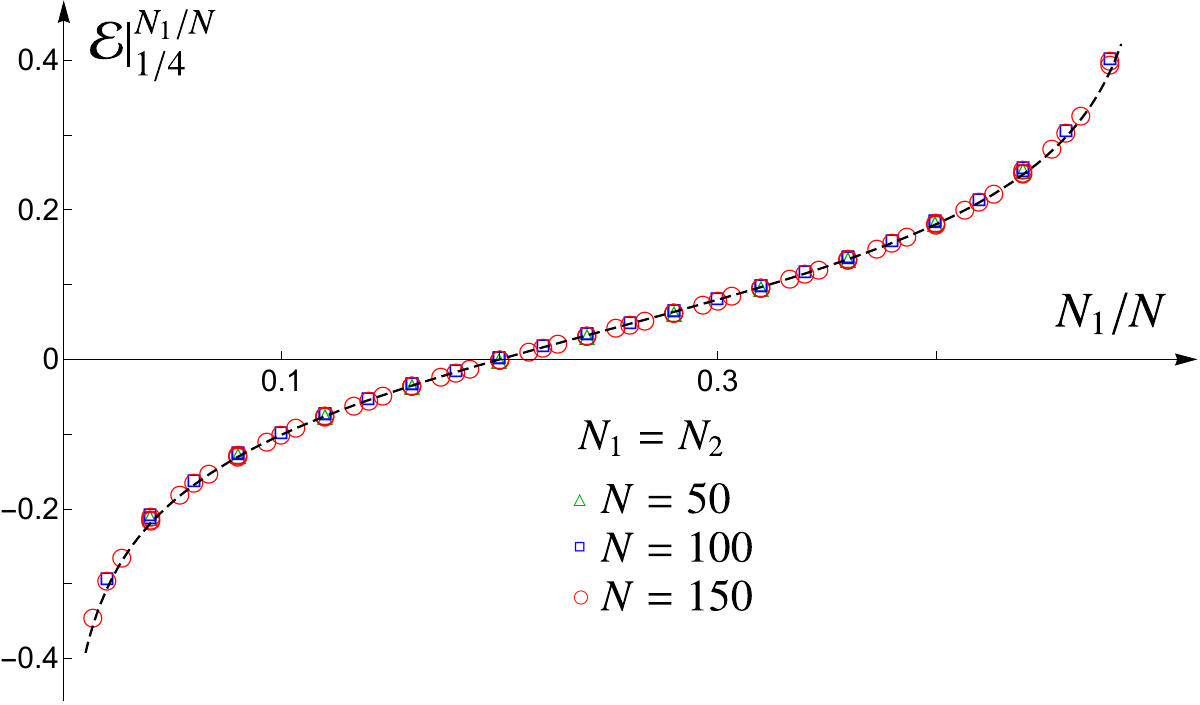}
        \label{fig:PBC_SubstractedLogNeg}
    }
    \hfill
    \subfigure{%
        \includegraphics[width=0.48\textwidth]{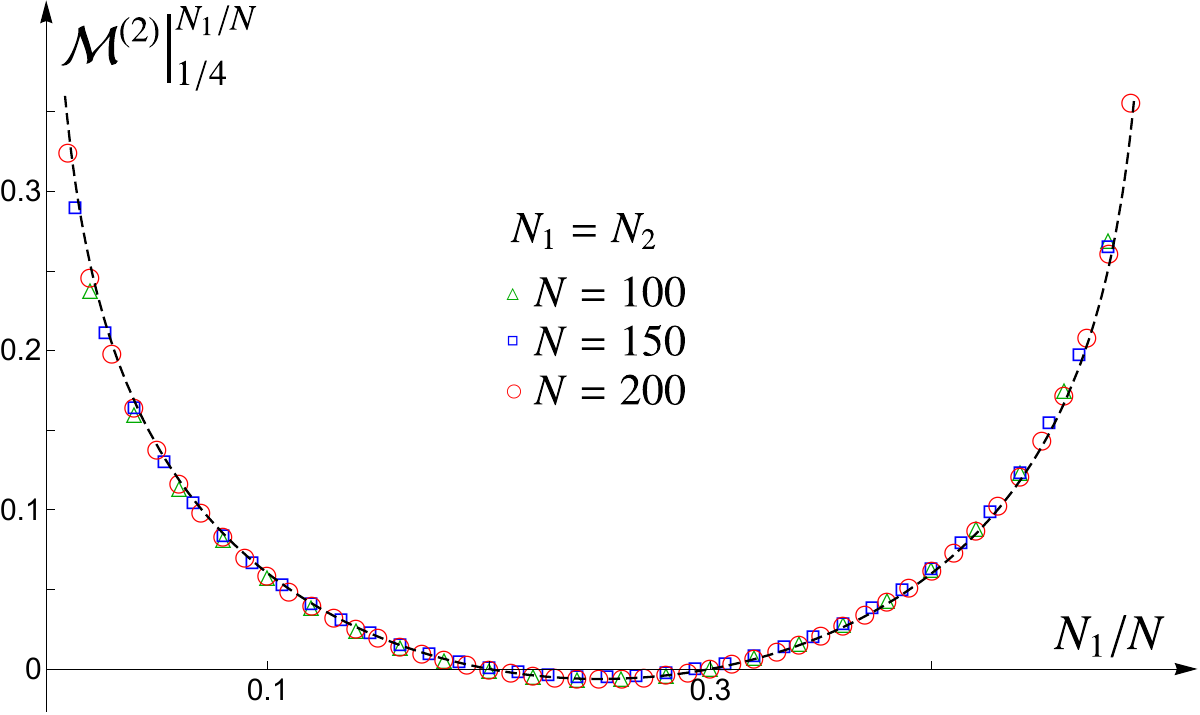}
        \label{fig:PBC_SubstractedRenyi2}
    }
    \hfill
    \subfigure{%
        \includegraphics[width=0.48\textwidth]{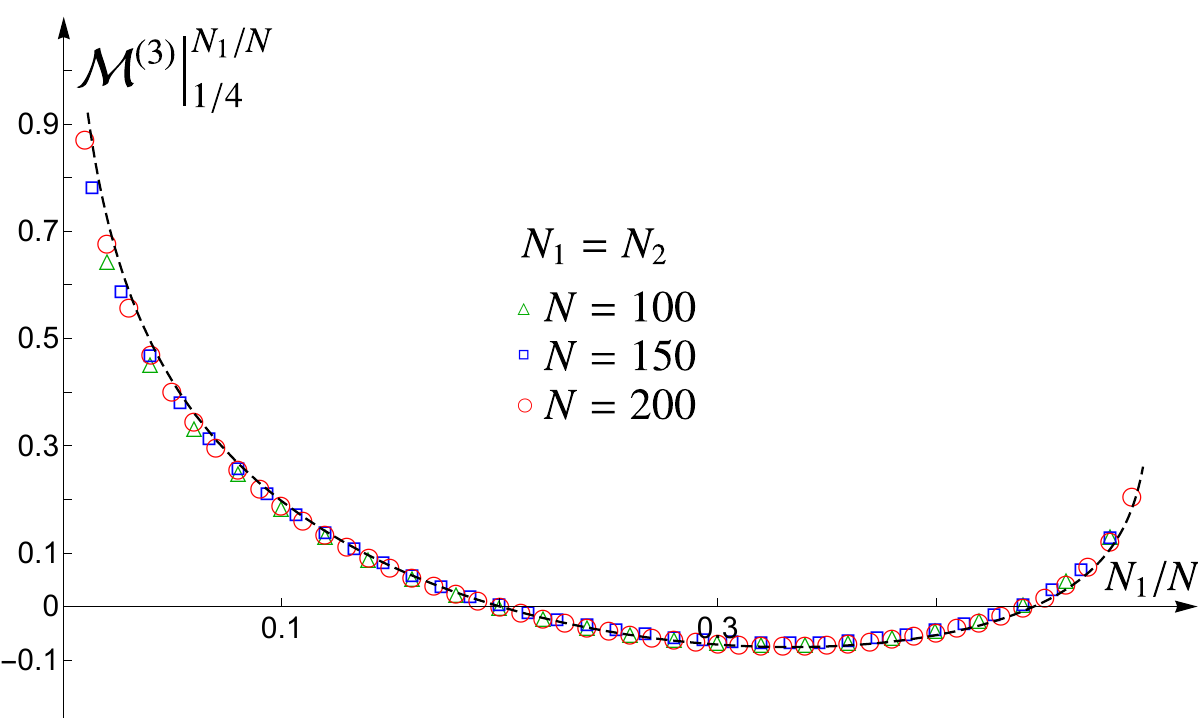}
        \label{fig:PBC_SubstractedRenyi3}
    }
    \subfigure{%
        \includegraphics[width=0.48\textwidth]{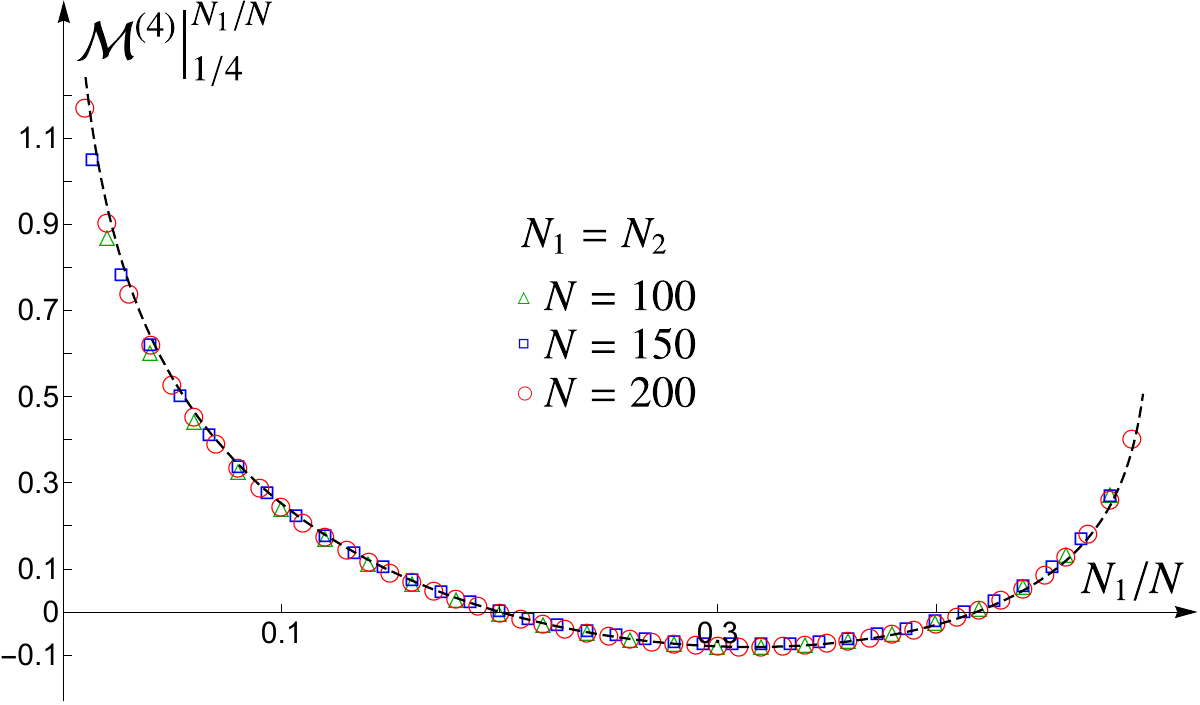}
        \label{fig:PBC_SubstractedRenyi4}
    }
    \hfill
    \subfigure{%
        \includegraphics[width=0.48\textwidth]{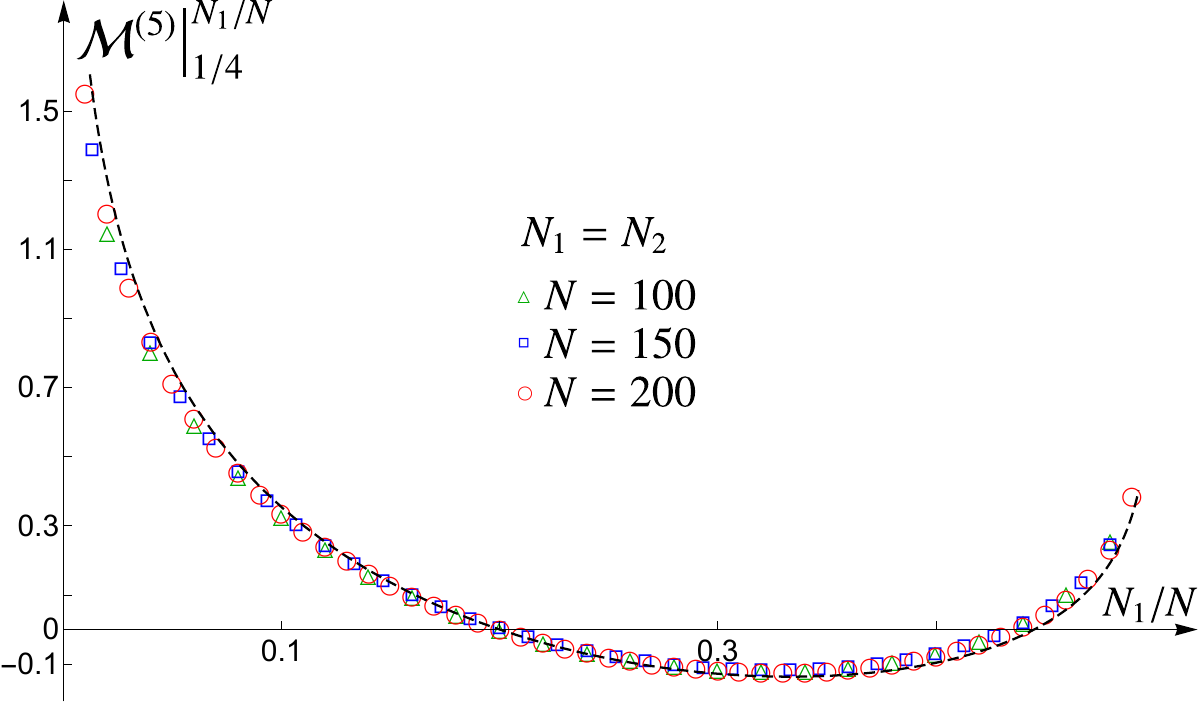}
        \label{fig:PBC_SubstractedRenyi5}
    }    
  \subfigure{%
        \includegraphics[width=0.48\textwidth]{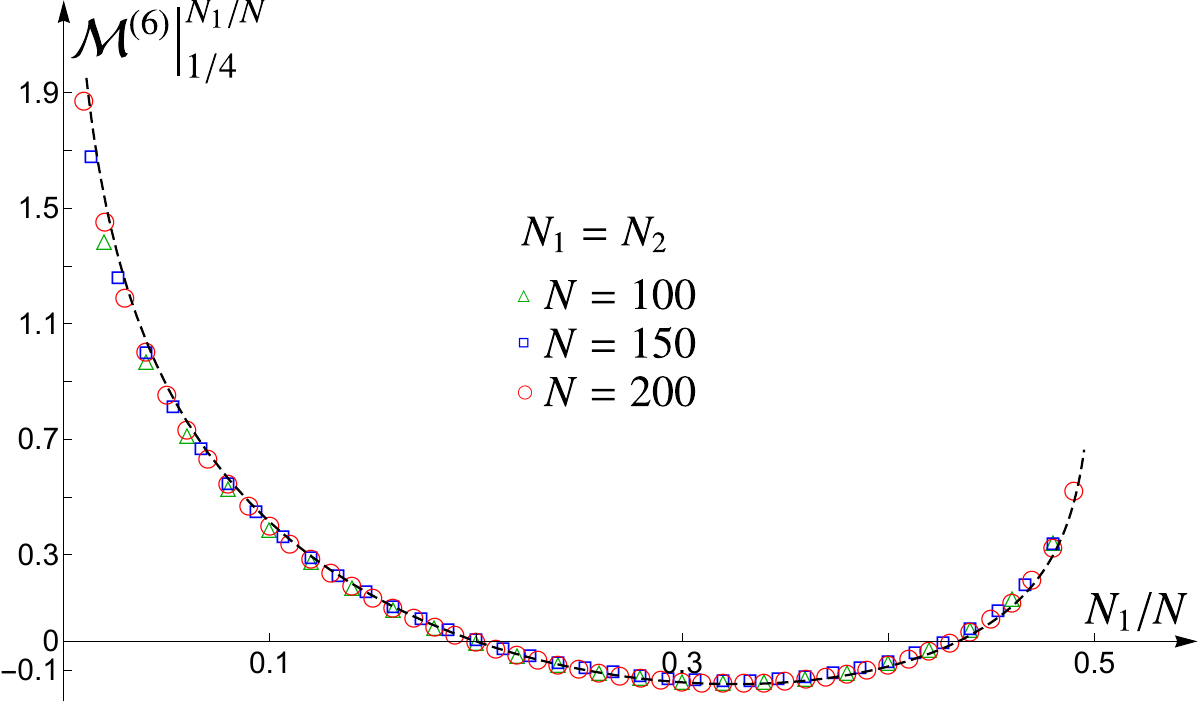}
        \label{fig:PBC_SubstractedRenyi6}
    }    
    \caption{Adjacent blocks in the circle:
    The quantities 
    (\ref{eq:logneg_lattice_adjacent_PBC}) (top left panel)
    and (\ref{eq:moments_lattice_adjacent_PBC}) 
    for various values of $N$ and 
     $2 \leqslant n \leqslant 6$.
    The black dashed curves are obtained from 
    the analytic expressions in
    \eqref{eq:log_moments_odd_z_periodic}-\eqref{eq:logneg_adjacent_periodic}.}
    \label{fig:Substract_Vdjacent_PBC}
\end{figure}

On the circle, we have obtained our numerical results 
by considering two blocks having equal size $N_1 = N_2$,
which varies while 
the total number of sites $N$ in the circle is kept fixed. 
For the circle, 
the analytic formulas in the continuum
are given by
(\ref{moments-adj-odd-temp}), (\ref{moments-adj-even-temp}) 
and (\ref{log-neg-adj-temp}).
It is convenient to rewrite the logarithm of 
(\ref{moments-adj-odd-temp}) and (\ref{moments-adj-even-temp}) 
in terms of the ratio $\tilde{\chi}_1\equiv N_1/N$ and $N$
respectively as 
\begin{eqnarray}
\label{eq:log_moments_odd_z_periodic}
\mathcal{M}_V^{( n_{\textrm{\tiny o}})}(\tilde{\chi}_1;N)
&=&
\frac{1}{24}\left(\frac{1}{ n_{\textrm{\tiny o}}}- n_{\textrm{\tiny o}} \right)
\log \! \big[ 
\big(\sin(\pi \tilde{\chi}_1) \big)^2 \sin(2\pi \tilde{\chi}_1)\big] 
\nonumber
\\ 
\rule{0pt}{.7cm}
& &
+\,\frac{3}{24}\left(\frac{1}{ n_{\textrm{\tiny o}}}- n_{\textrm{\tiny o}} \right)\log(N/\pi) 
+2\log ( c_{ n_{\textrm{\tiny o}}} )
\end{eqnarray}
and
\begin{eqnarray}
\label{eq:log_moments_even_z_periodic}
\mathcal{M}_{V}^{( n_{\textrm{\tiny e}})}(\widetilde{\chi}_1;N)
&=&
\frac{1}{6}
\left(\frac{2}{ n_{\textrm{\tiny e}}}-\frac{ n_{\textrm{\tiny e}}}{2}\right)
\log \! \big[ \sin (\pi \widetilde{\chi}_1) \big]
-
\frac{1}{12}
\left(\frac{ n_{\textrm{\tiny e}}}{2}+\frac{1}{ n_{\textrm{\tiny e}}}\right)
\log \!\big[ \sin (2\pi \widetilde{\chi}_1) \big]
\nonumber 
\\
\rule{0pt}{.7cm}
& &
+\,
\frac{2- n_{\textrm{\tiny e}}^2}{8  n_{\textrm{\tiny e}}}\,
\log (N/\pi) +2\log ( c_{ n_{\textrm{\tiny e}}}) \,.
\end{eqnarray}
Taking the replica limit $ n_{\textrm{\tiny e}}\rightarrow 1$
of (\ref{eq:log_moments_even_z}), 
for the logarithmic negativity we find 
\begin{equation}
\label{eq:logneg_adjacent_periodic}
\mathcal{E}(\widetilde{\chi}_1; N)
=
\frac{1}{8} \log \! 
\left[\, \frac{N}{2\pi}\tan(\pi \widetilde{\chi}_1) \,\right] \,.
\end{equation}

Also on the circle it is convenient to explore quantities where 
only the dependence on $\widetilde{\chi}_1$ occurs, while 
$c_n$ and $N_V$ simplify. 
In particular, let us consider 
\begin{equation}
\label{eq:moments_lattice_adjacent_PBC}
\mathcal{M}^{(n)}\Big|^{N_1/N}_{1/4}
\equiv \,
\mathcal{M}_V^{(n)}(N_1/N;N)-\mathcal{M}_V^{(n)}(1/4;N)
\end{equation}
and
\begin{equation}
\label{eq:logneg_lattice_adjacent_PBC}
\mathcal{E} \big|^{N_1/N}_{1/4}
\equiv \,\mathcal{E}(N_1/N;N)-\mathcal{E}(1/4;N) \,.
\end{equation}

In Fig.\,\ref{fig:Substract_Vdjacent_PBC}
we show our numerical results for (\ref{eq:moments_lattice_adjacent_PBC})
and (\ref{eq:logneg_lattice_adjacent_PBC}) 
in the lattice model for the chiral current, comparing them with the corresponding analytic expressions in the continuum (dashed curves)
that are obtained from 
\eqref{eq:log_moments_odd_z_periodic}-\eqref{eq:logneg_adjacent_periodic}.
The agreement is remarkably good and the finite size effects are smaller than the ones observed for the corresponding quantities on the line (see Fig.\,\ref{fig:Substract_Vdjacent_line}).
Notice that, in this figure, we set $N_1=N_2$
and keep $N$ fixed, while the ratio $N_1/N$ changes. 
Hence, $\mathcal{M}^{(2)}\big|^{N_1/N}_{1/4}$ 
displays a non-constant profile. 
Instead, the quantity $\mathcal{M}^{(2)}\big|^{N_1/N}_{1/4}$ 
remains constant
when $N_V/N$ is kept fixed while the ratio $N_1/N_V$ changes,
as one can observe from \eqref{moments-adj-even-temp}.

\subsubsection{Numerical results for disjoint blocks}
\label{sec-numerics-disjoint-blocks}

When the subsystem $V$ is the union of two disjoint blocks in the lattice model, either in the infinite line or in the circle, 
the important variable to consider 
is the cross ratio $\eta\in(0,1)$, 
whose explicit expression depends on whether the lattice model is on the line or on the circle
and reads
\begin{equation}
\label{eta-def-lattice}
\eta=
\left\{\begin{array}{ll}
\displaystyle
     \frac{N_1 N_2}{(N_1+D)\,(N_2+D)} 
      & \text{line} \vspace{0.3cm}
     \\ 
     \rule{0pt}{.6cm}
\displaystyle     
     \frac{\sin (\pi N_1/N) \,\sin (\pi N_2/N)}{
     \sin \!\big([\pi(N_1+D)/N\big)\,
     \sin \!\big([\pi(N_2+D)/N\big)} 
     \hspace{1.cm} &\text{circle} \,.
\end{array}\right.
\end{equation}

On the infinite line, 
our numerical data points have been obtained 
by considering equal blocks of given
length $N_1 = N_2$ and varying the number of sites $D$ separating them. 
Instead, on the circle, 
the numerical results correspond to subsystems where 
the length of the second block $N_2$ 
and the separation distance $D$ have been fixed,
while the length of the first block $N_1$ changes. 
In both these setups, we have that the corresponding 
cross ratio $\eta$ spans the whole domain $(0,1)$.

\begin{figure}[t!]
    \centering
      \subfigure{%
        \includegraphics[width=0.48\textwidth]{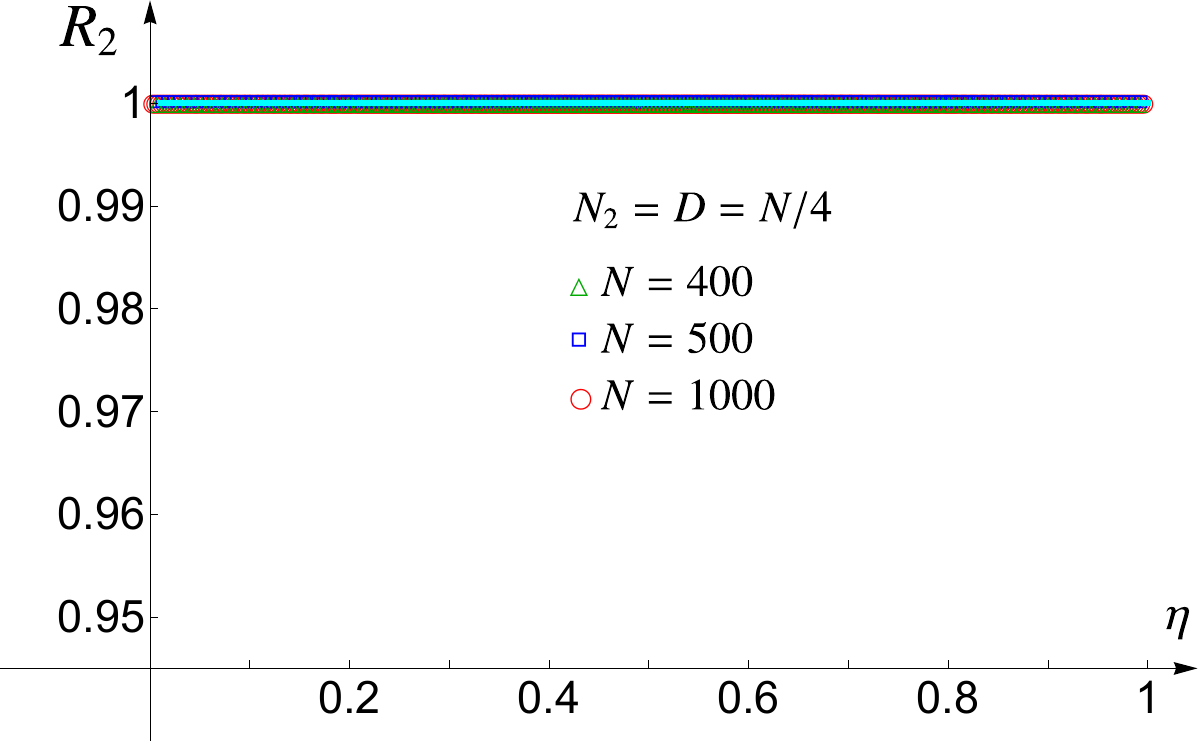}
        \label{fig:periodicRn2}
    }
    \hfill
          \subfigure{%
        \includegraphics[width=0.48\textwidth]{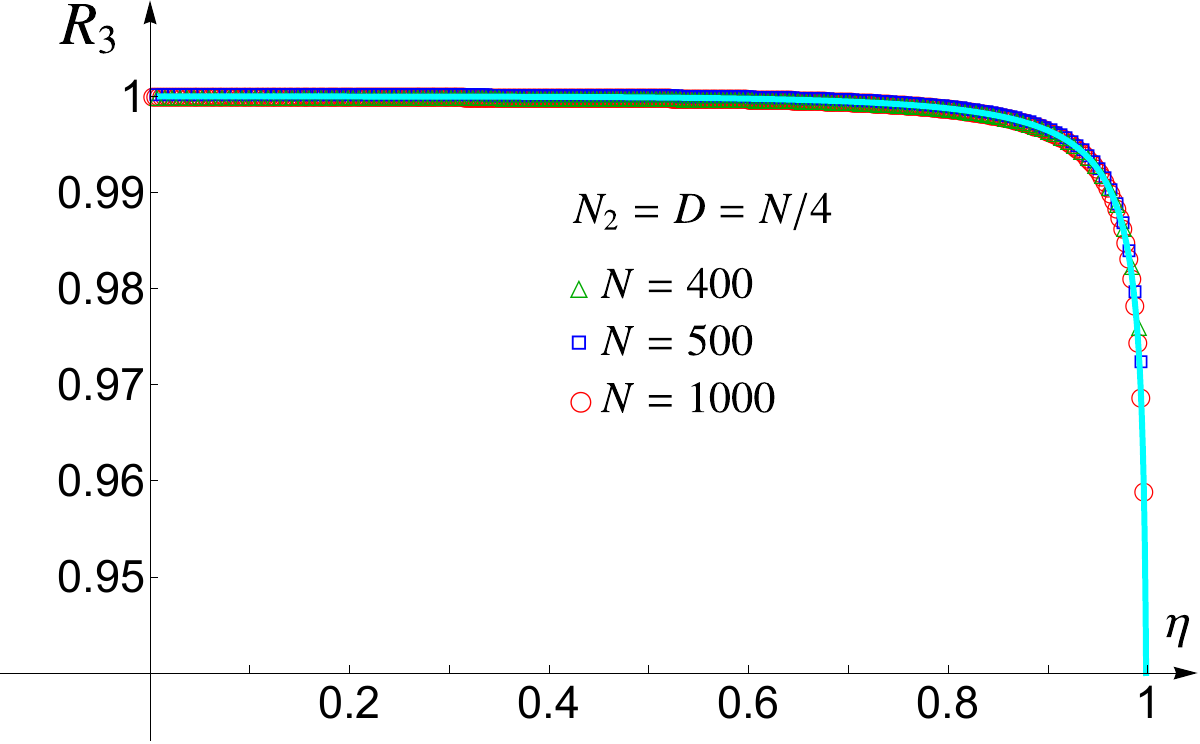}
        \label{fig:periodicRn3}
    }
    \hfill
    \subfigure{%
        \includegraphics[width=0.48\textwidth]{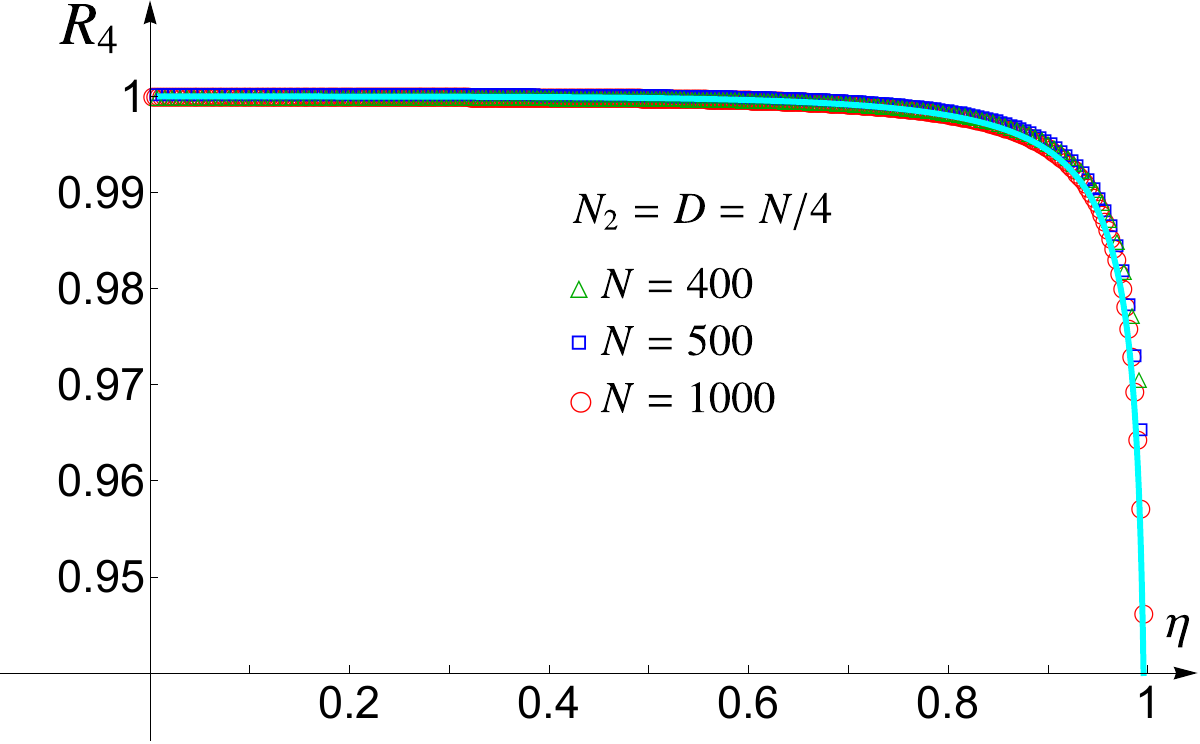}
        \label{fig:periodicRn4}
    }
    \hfill
    \subfigure{%
        \includegraphics[width=0.48\textwidth]{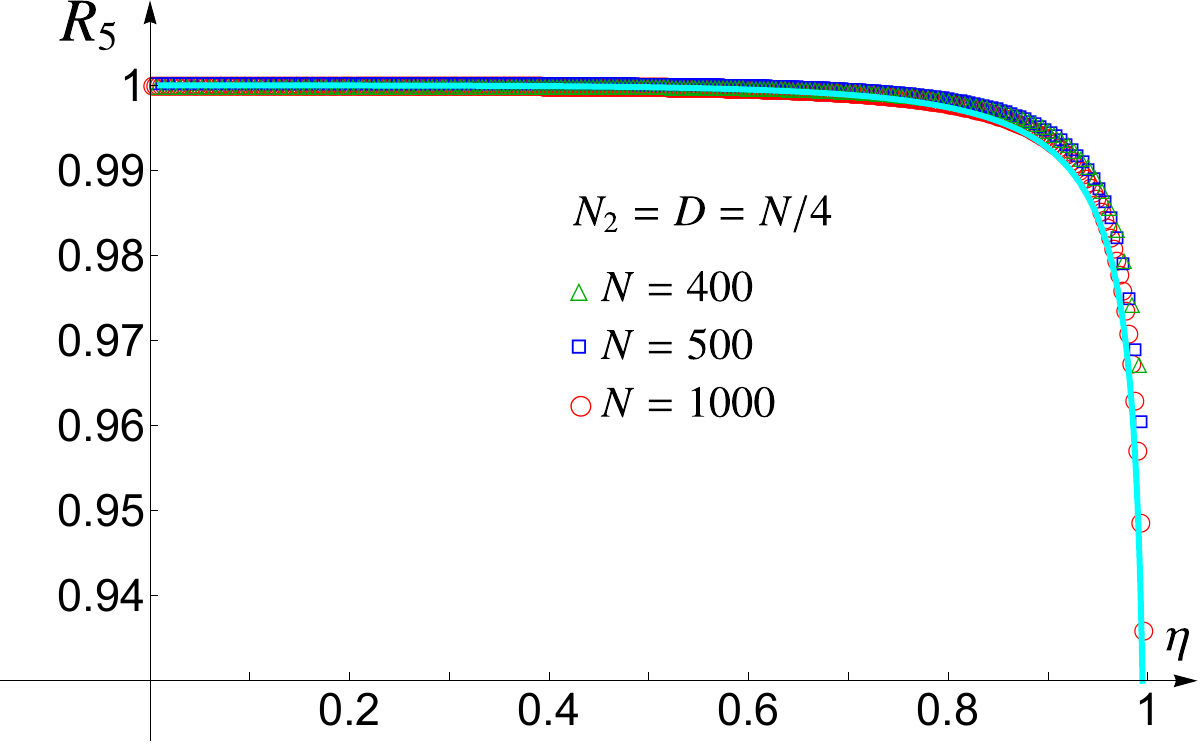}
        \label{fig:periodicRn5}
    }
\caption{Disjoint blocks in the circle:
The ratio $R_n $ for $n\in\{2,3,4,5\}$. 
The solid lines correspond to 
\eqref{Rn-CFT-2intervals-explicit-2F1}.
}
\label{fig:PeriodicRn}
\end{figure}

In Fig.\,\ref{fig:LineRn} and Fig.\,\ref{fig:PeriodicRn} 
we consider the ratio \eqref{Rn-ratio-def} for $n\in\{2,3,4,5\}$,
whose continuum limit is given by the function of $\eta \in (0,1)$ 
in \eqref{Rn-CFT-2intervals-explicit-2F1},
which does not contain normalization constants and 
provides the solid cyan lines. 
The agreement between the lattice data points 
and the CFT formula \eqref{Rn-CFT-2intervals-explicit-2F1} 
is quite remarkable. 
A first important check corresponds to the case $n=2$, where 
$R_2 = 1$ identically, which is confirmed also by the lattice points with high precision. 
This suggests that a nontrivial identity 
involving the matrices entering in this quantity 
might occur. 
The finite size effects increase with $n$, 
as already observed in previous analyses 
of this quantity in other models 
\cite{Calabrese:2012nk, Calabrese:2013mi, Alba:2013mg}.

\begin{figure}[t!]
    \centering
        \includegraphics[width=0.9\textwidth]{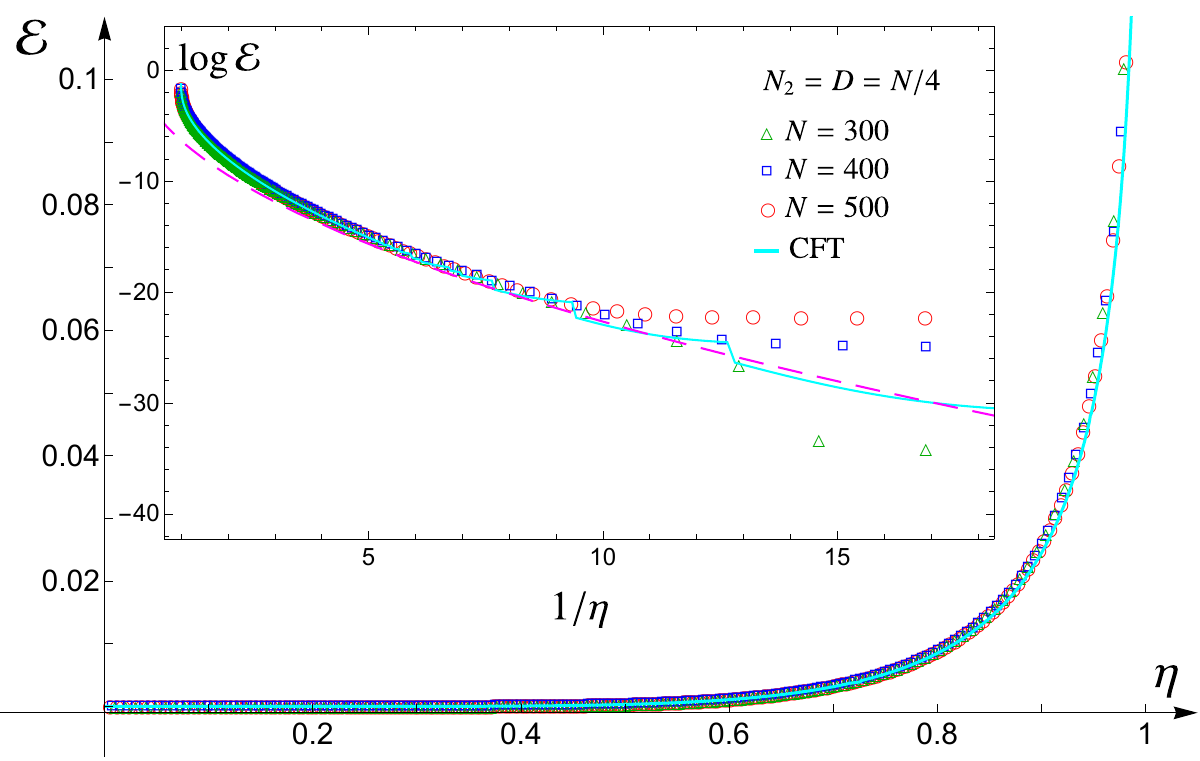}
    \caption{
    Logarithmic negativity for two disjoint blocks on the circle.
    The cyan solid line 
    corresponds to (\ref{log-neg-from-AbelPlana}). 
    The inset highlights the large separation regime $\eta \to 0^+$, where an exponential decay \eqref{eq:EN_decay_exp} occurs (magenta dashed line). 
    } 
    \label{fig:EnegAlmostAdj_circle}
\end{figure}

The main results of our numerical analyses 
correspond to the logarithmic negativity for two disjoint blocks, either on the line or on the circle, 
in terms of the cross ratio (\ref{eta-def-lattice}).
They are reported in Fig.\,\ref{fig:EnegAlmostAdj_line} 
and Fig.\,\ref{fig:EnegAlmostAdj_circle} respectively.
where the data points have been obtained numerically 
from the lattice model, while the cyan solid curves 
correspond 
to the CFT expressions (\ref{log-neg-from-AbelPlana}) 
for the chiral current in the continuum and it is the same one
in both these figures.
The agreement between the lattice data points 
and the analytic curve for the chiral current 
is excellent. 
In the insets of these figures we highlight the asymptotic 
regime of large separation distance, where $\eta \to 0^+$.
In this regime, an exponential decay is observed and 
the lattice data points nicely agree with the CFT result (\ref{eq:EN_decay_exp}),
that correspond to the magenta dashed line in both these figures. 

The exponential decay of the logarithmic negativity of two disjoint blocks for large separation distance has already been observed 
in the data points obtained numerically 
in the harmonic chain 
\cite{Marcovitch:2008sxc,Calabrese:2012nk,Klco:2021cxq,Klco:2020rga}
and in the critical Ising chain 
\cite{Calabrese:2013mi}
(see also \cite{Parez:2023xpj}).


\subsection{Comparison with the fermionic partial time-reversal}
\label{sec:fermionic_negativity_difference}

Consider the one dimensional tight-binding model,
whose Hamiltonian is 
\begin{equation}
\label{eq:tight_binding_model}
\hat{H} = -\frac{1}{2} \sum_j 
\left( \hat{c}_j^{\dagger} \,\hat{c}_{j+1} 
+ \hat{c}_{j+1}^{\dagger} \,\hat{c}_j 
\right)
\end{equation}
where $\hat{c}_i$ and $\hat{c}_i^{\dagger}$ 
denote the fermionic annihilation and creation operators 
satisfying the canonical anticommutation relations
\begin{equation}
\big\{\,
\hat{c}_i \, , \hat{c}_j^{\dagger}
\,\big\} = \delta_{i,j}
\;\;\;\qquad \;\;\;
\big\{\,\hat{c}_i \, , \hat{c}_j \,\big\} 
= 
\big\{\, \hat{c}_i^{\dagger}\, , \hat{c}_j^{\dagger} \,\big\} 
= 0 \,.
\end{equation}
The reduced density matrix for a subsystem $V$ of this quadratic model in its ground state
is fully characterised by the reduced correlation matrix $C_V$,
whose generic element is defined as 
\begin{equation}
\label{eq:fermion_CA}
\big(C_V\big)_{i,j} 
= \,
\big\langle 
\hat{c}_i^{\dagger} \,\hat{c}_j 
\big\rangle \Big|_{i,j \in V} 
= 
\frac{\sin \!\big[k_{\textrm{\tiny F}} (i-j)\big]}{\pi (i-j)}
\;\;\;\;\qquad\;\;\;
i,j \in V
\end{equation}
where $k_{\textrm{\tiny F}} = \pi/2$ at half-filling. 
The continuum limit is described by the massless Dirac fermion, 
that has been already considered in Sec.\,\ref{sec:Dirac_fermion} . 
Hence, in the scaling limit, 
the lattice results obtained for the tight-binding model
divided by $2$ can be compared with the ones for a chiral 
component of the massless Dirac field 
(see Sec.\,\ref{sec-comparison-chiral-fermion}).

\begin{figure}[t!]
\centering
\begin{subfigure}
    \centering
    \includegraphics[width=0.48\textwidth]{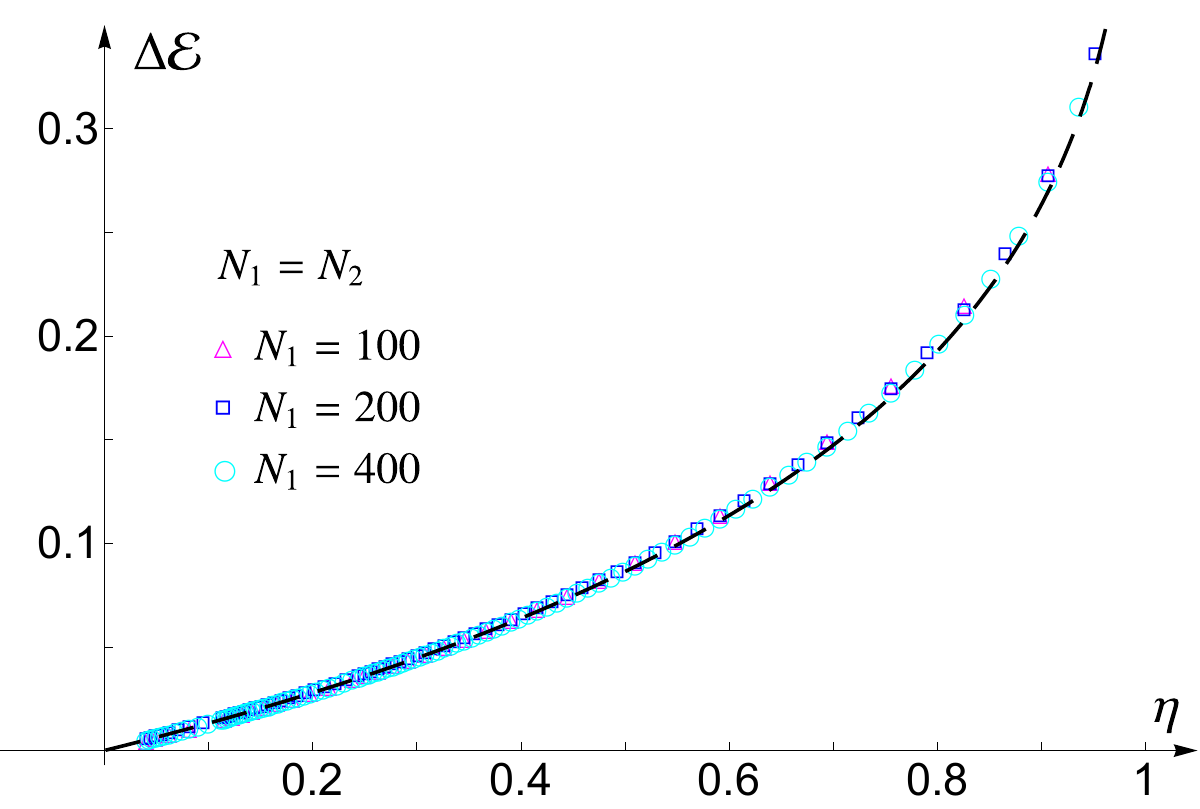}
\end{subfigure}
\hfill
\begin{subfigure}
    \centering
    \includegraphics[width=0.48\textwidth]{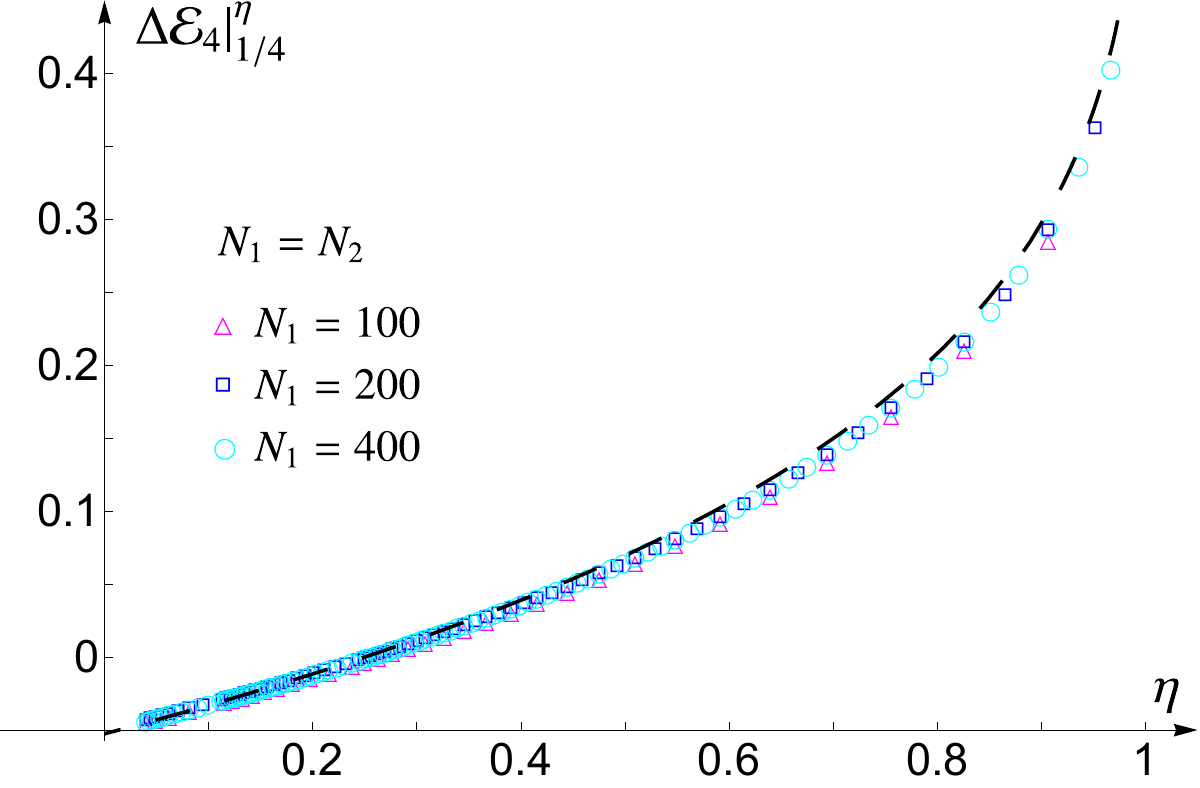}
\end{subfigure}
\caption{Difference in logarithmic negativities between the chiral current and its completion, the chiral Dirac CFT, for two equal blocks on the line, whose size in the lattice model (\ref{eq:tight_binding_model}) is parametrised by $N_1$
as we vary the distance between them
(see also Appendix\;\ref{app:fermion_numerics} for the determination of the lattice data points).
The dashed curve in the right and left panel is obtained from 
\eqref{Delta-E-ne-loglog} for $n_{\textrm{\tiny e}}=4$
and its analytic continuation to $n_{\textrm{\tiny e}}=1$ respectively. 
}
\label{fig:diff_neg}
\end{figure}

In this model, it is worth considering 
the partial time-reversal of the reduced density matrix 
\cite{Shapourian:2016cqu}, which is again a Gaussian operator, 
in contrast with its partial transpose \cite{Eisler:2015tgq}.
The moments of these Gaussian operators provide the quantities 
$\widetilde{\mathcal{E}}_{n_{\textrm{\tiny e}}}$
and $\widetilde{\mathcal{E}}$,
whose evaluation is discussed in
Appendix\;\ref{app:fermion_numerics}
(see (\ref{eq:Ryu_Log_Moments_Xi})).

The comparison with the corresponding quantities 
$\mathcal{E}_{n_{\textrm{\tiny e}}}$
and $\mathcal{E}$ for the lattice model of the chiral current 
is performed by considering the corresponding differences
$\Delta\mathcal{E}_{ n_{\textrm{\tiny e}}} 
\equiv   
\widetilde{\mathcal{E}}_{n_{\textrm{\tiny e}}}
- \mathcal{E}_{ n_{\textrm{\tiny e}}}$ 
and 
$\Delta\mathcal{E} 
\equiv \widetilde{\mathcal{E}} - \mathcal{E}$,
whose expression in CFT
is obtained from \eqref{Delta-E-ne-loglog}.
When $n_{\textrm{\tiny e}} \geqslant 2$,
in order to get rid of the normalisation constants occurring in the CFT expression, it is convenient to consider 
\begin{equation}
\label{Delta-mathcalE-TB-model}
  \Delta \mathcal{E}_{ n_{\textrm{\tiny e}}} \big|^{\eta}_{1/4} \equiv  
  \Delta \mathcal{E}_{ n_{\textrm{\tiny e}}}
  -  
  \Delta \mathcal{E}_{ n_{\textrm{\tiny e}}}
  \big|_{\eta=1/4}  \,.
\end{equation}

In Fig.\,\ref{fig:diff_neg} we show our results for 
$\Delta\mathcal{E} $ and 
(\ref{Delta-mathcalE-TB-model}) 
specialised to $ n_{\textrm{\tiny e}} = 4$,
in the left and right panel respectively. 
The corresponding CFT predictions (black dashed lines) are
obtained from \eqref{Delta-E-ne-loglog}.
The agreement between the lattice data points in the scaling limit and the CFT formulas
is excellent over the whole domain 
of the cross ratio $\eta \in (0,1)$.


\section{Conclusions}

In this work, we studied the logarithmic negativity 
for the free scalar 
chiral current in two spacetime dimensions, 
gaining new insights into the quantum nature  
of the topological entanglement term.
In particular, considering the asymptotic regime of close intervals,
where $\eta \to 1^{-}$,
we have found that the same topological subleading double-logarithmic contribution that appears in the mutual information for the chiral current in the configuration of two close intervals also emerges in the logarithmic negativity, with exactly the same coefficient
(see (\ref{I-n-loglog-term}) and (\ref{log-neg-2int-expansion-eta-1}) respectively).

In the chiral current model,
our main result is the analytic expression 
(\ref{log-neg-from-AbelPlana}) 
for the logarithmic negativity of two disjoint intervals 
valid for the whole range $\eta \in (0,1)$,
when the whole system is in the ground state,
which has been found by combining the R\'enyi entropies 
obtained through the entanglement spectrum  \cite{Arias:2018tmw} 
with the replica limit for the logarithmic negativity
\cite{Calabrese:2012ew, Calabrese:2012nk}.
In the case of the mutual information, 
this approach provides a nontrivial connection between the contributions to the entanglement coming from the modular Hamiltonian modes in the integral expression (\ref{Dn-integral-ACHP}) and those arising from the replica copies in the discrete sum (\ref{Dn-def-2F1}). 
This becomes crucial in the calculation of the logarithmic negativity discussed in Sec.\,\ref{sec:neg-cc-2int}, where, after performing the partial transposition, it is necessary to return to an integral representation to perform the analytic continuation that leads to the main result (\ref{log-neg-from-AbelPlana}). 
As already mentioned above, while a topological contribution appears in the limit of close intervals, when the intervals become exactly adjacent
and therefore $V_1 \cup V_2$ becomes the whole system,
the identification between the logarithmic negativity 
and the Rényi entropy of index $1/2$ holds exactly to all orders. 
This configuration exhibits no additional topological term, as expected, 
in agreement with the absence of Haag duality violations in single-interval setups. Furthermore, our analysis of the negativity moments is consistent with general CFT predictions 
(see (\ref{moments-rhoA-T2-current-explicit}) and Sec.\,\ref{subsec:two-intervals-small-distance}).

Remarkably, in the opposite regime of large separation distance between the two intervals, the logarithmic negativity exhibits 
a characteristic exponential decay,
in contrast with the mutual information, 
which follows a power law decay instead.
This exponential decay has been already observed in the literature,
but only through numerical computations in different free lattice models \cite{Marcovitch:2008sxc, Wichterich:2008vfx, Calabrese:2012nk, Calabrese:2013mi,Klco:2021cxq,Klco:2020rga, Parez:2022ind, Parez:2023xpj}.
In the CFT model given by the chiral current, 
we have derived the exponential decay analytically
(see (\ref{eq:EN_decay_exp})), showing that it
originates from the presence of branch cuts.
The coefficient governing the decay is determined 
numerically by studying the position of the first branch cut 
of the logarithm of the hypergeometric function appearing in the negativity expression,
as discussed in the last part of Sec.\,\ref{subsec:two-intervals-large-distance}.

From an algebraic perspective in line with previous studies on entanglement in theories with symmetries \cite{Casini:2019kex,Casini:2020rgj} and considering  that the chiral Dirac fermion provides the only well-defined completion of the chiral current, 
we observe that the natural quantity to consider is the negativity difference, rather than the negativity itself.  
Remarkably, in the limit where the bipartite subsystem becomes the whole system in its ground state, 
we find that $\Delta I =\Delta \mathcal{E}$
(see (\ref{eq:DeltaE_intro})).
We conjecture that this behaviour is universal for theories with symmetries and, in turn, with Haag duality violation. 
In Sec.\;\ref{thermofield}, this claim is further supported by a general example based on TFD states in the high-temperature regime, a model originally designed to probe differences between mutual information and negativity when comparing the full and neutral algebras in a simpler setting. 
Since the negativity captures only quantum entanglement, 
the appearance of the extra contribution that gives rise to $\Delta \mathcal{E}$ demonstrates the genuinely quantum nature of the topological term, which is expected to appear in any theory exhibiting a violation of Haag duality, as in the case of the chiral current. 
Essentially, the topological contribution arises from the loss of quantum correlations when reducing the full theory to the neutral model. In turn, the classical correlations,
which are also present in the mutual information, 
are exactly cancelled in the mutual information difference.

A crucial support to our analytic results is provided 
by the numerical computations 
in the lattice model for the chiral current 
already employed in \cite{Arias:2018tmw} for the mutual information, 
which are discussed in Sec.\,\ref{renyisvscorrelators}.
An excellent agreement has been found in the continuum limit, 
as shown in Fig.\,\ref{fig:EnegAlmostAdj_line}
and Fig.\,\ref{fig:EnegAlmostAdj_circle} 
for the logarithmic negativity, 
and in Fig.\,\ref{fig:LineRn} and Fig.\,\ref{fig:PeriodicRn}
for the moments of the partial transpose. 
We find it worth highlighting also the remarkable agreement 
with the lattice data points 
observed in Fig.\,\ref{fig:Substract_Vdjacent_line}
and Fig.\,\ref{fig:Substract_Vdjacent_PBC}
for adjacent intervals.

Both for two adjacent and disjoint intervals
(see (\ref{moments-adj-odd})-(\ref{log-neg-adj-temp})
and (\ref{moments-rhoA-T2-current-explicit})-(\ref{neg-replica-limit-current-2intervals}) respectively),
the analytic expressions that we have obtained for the logarithmic negativity and the moments of the partial transpose
are compatible with the approach of \cite{Calabrese:2012ew,Calabrese:2012nk}, based on the identification between the moments of the reduced density matrix and correlation functions of local branch-point twist fields supported at the endpoints of the two intervals.
On the other hand, in the algebraic language, the present model corresponds to an incomplete theory with Haag duality violation. According to \cite{Benedetti:2024dku}, only complete theories possess well-defined local twist fields. However, since our analysis involves only a single chiral sector, its relation to the framework of \cite{Benedetti:2024dku}, where completeness is tied to the modular invariance of the torus partition function expressed as a correlation function of local twist operators, is not straightforward. This issue deserves further investigation, as a proper notion of modular invariance in two-dimensional CFT requires the presence of  two chiral copies \cite{Rehren:2000ti}.

The analyses discussed in this work can be extended 
in various directions. 
For instance, by studying 
the operator $\rho_V^{\textrm{\tiny $\Gamma_2$}} $ 
in the vacuum
(see e.g. \cite{Murciano:2022vhe,Rottoli:2022plr}),
or other mixed states 
(e.g. the thermal state \cite{Calabrese:2014yza}),
even in out-of-equilibrium scenarios 
\cite{Coser:2014gsa}.

\subsection*{Acknowledgments}

We thank Valentin Benedetti, 
Pasquale Calabrese, Horacio Casini, 
Christopher Herzog, Javier Magan and Shinsei Ryu 
for useful discussions. 
This work was supported by CONICET, CNEA and Universidad Nacional de Cuyo, Instituto Balseiro, Argentina.
E.T. acknowledges the Isaac Newton Institute (Cambridge),
within the program {\it Quantum field theory with boundaries, impurities, and defects}
(supported by EPSRC grant EP/Z000580/1),
and the Yukawa Institute for Theoretical Physics (Kyoto),
within the workshop {\it Extreme Universe 2025} (YITP-T-25-01)
and the long-term workshop {\it Progress of Theoretical Bootstrap},
for hospitality and financial support 
during the last part of this work.
This work was funded by the European Union -- NextGenerationEU, Mission 4, Component 2, Inv.1.3, in the framework of the PNRR Project National Quantum Science and Technology Institute (NQSTI) PE00023; CUP: G93C22001090006.

\appendix

\section{Partial time-reversal in the tight-binding model}
\label{app:fermion_numerics}

In this appendix, we discuss some details about the numerical evaluation of the partial time reversal 
of the reduced density matrix
and the corresponding quantities $\widetilde{\mathcal{E}}_n $ 
for a subsystem $V=V_1\cup V_2$ made by two blocks 
in the lattice model defined by \eqref{eq:tight_binding_model}. 
The forthcoming expressions are crucial 
to understand the corresponding field theory expressions 
$\widetilde{\mathcal{E}}^{\,\textrm{\tiny (W)}}_n$
in the continuum, discussed in Sec.\,\ref{sec-chiral-dirac-fermion}.

It is convenient to introduce the Majorana basis, defined by
\begin{equation}
a_{2m-1} = \hat{c}_m + \hat{c}_m^{\dagger}
\;\;\;\qquad \;\;\;
a_{2m} = \ri\, \big(\hat{c}_m - \hat{c}_m^{\dagger} \big)
\end{equation}
which satisfy $\big\{a_k \,, a_l \big\} = 2\delta_{k,l}$.
We consider the ground state of the system, which is a Gaussian state.
In the Majorana basis, the generic element of the 
covariance matrix $\Gamma$ is defined as
\begin{equation}
\label{cov-mat-fermion}
\Gamma_{i,j} \equiv \frac{1}{2}\,
\big\langle \big[ a_i \,, a_j \big] \big\rangle \,.
\end{equation}

The reduced density matrix $\rho_V$ of a subsystem $V$
is fully determined by the corresponding reduced covariance 
matrix $\Gamma_V$ \cite{peschel_reduced_2009},
which is the $(2N_V) \times (2N_V)$ matrix 
obtained by restricting (\ref{cov-mat-fermion}) 
to $i,j \in V$.
When the subsystem $V = V_1 \cup V_2$ is the union of two blocks 
$V_1$ and $V_2$, made by $N_1$ and $N_2$ consecutive sites respectively,
it is convenient to consider the corresponding 
block decomposition of $\Gamma_V$, namely
\begin{equation}
\label{eq:blockform_GammaA}
\Gamma_V =
\left(
\begin{array}{cc}
\Gamma_{11} \; & \Gamma_{12} 
\\
\Gamma_{21} \;& \Gamma_{22}
\end{array}
\right)
\end{equation}
where the blocks $\Gamma_{IJ}$ with $I,J\in\{1,2\}$ are $(2N_I) \times (2N_J)$ matrices. 
In our case, the conservation of the particle number implies that 
the diagonal blocks $\Gamma_{11}$ and $\Gamma_{22}$ vanish. 
In terms of the matrix
$C_V$  defined in \eqref{eq:fermion_CA},
we have that (\ref{eq:blockform_GammaA}) reads \cite{Shapourian:2019xfi} 
\begin{equation}
\label{eq:GammaV_corr_mat}
\Gamma_V =
\left(
\begin{array}{cc}
\boldsymbol{0}_{N_V} & -\ri(\mathbb{I}_{N_V} - 2C_V) \; 
\\
\ri\big(\mathbb{I}_{N_V} - 2C_V\big) & \boldsymbol{0}_{N_V} 
\end{array}
\right)
\end{equation}
where $\boldsymbol{0}_{N_V}$ is the $N_V\times N_V$ null matrix.
The eigenvalues of $\Gamma_V$ are $\pm \nu_i$, 
with $ 1\leqslant i\leqslant N_V$,
and they are related to the 
eigenvalues  $\zeta_i \in (0,1)$ of $C_V$ 
by the relation $\nu_i = 1 - 2\zeta_i$.

From the blocks of \eqref{eq:blockform_GammaA},
let us construct the following matrices
\cite{Eisler:2015tgq,Shapourian:2016cqu}
\begin{equation}
\label{Gamma-pm-def}
\Gamma_{\pm} \equiv
\left(
\begin{array}{cc}
-\Gamma_{11} & \pm \ri\,\Gamma_{12} 
\\
\pm \ri\,\Gamma_{21} & \Gamma_{22}
\end{array}
\right) \,.
\end{equation}
Consider the operator $O_{\pm}$ whose covariance matrices are given by (\ref{Gamma-pm-def}), 
namely
\begin{equation}
\label{eq:Gammapm}
\big(\Gamma_{\pm}\big)_{i,j} 
= 
\frac{1}{2}\,\mathrm{Tr}\Big(\big[a_i\,, a_j\big] \,O_{\pm}
\Big) \,.
\end{equation}
The operators $O_{\pm}$ 
also admit the following expansions 
in the Majorana basis 
\cite{Eisler:2015tgq}
\begin{equation}
\label{eq:GaussianOperators_majorana}
O_{\pm} = 
\sum_{\tau, \sigma} c_{\tau, \sigma}
\Bigg(\,
\prod_{j\in V_1} a_j^{\tau_j}
\Bigg)
\Bigg(\,
\prod_{j\in V_2} \!\! \big(\! \pm \!\ri \,a_j \big)^{\sigma_j}
\Bigg)
\;\;\;\qquad\;\;\;
\tau_j\, , \sigma_j \in \big\{0\,,1\big\}
\end{equation}
which have been employed in \cite{Coser:2015eba}
to make contact with explicit field theory expressions in the continuum limit
(see \eqref{eq:moments_Ryu_Dirac}
and (\ref{eq:CFT_equation})). 
Some properties of the coefficients 
$c_{\tau, \sigma}$ have been discussed in 
\cite{Eisler:2015tgq}.
The Gaussian operators $O_{\pm}$ are related as follows
\begin{equation}
\label{O-pm-relation-app}
O_+ = \mathsf{S}\; O_-\, \mathsf{S}
\end{equation}
where the string operator $\mathsf{S}$ acts only on 
the block $V_2$ and is defined by \cite{Fagotti:2010yr,Coser:2015eba}
\begin{equation}
\label{S-matrix-def-app}
\mathsf{S} \equiv 
\, \ri^{N_2} \!\! \prod_{j\in V_2} \!\! a_j \,.
\end{equation}

In \cite{Eisler:2015tgq} 
it has been found that 
\begin{equation}
\label{eq:PT_Eisler}
\rho_V^{\textrm{\tiny $\Gamma_2$}} 
= \frac{1 - \ri}{2} \; O_+ 
+ 
\frac{1 + \ri}{2} \; O_-
\end{equation}
which implies that  
$\tr (\rho_V^{\textrm{\tiny $\Gamma_2$}})^{n}$ are traces of polynomials 
built from the Gaussian operators $O_{\pm}$. 
The property given by 
(\ref{O-pm-relation-app}) and (\ref{S-matrix-def-app})
ensures that any trace of a product of $O_{\pm}$ operators is invariant under the exchange $O_+ \leftrightarrow O_-\,$.

The operators introduced in (\ref{eq:Gammapm}) and (\ref{eq:GaussianOperators_majorana}) 
are employed to define the partial-time reversal on the lattice \cite{Shapourian:2016cqu}.
In particular, one first constructs 
the following density operator
\begin{equation}
\Upsilon \equiv 
\frac{O_+ \, O_-}{\tr\!\big(O_+ \,O_- \big)}
\end{equation}
whose normalization factor $\mathcal{Z}_{\Upsilon}$ satisfies \cite{Shapourian:2018lsz}
\begin{equation}
\label{Z-cal-Upsilon-def}
\mathcal{Z}_{\Upsilon} \equiv 
\tr\!\big(O_+ \,O_- \big)
= \tr \rho_V^2 \,.
\end{equation}
The $N_V\times N_V$ correlation matrix associated with the composite operator $\Upsilon$ can be written as follows
\begin{equation}
\label{eq:fermion_CXi}
C_{\Upsilon} = \frac{1}{2}
\left[\,
\mathbb{I} - 
\big( \,\mathbb{I} + \Gamma_{+} \Gamma_{-}\big)^{-1}
\big(\, \Gamma_{+} + \Gamma_{-}\big)
\,\right]
\end{equation}
where $\Gamma_{\pm}$ are the covariance matrices defined in \eqref{Gamma-pm-def}. 
The eigenvalues of $C_{\Upsilon}$ are denoted by  
$\xi_i$ with $1\leqslant i\leqslant N_V$ in the following.

The scalar quantities \eqref{eq:even_moments_Ryu_Dirac} 
can be expressed through the moments of the unnormalized operator $\Upsilon$ as follows
\begin{equation}
\label{eq:Ryu_Log_Moments_Xi}
\widetilde{\mathcal{E}}_{n_{\textrm{\tiny e}}}  
= 
\log \!\big[\,\mathcal{Z}_{\Upsilon}^{n_{\textrm{\tiny e}} / 2} 
\tr\!\big(\Upsilon^{n_{\textrm{\tiny e}} / 2}\big)\,\big]
= 
\log \! \big[ \tr\!\big(\Upsilon^{n_{\textrm{\tiny e}} / 2}\big) \,\big] 
+ \frac{n_{\textrm{\tiny e}}}{2} \,\log\! \big( \tr \rho_V^2 \big)
\end{equation}
where the last expression has been obtained by using (\ref{Z-cal-Upsilon-def}) and it can be written in terms of the eigenvalues 
$\xi_j$ and $\zeta_j$ of the reduced correlation matrices \eqref{eq:fermion_CXi} and \eqref{eq:fermion_CA} respectively 
(associated with $\Upsilon$ and $\rho_V$ respectively)
as follows 
\begin{equation}
\label{eq:moments_Dirac_eigenspectra}
\widetilde{\mathcal{E}}_{n_{\textrm{\tiny e}}}
= 
\sum_{j=1}^{N_V} \log \!\big[\,\xi_j^{n_{\textrm{\tiny e}}/ 2} + (1-\xi_j)^{n_{\textrm{\tiny e}} / 2}\,\big]
+ 
\frac{n_{\textrm{\tiny e}}}{2} 
\sum_{j=1}^{N_V} \log \!\big[\,\zeta_j^2 + (1-\zeta_j)^2\,\big]\,.
\end{equation}

The continuum limit of the lattice results 
obtained through this expression is given by \eqref{eq:Ryu_CFT_LogNegMoments} for the Dirac field,
while the corresponding results for the 
complex Weyl fermion are simply given by (\ref{eq:moments_Dirac_eigenspectra}) multiplied by $1/2$ for the lattice and by \eqref{eq:Ryu_CFT_LogNegMoments_Weyl} in the continuum limit. 
Combining the latter results with 
\eqref{eq:logarithmic_moments_symplectic_spectrum} and \eqref{eq:negativity_symplectic_spectrum},
we find the lattice data points for 
\eqref{Delta-mathcalE-TB-model} 
(also for $n_{\textrm{\tiny e}} \rightarrow 1$),
which must be compared with the corresponding expressions 
coming from $\Delta\mathcal{E}_{ n_{\textrm{\tiny e}}} $ and $\Delta \mathcal{E} $ in the continuum limit, 
given by \eqref{Delta-E-ne-loglog} and its  $n_{\textrm{\tiny e}}\rightarrow 1$ limit respectively. 
The outcomes of this analysis are reported 
in Fig.\,\ref{fig:diff_neg}, where we show some results for 
$\Delta\mathcal{E} $ and 
(\ref{Delta-mathcalE-TB-model}), discussed in 
Sec.\,\ref{sec:fermionic_negativity_difference}.

\section{Small separation distance regime for
$\Delta\mathcal{E}_{ n_{\textrm{\tiny e}}}^{\textrm{($\Gamma_2$)}}$}
\label{app:negativity-difference-Eisler}

In this appendix, we argue the validity of the asymptotic behaviour \eqref{eq:DeltaEprime_asymptotics}
for (\ref{eq:DeltaEprime_def}).

The argument relies on the bounds for the ground state of the massless Dirac field 
conjectured in \cite{Herzog:2016ohd}.
Since the logarithm is strictly increasing, 
for even integers $n_{\textrm{\tiny e}} \geqslant 2$ they read
\begin{equation}
\label{eq:Upperlowerbound}
\log \!\Big[ \tr\!
\big(O_{+}O_{-}\big)^{ n_{\textrm{\tiny e}}/2}
\,\Big]
\leqslant
\,\log \!\Big[ \tr\!
\big(\rho^{\textrm{\tiny $\Gamma_2$}}_V \big)^{ n_{\textrm{\tiny e}}}
\Big] 
\leqslant
\,\log\!\Big[
\left(1 -b_{ n_{\textrm{\tiny e}}} \right)
\tr O_{+}^{ n_{\textrm{\tiny e}}}
+
b_{ n_{\textrm{\tiny e}}}
\textrm{Tr}
\big(O_{+}O_{-}\big)^{ n_{\textrm{\tiny e}}/2}
\,\Big]
\end{equation}
where 
\begin{equation}
b_{ n_{\textrm{\tiny e}}}
\equiv 
\frac{1}{2^{n_{\textrm{\tiny e}}/2}}\,
\bigg(
\begin{array}{c}
n_{\textrm{\tiny e}}
\\
n_{\textrm{\tiny e}}/2
\end{array}
\bigg)
\end{equation}
and the traces occurring in the 
leftmost and rightmost expressions in 
(\ref{eq:Upperlowerbound}) 
are evaluated through 
(\ref{eq:CFT_equation})
and (\ref{eq:evenTr_string}). Notice that,
since $b_{2} = 1$,
the 
leftmost and rightmost expressions in 
(\ref{eq:Upperlowerbound}) coincide 
for $ n_{\textrm{\tiny e}} = 2$. 
By using 
\eqref{eq:even_moments_Ryu_Dirac}
and (\ref{eq:evenTr_string})
the inequalities in (\ref{eq:Upperlowerbound})
can be written in the following equivalent form
\begin{equation}
\label{app-aymp-eta1-PT}
   \widetilde{\mathcal{E}}_{ n_{\textrm{\tiny e}}}
   \leqslant\,
   \log \!\Big[ \tr\!
   \big(\rho^{\textrm{\tiny $\Gamma_2$}}_V
   \big)^{ n_{\textrm{\tiny e}}}
   \Big]
   \leqslant  \,
   \widetilde{\mathcal{E}}_{ n_{\textrm{\tiny e}}}
   +
   \log b_{ n_{\textrm{\tiny e}}}
   +
   \log \!\left[\,
   1+
   \frac{1-b_{ n_{\textrm{\tiny e}}}
   }{
   b_{ n_{\textrm{\tiny e}}}
   }\,
   (1-\eta)^{ n_{\textrm{\tiny e}}/4}\,
   \right]  \,.
\end{equation}
In the asymptotic regime defined by $\eta\rightarrow 1^{-}$, 
the leading divergence of both 
the functions in the leftmost side 
and the rightmost side of this 
sequence of two inequalities
is provided by 
$\widetilde{\mathcal{E}}_{ n_{\textrm{\tiny e}}}$.
Hence, 
combining (\ref{eq:Ryu_CFT_LogNegMoments})
and (\ref{app-aymp-eta1-PT}),
we obtain 
the following asymptotic behaviour  
\begin{equation}
\label{eq:divergent_Dirac_asymptotic}
   \log \!\Big[ \tr\!
   \big(\rho^{\textrm{\tiny $\Gamma_2$}}_V
   \big)^{ n_{\textrm{\tiny e}}}
   \Big]
\sim
-\frac{1}{12}
\left( n_{\textrm{\tiny e}} + \frac{2}{n_{\textrm{\tiny e}}} \right)
\log(1 - \eta)
\;\;\qquad\;\;\;
\eta\rightarrow 1^{-} \,.
\end{equation}
Multiplying this result by $1/2$, 
we obtain the leading divergence of 
the first term in the r.h.s. of 
(\ref{eq:DeltaEprime_def})
and it simplifies 
with the leading divergence
of \eqref{neg-ren-even-cc-extended}.
Then, by using \eqref{eta-to-one-asymptotics}
in the resulting expression, 
the asymptotic behaviour \eqref{eq:DeltaEprime_asymptotics}
is obtained. 

\section{ Pure state algebras  
for blocks in the line and the circle}
\label{app:purestatea_links}

In this appendix, we describe the algebra assignments in pure global states, on the circle and on the line,  through the generalized real-time approach of 
Sec.\,\ref{sec:realtimeapproach}.
The occurrence of zero modes in the lattice model makes the evaluation of the entanglement quantifiers dependent on the choice of the center \cite{Casini:2014aia}. In this section, we work with pure algebras without center.

Consider the canonical position and momentum operators satisfying \eqref{CCR-relations} on a circle made by $N$ sites and construct
\begin{equation}
\label{eq:operatoralgebra_circle_app}
\hat{\mathsf{q}}_i=\hat{q}_{i+1}-\hat{q}_i 
\;\;\;\qquad\;\;\;
\hat{\mathsf{p}}_i=\dot{\hat{q}}_i  
\;\;\;\qquad\;\;\;
1\leqslant i\leqslant N
\end{equation}
where the periodic boundary condition leads us to impose 
that $\hat{\mathsf{q}}_N\equiv q_{1}-q_{N}$.
The $\hat{\mathsf{p}}_i$ is the site operator associated to the $i$-th site, 
while $\hat{\mathsf{q}}_i$ is the link operator associated 
to the link connecting the $i$-th and the $(i+1)$-th sites. 
These operators satisfy the  non-canonical commutation relations of the type \eqref{eq:corr_noncan} given by 
\begin{equation}
\label{eq:commutator_appendix_noncan}
\big[\,
\hat{\mathsf{q}}_i\, ,\hat{\mathsf{p}}_j
\, \big]
=
\ri\big(\delta_{i,j+1}-\delta_{i,j}+\delta_{i,N}\delta_{j,1}\big)
\end{equation}
which is equivalent to the one introduced in \eqref{eq:chiralcurrent_M_T}.

For the ground state of the system $V$ 
and the bipartition given by 
$V_1$ and $V_2$, 
made by $N_1$ and $N_2$ consecutive sites respectively, 
we choose the algebras given by  
\begin{equation}
\label{eq:V1-V2-algebra_circle}
\{(\hat{\mathsf{q}}_1,\hat{\mathsf{p}}_1),....,(\hat{\mathsf{q}}_{N_1},\hat{\mathsf{p}}_{N_1})\} 
\;\;\;\qquad\;\;\;
\{(\hat{\mathsf{q}}_{N_1+1},\hat{\mathsf{p}}_{N_1+2}),....,(\hat{\mathsf{q}}_{N-1},\hat{\mathsf{p}}_{N})\} 
\end{equation} 
on $V_1$ and $V_2$ respectively. 
Hence, the link operator 
$\hat{\mathsf{q}}_{N}$ and site operator $\hat{\mathsf{p}}_{N_1+1}$ are not included 
in the algebra provided by 
the second set of operators in 
\eqref{eq:V1-V2-algebra_circle}.

\begin{figure}[t!]
    \centering
    \subfigure{%
        \includegraphics[width=0.28\textwidth]{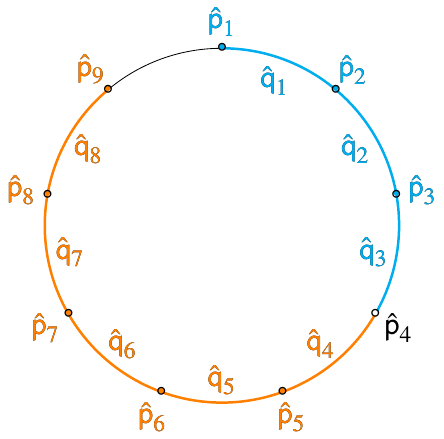}
        \label{fig:algebra_circle}
    }
    \hfill
    \subfigure{%
        \includegraphics[width=0.64\textwidth]{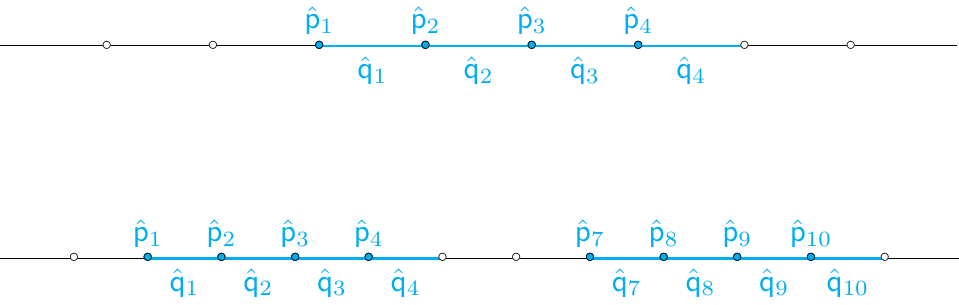}
        \label{fig:algebra_line}
    }
    \caption{
    Choices of algebras for a bipartition
    of the circle (left panel)
    and for two bipartitions of the line 
    (right panels) corresponding to either a single block
    (top right panel) or a union of two disjoint blocks (bottom right panel).}
    \label{fig:purestate_algebra}
\end{figure}

    Considering the three algebras 
 associated to the regions $V_1$, $V_2$ and $V$,
 the number of field operators must be equal to the number of momentum operators within each of them, 
 in order to avoid the existence of a center \cite{Casini:2014aia}.
 This is guaranteed by the removal of both $\hat{\mathsf{q}}_{N}$ and $\hat{\mathsf{p}}_{N_1+1}$.
Thus, for the circle made by $N$ sites, 
the commutant of the algebra associated with a block 
made by $N_1$ consecutive sites is a block made by 
$N-N_1-1$ consecutive sites.
An explicit case is illustrated in the left panel of Fig.\,\ref{fig:purestate_algebra}, for $N=9$ and $N_1=3$.
By using \eqref{CVr-def} and \eqref{EE-from-ss}, 
we also checked numerically, 
that $S(V_1)=S(V_2)$ for various values of $N$ and $N_1$, as expected for a pure global state.

Similar considerations can be applied  for 
the null line algebra discussed in Sec.\,\ref{sec:staggered_boson_review}, 
for a pure global state on a circle made by an even number $N_0$ of sites. 

 When the whole system is in a pure state, complementary subsystems must have the same entanglement entropy. For this algebra choice,
the size of the commutant is modified by the structure of the null line commutation relations 
\eqref{eq:commutator-chiral}.
This implies that, 
for a block made by $2k$ consecutive sites, 
the commutant is a block made by $N_0 - 2k - 2$ sites,
as discussed in \cite{Arias:2018tmw}. 
The additional subtraction of two sites 
corresponds to the occurrence of two sites in the complementary domain adjacent to the block 
that fail to commute with it. 
This counting is fully consistent with the analysis discussed above upon identifying $N = N_0/2$ and $N_1 = k$, which again yields $N - N_1 - 1$ operators in the algebra associated with the complementary domain.

\begin{figure}[t!]
\begin{center}
\includegraphics[width=0.65\textwidth]{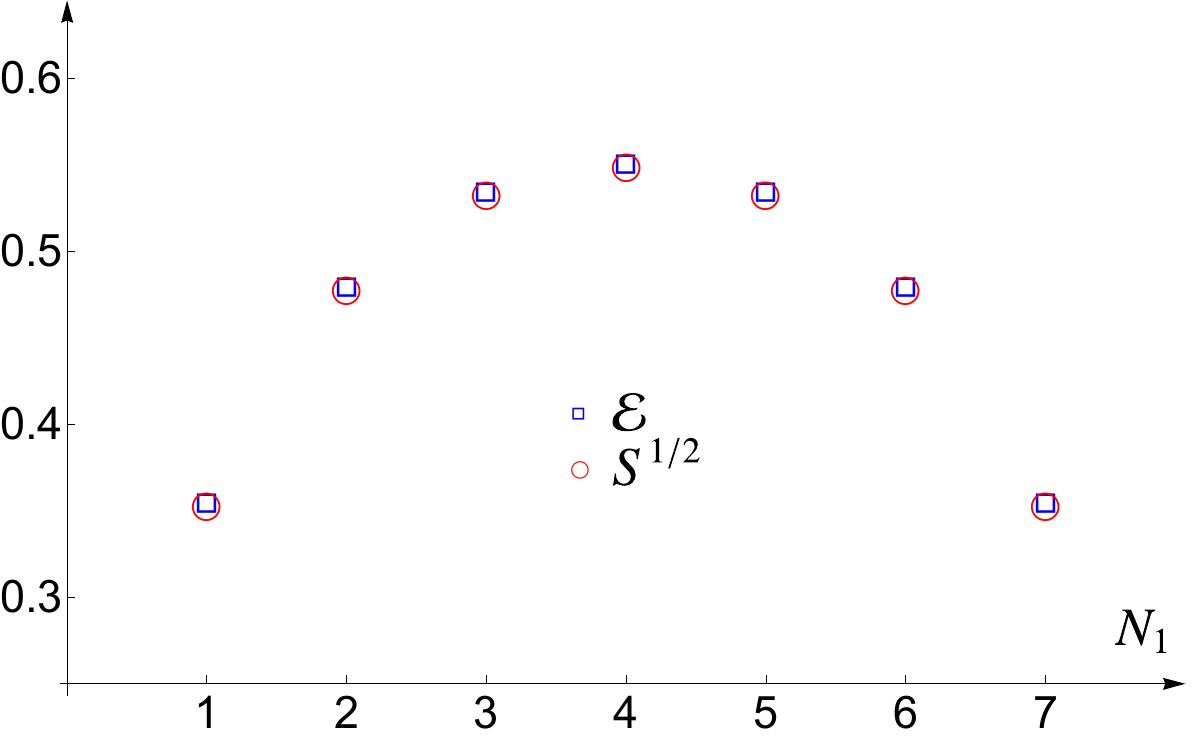}
\caption{Check of the identity ${\cal E}=S^{(1/2)}_{V_1}$,
for a block made by $N_1$ sites
on the circle with $N=9$ 
(see the left panel of 
Fig.\,\ref{fig:purestate_algebra}).
}
\label{renyineg}
\end{center}
\end{figure}

The algebra  \eqref{eq:V1-V2-algebra_circle} 
can be employed to verify the well-known identity 
${\cal E}=S^{(1/2)}_{V_1}$,
found in \cite{Vidal:2002zz}. 
Indeed, the commutation relation \eqref{eq:commutator_appendix_noncan} 
combined with the exclusion of $\hat{\mathsf{q}}_{N}$ 
and $\hat{\mathsf{p}}_{N_1+1}$ discussed above
lead to an invertible matrix $T_V$ in 
\eqref{eq:Ctilde_matrix} 
and therefore the logarithmic negativity \eqref{eq:negativity_symplectic_spectrum}
for the whole system $V$ is well-defined. 
The numerical results for ${\cal E}$
can be compared with the ones for $S^{(1/2)}_{V_1}$,
obtained from \eqref{CVr-def} and \eqref{renyi-from-ss}.
In Fig.\,\ref{renyineg} we report an explicit example of this comparison, showing that the numerical values of these two quantities coincide.
In a similar way, we have verified numerically the validity of the identities 
(\ref{pure-states-moments-relation}) 
for pure states 
\cite{Calabrese:2012ew, Calabrese:2012nk}.

The case of the line corresponds to the limit
$N\rightarrow \infty$ 
of \eqref{eq:operatoralgebra_circle_app}. 
In this setting, one first introduces the operators
$\hat{\mathsf{q}}_i=\hat{q}_{i+1}-\hat{q}_i$
and $\hat{\mathsf{p}}_i=\dot{\hat{\mathsf{q}}}_i$,
with $i\in\mathbb{Z}$.
Then, when the whole system on the line is in a pure state, the assignment of the local algebra for the block $V$ made by $\widetilde{N}_V$ consecutive sites is given by
$\{(\hat{\mathsf{q}}_1,\hat{\mathsf{p}}_1)\, , \dots ,(\hat{\mathsf{q}}_{N_V},\hat{\mathsf{p}}_{N_V})\} $,
where $N_V = \widetilde{N}_V - 1$,
as shown for $N_V = 4$
 in the top panel of the right part of Fig.\,\ref{fig:purestate_algebra}.
Similarly, we can consider the local algebra of two disjoint blocks of sizes $N_1$ and $N_2$ separated by a distance $D$ in the line, 
where the cross ratio \eqref{eta-def-lattice} can be evaluated. 
A possible choice of algebra reads
\begin{equation} 
\big\{
(\hat{\mathsf{q}}_1,\hat{\mathsf{p}}_1), 
\dots ,(\hat{\mathsf{q}}_{N_1},\hat{\mathsf{p}}_{N_1})
\big\}
\,\cup\,
\big\{
(\hat{\mathsf{q}}_{N_1+D+1},\hat{\mathsf{p}}_{N_1+D+1}),
\dots ,(\hat{\mathsf{q}}_{N_1+N_2+D},\hat{\mathsf{p}}_{N_1+N_2+D})
\big\} \,.
\end{equation}
In the lower panel of the right part of Fig.\,\ref{fig:purestate_algebra}, we have illustrated
the case given by $N_1=N_2=4$ and $D=2$, 
where $\eta=4/9$.

\newpage
\bibliography{refs}

@article{Agon:2015ftl,
    author = "Ag{\'o}n, Cesar and Faulkner, Thomas",
    title = "{Quantum Corrections to Holographic Mutual Information}",
    eprint = "1511.07462",
    archivePrefix = "arXiv",
    primaryClass = "hep-th",
    doi = "10.1007/JHEP08(2016)118",
    journal = "JHEP",
    volume = "08",
    pages = "118",
    year = "2016"
}

@article{Huerta:2022cqw,
    author = "Huerta, Marina and van der Velde, Guido",
    title = "{Instability of universal terms in the entanglement entropy}",
    eprint = "2204.09464",
    archivePrefix = "arXiv",
    primaryClass = "hep-th",
    doi = "10.1103/PhysRevD.105.125021",
    journal = "Phys. Rev. D",
    volume = "105",
    number = "12",
    pages = "125021",
    year = "2022"
}

@article{Shiozaki:2017ive,
    author = "Shiozaki, Ken and Shapourian, Hassan and Gomi, Kiyonori and Ryu, Shinsei",
    title = "{Many-body topological invariants for fermionic short-range entangled topological phases protected by antiunitary symmetries}",
    eprint = "1710.01886",
    archivePrefix = "arXiv",
    primaryClass = "cond-mat.str-el",
    doi = "10.1103/PhysRevB.98.035151",
    journal = "Phys. Rev. B",
    volume = "98",
    number = "3",
    pages = "035151",
    year = "2018"
}

@article{Shapourian:2016kvr,
    author = "Shapourian, Hassan and Shiozaki, Ken and Ryu, Shinsei",
    title = "{Many-Body Topological Invariants for Fermionic Symmetry-Protected Topological Phases}",
    eprint = "1607.03896",
    archivePrefix = "arXiv",
    primaryClass = "cond-mat.str-el",
    doi = "10.1103/PhysRevLett.118.216402",
    journal = "Phys. Rev. Lett.",
    volume = "118",
    number = "21",
    pages = "216402",
    year = "2017"
}

@article{Klco:2020rga,
    author = "Klco, Natalie and Savage, Martin J.",
    title = "{Geometric quantum information structure in quantum fields and their lattice simulation}",
    eprint = "2008.03647",
    archivePrefix = "arXiv",
    primaryClass = "quant-ph",
    reportNumber = "INT-PUB-20-031",
    doi = "10.1103/PhysRevD.103.065007",
    journal = "Phys. Rev. D",
    volume = "103",
    number = "6",
    pages = "065007",
    year = "2021"
}

@article{Weedbrook:2011wxo,
    author = "Weedbrook, Christian and Pirandola, Stefano and Garc{\'\i}a-Patr{\'o}n, Ra{\'u}l and Cerf, Nicolas J. and Ralph, Timothy C. and Shapiro, Jeffrey H. and Lloyd, Seth",
    title = "{Gaussian quantum information}",
    eprint = "1110.3234",
    archivePrefix = "arXiv",
    primaryClass = "quant-ph",
    doi = "10.1103/RevModPhys.84.621",
    journal = "Rev. Mod. Phys.",
    volume = "84",
    number = "2",
    pages = "621",
    year = "2012"
}

@book{FarkasZemel:2011Thomae,
  author        = "Farkas, H. M. and Zemel, S.",
  title         = "Generalizations of Thomae's Formula for ZN Curves",
  publisher     = "Springer",
  series        = "Developments in Mathematics",
  volume        = "21",
  pages         = "1--354",
  year          = "2011",
  isbn          = "978-1-4419-7846-2",
  isbnElectronic= "978-1-4419-7847-9",
  doi           = "10.1007/978-1-4419-7847-9"
}

@article{Nakayashiki:1997Thomae,
  author        = "Nakayashiki, Atsushi",
  title         = "On the Thomae Formula for ZN Curves",
eprint = "alg-geom/9608016",
archivePrefix = "arXiv",
primaryClass = "alg-geom",
  journal       = "Publ. Res. Inst. Math. Sci.",
  volume        = "33",
  number        = "6",
  pages         = "987--1015",
  year          = "1997",
  doi           = "10.2977/PRIMS/1195144885"
}

@article{Abate:2025ywp,
    author = "Abate, Nicol{\'a}s and Casini, Horacio and Huerta, Marina and Martinek, Leandro",
    title = "{Exact Mutual Information Difference: Scalar vs. Maxwell Fields}",
    eprint = "2511.04742",
    archivePrefix = "arXiv",
    primaryClass = "hep-th",
    month = "11",
    year = "2025"
}

@article{Parez:2022ind,
    author = "Parez, Gilles and Berthiere, Cl{\'e}ment and Witczak-Krempa, William",
    title = "{Separability and entanglement of resonating valence-bond states}",
    eprint = "2212.11740",
    archivePrefix = "arXiv",
    primaryClass = "cond-mat.str-el",
    doi = "10.21468/SciPostPhys.15.2.066",
    journal = "SciPost Phys.",
    volume = "15",
    number = "2",
    pages = "066",
    year = "2023"
}

@article{Botero:2004vpl,
    author = "Botero, Alonso and Reznik, Benni",
    title = "{Spatial structures and localization of vacuum entanglement in the linear harmonic chain}",
    eprint = "quant-ph/0403233",
    archivePrefix = "arXiv",
    doi = "10.1103/PhysRevA.70.052329",
    journal = "Phys. Rev. A",
    volume = "70",
    number = "5",
    pages = "052329",
    year = "2004"
}

@article{Simon:1999lfr,
    author = "Simon, R.",
    title = "{Peres-Horodecki Separability Criterion for Continuous Variable Systems}",
    eprint = "quant-ph/9909044",
    archivePrefix = "arXiv",
    doi = "10.1103/PhysRevLett.84.2726",
    journal = "Phys. Rev. Lett.",
    volume = "84",
    pages = "2726--2729",
    year = "2000"
}

@article{Parez:2023xpj,
    author = "Parez, Gilles and Witczak-Krempa, William",
    title = "{Entanglement negativity between separated regions in quantum critical systems}",
    eprint = "2310.15273",
    archivePrefix = "arXiv",
    primaryClass = "cond-mat.str-el",
    doi = "10.1103/PhysRevResearch.6.023125",
    journal = "Phys. Rev. Res.",
    volume = "6",
    number = "2",
    pages = "023125",
    year = "2024"
}

@book{WhittakerWatson:ModernAnalysis,
  author       = "E. T. Whittaker and G. N. Watson",
  title        = "A Course of Modern Analysis",
  publisher    = "Cambridge University Press",
  year         = "1920",
  address      = "Cambridge, England",
  pages        = "200",
  note         = "See p. 200",
}

@article{Saharian:2007ph,
    author = "Saharian, Aram A.",
    title = "{The Generalized Abel-Plana formula with applications to Bessel functions and Casimir effect}",
    eprint = "0708.1187",
    archivePrefix = "arXiv",
    primaryClass = "hep-th",
    month = "8",
    year = "2007"
}

@article{Kaplan:1992bt,
    author = "Kaplan, David B.",
    title = "{A Method for simulating chiral fermions on the lattice}",
    eprint = "hep-lat/9206013",
    archivePrefix = "arXiv",
    reportNumber = "UCSD-PTH-92-16",
    doi = "10.1016/0370-2693(92)91112-M",
    journal = "Phys. Lett. B",
    volume = "288",
    pages = "342--347",
    year = "1992"
}

@article{Ginsparg:1981bj,
    author = "Ginsparg, Paul H. and Wilson, Kenneth G.",
    title = "{A Remnant of Chiral Symmetry on the Lattice}",
    reportNumber = "CLNS-81-520, HUTP-81-A060",
    doi = "10.1103/PhysRevD.25.2649",
    journal = "Phys. Rev. D",
    volume = "25",
    pages = "2649",
    year = "1982"
}

@article{Nielsen:1981hk,
    author = "Nielsen, Holger Bech and Ninomiya, M.",
    title = "{No Go Theorem for Regularizing Chiral Fermions}",
    reportNumber = "RL-81-052",
    doi = "10.1016/0370-2693(81)91026-1",
    journal = "Phys. Lett. B",
    volume = "105",
    pages = "219--223",
    year = "1981"
}

@article{Gentile:2025koe,
    author = "Gentile, Francesco and Rotaru, Andrei and Tonni, Erik",
    title = "{Entanglement Hamiltonian of two disjoint blocks in the harmonic chain}",
    eprint = "2503.19644",
    archivePrefix = "arXiv",
    primaryClass = "cond-mat.stat-mech",
    doi = "10.1088/1742-5468/addaa7",
    journal = "J. Stat. Mech.",
    volume = "2025",
    number = "7",
    pages = "073102",
    year = "2025"
}

@article{Seiberg:2023cdc,
    author = "Seiberg, Nathan and Shao, Shu-Heng",
    title = "{Majorana chain and Ising model - (non-invertible) translations, anomalies, and emanant symmetries}",
    eprint = "2307.02534",
    archivePrefix = "arXiv",
    primaryClass = "cond-mat.str-el",
    reportNumber = "YITP-SB-2023-14",
    doi = "10.21468/SciPostPhys.16.3.064",
    journal = "SciPost Phys.",
    volume = "16",
    number = "3",
    pages = "064",
    year = "2024"
}

@article{Shapourian:2019xfi,
    author = "Shapourian, Hassan and Ruggiero, Paola and Ryu, Shinsei and Calabrese, Pasquale",
    title = "{Twisted and untwisted negativity spectrum of free fermions}",
    eprint = "1906.04211",
    archivePrefix = "arXiv",
    primaryClass = "cond-mat.stat-mech",
    doi = "10.21468/SciPostPhys.7.3.037",
    journal = "SciPost Phys.",
    volume = "7",
    number = "3",
    pages = "037",
    year = "2019"
}

@article{Herzog:2016ohd,
    author = "Herzog, Christopher P. and Wang, Yihong",
    title = "{Estimation for Entanglement Negativity of Free Fermions}",
    eprint = "1601.00678",
    archivePrefix = "arXiv",
    primaryClass = "hep-th",
    reportNumber = "YITP-SB-15-17",
    doi = "10.1088/1742-5468/2016/07/073102",
    journal = "J. Stat. Mech.",
    volume = "1607",
    number = "7",
    pages = "073102",
    year = "2016"
}

@article{Shapourian:2018lsz,
    author = "Shapourian, Hassan and Ryu, Shinsei",
    title = "{Finite-temperature entanglement negativity of free fermions}",
    eprint = "1807.09808",
    archivePrefix = "arXiv",
    primaryClass = "cond-mat.stat-mech",
    doi = "10.1088/1742-5468/ab11e0",
    journal = "J. Stat. Mech.",
    volume = "1904",
    pages = "043106",
    year = "2019"
}

@article{DeNobili:2016nmj,
    author = "De Nobili, Cristiano and Coser, Andrea and Tonni, Erik",
    title = "{Entanglement negativity in a two dimensional harmonic lattice: Area law and corner contributions}",
    eprint = "1604.02609",
    archivePrefix = "arXiv",
    primaryClass = "cond-mat.stat-mech",
    doi = "10.1088/1742-5468/2016/08/083102",
    journal = "J. Stat. Mech.",
    volume = "1608",
    number = "8",
    pages = "083102",
    year = "2016"
}

@article{Eisler:2016Teo,
    author = "Eisler, Viktor and Zimborás, Zoltán",
    title = "{Entanglement negativity in two-dimensional free lattice models}",
    eprint = "10.1103/PhysRevB.93.115148",
    doi = "10.1103/PhysRevB.93.115148",
    journal = "Phys. Rev. B",
    volume = "93",
    number = "11",
    pages = "115148",
    year = "2016"
}

@article{Rottoli:2022plr,
    author = "Rottoli, Federico and Murciano, Sara and Tonni, Erik and Calabrese, Pasquale",
    title = "{Entanglement and negativity Hamiltonians for the massless Dirac field on the half line}",
    eprint = "2210.12109",
    archivePrefix = "arXiv",
    primaryClass = "cond-mat.stat-mech",
    doi = "10.1088/1742-5468/acb262",
    journal = "J. Stat. Mech.",
    volume = "2301",
    pages = "013103",
    year = "2023"
}

@article{Murciano:2022vhe,
    author = "Murciano, Sara and Vitale, Vittorio and Dalmonte, Marcello and Calabrese, Pasquale",
    title = "{Negativity Hamiltonian: An Operator Characterization of Mixed-State Entanglement}",
    eprint = "2201.03989",
    archivePrefix = "arXiv",
    primaryClass = "cond-mat.stat-mech",
    doi = "10.1103/PhysRevLett.128.140502",
    journal = "Phys. Rev. Lett.",
    volume = "128",
    number = "14",
    pages = "140502",
    year = "2022"
}

@article{Shapourian:2016cqu,
    author = "Shapourian, Hassan and Shiozaki, Ken and Ryu, Shinsei",
    title = "{Partial time-reversal transformation and entanglement negativity in fermionic systems}",
    eprint = "1611.07536",
    archivePrefix = "arXiv",
    primaryClass = "cond-mat.str-el",
    doi = "10.1103/PhysRevB.95.165101",
    journal = "Phys. Rev. B",
    volume = "95",
    number = "16",
    pages = "165101",
    year = "2017"
}

@article{Coser:2015dvp,
    author = "Coser, Andrea and Tonni, Erik and Calabrese, Pasquale",
    title = "{Spin structures and entanglement of two disjoint intervals in conformal field theories}",
    eprint = "1511.08328",
    archivePrefix = "arXiv",
    primaryClass = "cond-mat.stat-mech",
    doi = "10.1088/1742-5468/2016/05/053109",
    journal = "J. Stat. Mech.",
    volume = "1605",
    number = "5",
    pages = "053109",
    year = "2016"
}

@article{Coser:2015mta,
    author = "Coser, Andrea and Tonni, Erik and Calabrese, Pasquale",
    title = "{Partial transpose of two disjoint blocks in XY spin chains}",
    eprint = "1503.09114",
    archivePrefix = "arXiv",
    primaryClass = "cond-mat.stat-mech",
    doi = "10.1088/1742-5468/2015/08/P08005",
    journal = "J. Stat. Mech.",
    volume = "1508",
    number = "8",
    pages = "P08005",
    year = "2015"
}

@article{Coser:2015eba,
    author = "Coser, Andrea and Tonni, Erik and Calabrese, Pasquale",
    title = "{Towards the entanglement negativity of two disjoint intervals for a one dimensional free fermion}",
    eprint = "1508.00811",
    archivePrefix = "arXiv",
    primaryClass = "cond-mat.stat-mech",
    doi = "10.1088/1742-5468/2016/03/033116",
    journal = "J. Stat. Mech.",
    volume = "1603",
    number = "3",
    pages = "033116",
    year = "2016"
}

@article{Eisler:2015tgq,
    author = "Eisler, Viktor and Zimbor{\'a}s, Zolt{\'a}n",
    title = "{On the partial transpose of fermionic Gaussian states}",
    eprint = "1502.01369",
    archivePrefix = "arXiv",
    primaryClass = "cond-mat.stat-mech",
    doi = "10.1088/1367-2630/17/5/053048",
    journal = "New J. Phys.",
    volume = "17",
    number = "5",
    pages = "053048",
    year = "2015"
}

@article{Calabrese:2013mi,
    author = "Calabrese, Pasquale and Tagliacozzo, Luca and Tonni, Erik",
    title = "{Entanglement negativity in the critical Ising chain}",
    eprint = "1302.1113",
    archivePrefix = "arXiv",
    primaryClass = "cond-mat.stat-mech",
    doi = "10.1088/1742-5468/2013/05/P05002",
    journal = "J. Stat. Mech.",
    volume = "1305",
    pages = "P05002",
    year = "2013"
}

@article{Alba:2013mg,
    author = "Alba, Vincenzo",
    title = "{Entanglement negativity and conformal field theory: a Monte Carlo study}",
    eprint = "1302.1110",
    archivePrefix = "arXiv",
    primaryClass = "cond-mat.stat-mech",
    doi = "10.1088/1742-5468/2013/05/P05013",
    journal = "J. Stat. Mech.",
    volume = "1305",
    pages = "P05013",
    year = "2013"
}

@article{Coser:2014gsa,
    author = "Coser, Andrea and Tonni, Erik and Calabrese, Pasquale",
    title = "{Entanglement negativity after a global quantum quench}",
    eprint = "1410.0900",
    archivePrefix = "arXiv",
    primaryClass = "cond-mat.stat-mech",
    doi = "10.1088/1742-5468/2014/12/P12017",
    journal = "J. Stat. Mech.",
    volume = "1412",
    number = "12",
    pages = "P12017",
    year = "2014"
}

@article{DeNobili:2015dla,
    author = "De Nobili, Cristiano and Coser, Andrea and Tonni, Erik",
    title = "{Entanglement entropy and negativity of disjoint intervals in CFT: Some numerical extrapolations}",
    eprint = "1501.04311",
    archivePrefix = "arXiv",
    primaryClass = "cond-mat.stat-mech",
    doi = "10.1088/1742-5468/2015/06/P06021",
    journal = "J. Stat. Mech.",
    volume = "1506",
    number = "6",
    pages = "P06021",
    year = "2015"
}

@article{Calabrese:2014yza,
    author = "Calabrese, Pasquale and Cardy, John and Tonni, Erik",
    title = "{Finite temperature entanglement negativity in conformal field theory}",
    eprint = "1408.3043",
    archivePrefix = "arXiv",
    primaryClass = "cond-mat.stat-mech",
    doi = "10.1088/1751-8113/48/1/015006",
    journal = "J. Phys. A",
    volume = "48",
    number = "1",
    pages = "015006",
    year = "2015"
}

@article{Lee:2000,
 author = "Lee, Jinhyoung and Kim, M. S. and Park, Y. J. and Lee, S.",
    title = "{Partial teleportation of entanglement in a noisy environment}",
    eprint = "quant-ph/0003060",
    archivePrefix = "arXiv",
    doi = "10.1080/09500340008235138",
    journal = "J. Mod. Opt.",
    volume = "47",
    number = "12",
    pages = "2151--2164",
    year = "2000"
}

@article{Plenio:2005cwa,
    author = "Plenio, M. B.",
    title = "{Logarithmic Negativity: A Full Entanglement Monotone That is not Convex}",
    eprint = "quant-ph/0505071",
    archivePrefix = "arXiv",
    doi = "10.1103/PhysRevLett.95.090503",
    journal = "Phys. Rev. Lett.",
    volume = "95",
    pages = "090503",
    year = "2005"
}

@mastersthesis{Eisert:2006kue,
    author = "Eisert, J.",
    title = "{Entanglement in quantum information theory}",
    eprint = "quant-ph/0610253",
    archivePrefix = "arXiv",
    type = "Other thesis",
    month = "10",
    year = "2006"
}

@article{Vidal:2002zz,
    author = "Vidal, G. and Werner, R. F.",
    title = "{Computable measure of entanglement}",
    eprint = "quant-ph/0102117",
    archivePrefix = "arXiv",
    doi = "10.1103/PhysRevA.65.032314",
    journal = "Phys. Rev. A",
    volume = "65",
    pages = "032314",
    year = "2002"
}

@article{Zyczkowski:1998yd,
    author = "Zyczkowski, Karol and Horodecki, Pawel and Sanpera, Anna and Lewenstein, Maciej",
    title = "{On the volume of the set of mixed entangled states}",
    eprint = "quant-ph/9804024",
    archivePrefix = "arXiv",
    doi = "10.1103/PhysRevA.58.883",
    journal = "Phys. Rev. A",
    volume = "58",
    pages = "883",
    year = "1998"
}

@article{Peres:1996dw,
    author = "Peres, Asher",
    title = "{Separability criterion for density matrices}",
    eprint = "quant-ph/9604005",
    archivePrefix = "arXiv",
    doi = "10.1103/PhysRevLett.77.1413",
    journal = "Phys. Rev. Lett.",
    volume = "77",
    pages = "1413--1415",
    year = "1996"
}

@article{Berenstein:2023ric,
    author = "Berenstein, David and Lloyd, P. N. Thomas",
    title = "{One dimensional staggered bosons, clock models, and their noninvertible symmetries}",
    eprint = "2311.00057",
    archivePrefix = "arXiv",
    primaryClass = "hep-th",
    doi = "10.1103/PhysRevD.110.054508",
    journal = "Phys. Rev. D",
    volume = "110",
    number = "5",
    pages = "054508",
    year = "2024"
}

@article{Berenstein:2023tru,
    author = "Berenstein, David",
    title = "{Staggered bosons}",
    eprint = "2303.12837",
    archivePrefix = "arXiv",
    primaryClass = "hep-th",
    doi = "10.1103/PhysRevD.108.074509",
    journal = "Phys. Rev. D",
    volume = "108",
    number = "7",
    pages = "074509",
    year = "2023"
}

@article{Cardy:2013nua,
    author = "Cardy, John",
    title = "{Some results on the mutual information of disjoint regions in higher dimensions}",
    eprint = "1304.7985",
    archivePrefix = "arXiv",
    primaryClass = "hep-th",
    doi = "10.1088/1751-8113/46/28/285402",
    journal = "J. Phys. A",
    volume = "46",
    pages = "285402",
    year = "2013"
}

@article{Chen:2016mya,
    author = "Chen, Bin and Long, Jiang",
    title = "{R{\'e}nyi mutual information for a free scalar field in even dimensions}",
    eprint = "1612.00114",
    archivePrefix = "arXiv",
    primaryClass = "hep-th",
    doi = "10.1103/PhysRevD.96.045006",
    journal = "Phys. Rev. D",
    volume = "96",
    number = "4",
    pages = "045006",
    year = "2017"
}

@article{Agon:2024xvs,
    author = {Agon, Cesar A. and Casini, Horacio and G{\"u}rsoy, Umut and Planella Planas, Guim},
    title = "{Mutual information from modular flow in CFTs}",
    eprint = "2409.01406",
    archivePrefix = "arXiv",
    primaryClass = "hep-th",
    doi = "10.1007/JHEP08(2025)176",
    journal = "JHEP",
    volume = "08",
    pages = "176",
    year = "2025"
}

@article{Benedetti:2024dku,
    author = "Benedetti, Valentin and Casini, Horacio and Kawahigashi, Yasuyuki and Longo, Roberto and Magan, Javier M.",
    title = "{Modular invariance as completeness}",
    eprint = "2408.04011",
    archivePrefix = "arXiv",
    primaryClass = "hep-th",
    doi = "10.1103/PhysRevD.110.125004",
    journal = "Phys. Rev. D",
    volume = "110",
    number = "12",
    pages = "125004",
    year = "2024"
}

@article{Benedetti:2023owa,
    author = "Benedetti, Valentin and Casini, Horacio and Magan, Javier M.",
    title = "{ABJ anomaly as a U(1) symmetry and Noether\textquoteright{}s theorem}",
    eprint = "2309.03264",
    archivePrefix = "arXiv",
    primaryClass = "hep-th",
    doi = "10.21468/SciPostPhys.18.2.041",
    journal = "SciPost Phys.",
    volume = "18",
    number = "2",
    pages = "041",
    year = "2025"
}

@article{Benedetti:2022zbb,
    author = "Benedetti, Valentin and Casini, Horacio and Magan, Javier M.",
    title = "{Generalized symmetries and Noether\textquoteright{}s theorem in QFT}",
    eprint = "2205.03412",
    archivePrefix = "arXiv",
    primaryClass = "hep-th",
    doi = "10.1007/JHEP08(2022)304",
    journal = "JHEP",
    volume = "08",
    pages = "304",
    year = "2022"
}

@article{Arias:2018tmw,
    author = "Arias, Ra\'ul E. and Casini, Horacio and Huerta, Marina and Pontello, Diego",
    title = "{Entropy and modular Hamiltonian for a free chiral scalar in two intervals}",
    eprint = "1809.00026",
    archivePrefix = "arXiv",
    primaryClass = "hep-th",
    doi = "10.1103/PhysRevD.98.125008",
    journal = "Phys. Rev. D",
    volume = "98",
    number = "12",
    pages = "125008",
    year = "2018"
}

@article{Casini:2013rba,
    author = "Casini, Horacio and Huerta, Marina and Rosabal, Jose Alejandro",
    title = "{Remarks on entanglement entropy for gauge fields}",
    eprint = "1312.1183",
    archivePrefix = "arXiv",
    primaryClass = "hep-th",
    doi = "10.1103/PhysRevD.89.085012",
    journal = "Phys. Rev. D",
    volume = "89",
    number = "8",
    pages = "085012",
    year = "2014"
}

@article{Casini:2019kex,
    author = "Casini, Horacio and Huerta, Marina and Mag\'an, Javier M. and Pontello, Diego",
    title = "{Entanglement entropy and superselection sectors. Part I. Global symmetries}",
    eprint = "1905.10487",
    archivePrefix = "arXiv",
    primaryClass = "hep-th",
    doi = "10.1007/JHEP02(2020)014",
    journal = "JHEP",
    volume = "02",
    pages = "014",
    year = "2020"
}

@article{Casini:2020rgj,
    author = "Casini, Horacio and Huerta, Marina and Magan, Javier M. and Pontello, Diego",
    title = "{Entropic order parameters for the phases of QFT}",
    eprint = "2008.11748",
    archivePrefix = "arXiv",
    primaryClass = "hep-th",
    doi = "10.1007/JHEP04(2021)277",
    journal = "JHEP",
    volume = "04",
    pages = "277",
    year = "2021"
}

@article{Sorkin:2012sn,
    author = "Sorkin, Rafael D.",
    title = "{Expressing entropy globally in terms of (4D) field-correlations}",
    eprint = "1205.2953",
    archivePrefix = "arXiv",
    primaryClass = "hep-th",
    doi = "10.1088/1742-6596/484/1/012004",
    journal = "J. Phys. Conf. Ser.",
    volume = "484",
    pages = "012004",
    year = "2014"
}

@article{Casini:2009sr,
    author = "Casini, H. and Huerta, M.",
    title = "{Entanglement entropy in free quantum field theory}",
    eprint = "0905.2562",
    archivePrefix = "arXiv",
    primaryClass = "hep-th",
    doi = "10.1088/1751-8113/42/50/504007",
    journal = "J. Phys. A",
    volume = "42",
    pages = "504007",
    year = "2009"
}

@article{Casini:2014aia,
    author = "Casini, Horacio and Huerta, Marina",
    title = "{Entanglement entropy for a Maxwell field: Numerical calculation on a two dimensional lattice}",
    eprint = "1406.2991",
    archivePrefix = "arXiv",
    primaryClass = "hep-th",
    doi = "10.1103/PhysRevD.90.105013",
    journal = "Phys. Rev. D",
    volume = "90",
    number = "10",
    pages = "105013",
    year = "2014"
}

@article{Rehren:2000ti,
    author = "Rehren, Karl-Henning",
    editor = "Longo, R.",
    title = "{Locality and modular invariance in 2-D conformal QFT}",
    eprint = "math-ph/0009004",
    archivePrefix = "arXiv",
    journal = "Fields Inst. Commun.",
    volume = "30",
    pages = "341--354",
    year = "2001"
}

@article{Fagotti:2010yr,
    author = "Fagotti, Maurizio and Calabrese, Pasquale",
    title = "{Entanglement entropy of two disjoint blocks in XY chains}",
    eprint = "1003.1110",
    archivePrefix = "arXiv",
    primaryClass = "cond-mat.stat-mech",
    doi = "10.1088/1742-5468/2010/04/P04016",
    journal = "J. Stat. Mech.",
    volume = "1004",
    pages = "P04016",
    year = "2010"
}

@article{Agon:2013iva,
    author = "Agon, Cesar A. and Headrick, Matthew and Jafferis, Daniel L. and Kasko, Skyler",
    title = "{Disk entanglement entropy for a Maxwell field}",
    eprint = "1310.4886",
    archivePrefix = "arXiv",
    primaryClass = "hep-th",
    reportNumber = "BRX-TH-672",
    doi = "10.1103/PhysRevD.89.025018",
    journal = "Phys. Rev. D",
    volume = "89",
    number = "2",
    pages = "025018",
    year = "2014"
}

@article{Calabrese:2004eu,
    author = "Calabrese, Pasquale and Cardy, John L.",
    title = "{Entanglement entropy and quantum field theory}",
    eprint = "hep-th/0405152",
    archivePrefix = "arXiv",
    doi = "10.1088/1742-5468/2004/06/P06002",
    journal = "J. Stat. Mech.",
    volume = "0406",
    pages = "P06002",
    year = "2004"
}

@article{Calabrese:2012nk,
    author = "Calabrese, Pasquale and Cardy, John and Tonni, Erik",
    title = "{Entanglement negativity in extended systems: A field theoretical approach}",
    eprint = "1210.5359",
    archivePrefix = "arXiv",
    primaryClass = "cond-mat.stat-mech",
    doi = "10.1088/1742-5468/2013/02/P02008",
    journal = "J. Stat. Mech.",
    volume = "1302",
    pages = "P02008",
    year = "2013"
}

@article{Calabrese:2012ew,
    author = "Calabrese, Pasquale and Cardy, John and Tonni, Erik",
    title = "{Entanglement negativity in quantum field theory}",
    eprint = "1206.3092",
    archivePrefix = "arXiv",
    primaryClass = "cond-mat.stat-mech",
    doi = "10.1103/PhysRevLett.109.130502",
    journal = "Phys. Rev. Lett.",
    volume = "109",
    pages = "130502",
    year = "2012"
}

@article{Calabrese:2009ez,
    author = "Calabrese, Pasquale and Cardy, John and Tonni, Erik",
    title = "{Entanglement entropy of two disjoint intervals in conformal field theory}",
    eprint = "0905.2069",
    archivePrefix = "arXiv",
    primaryClass = "hep-th",
    doi = "10.1088/1742-5468/2009/11/P11001",
    journal = "J. Stat. Mech.",
    volume = "0911",
    pages = "P11001",
    year = "2009"
}

@article{Calabrese:2010he,
    author = "Calabrese, Pasquale and Cardy, John and Tonni, Erik",
    title = "{Entanglement entropy of two disjoint intervals in conformal field theory II}",
    eprint = "1011.5482",
    archivePrefix = "arXiv",
    primaryClass = "hep-th",
    reportNumber = "NSF-KITP-10-152, MIT-CTP 4194",
    doi = "10.1088/1742-5468/2011/01/P01021",
    journal = "J. Stat. Mech.",
    volume = "1101",
    pages = "P01021",
    year = "2011"
}

@article{Audenaert:2002xfl,
    author = "Audenaert, K. and Eisert, J. and Plenio, M. B. and Werner, R. F.",
    title = "{Entanglement Properties of the Harmonic Chain}",
    eprint = "quant-ph/0205025",
    archivePrefix = "arXiv",
    doi = "10.1103/PhysRevA.66.042327",
    journal = "Phys. Rev. A",
    volume = "66",
    pages = "042327",
    year = "2002"
}

@article{Marcovitch:2008sxc,
    author = "Marcovitch, S. and Retzker, A. and Plenio, M. B. and Reznik, B.",
    title = "{Critical and noncritical long-range entanglement in Klein-Gordon fields}",
    eprint = "0811.1288",
    archivePrefix = "arXiv",
    primaryClass = "quant-ph",
    doi = "10.1103/PhysRevA.80.012325",
    journal = "Phys. Rev. A",
    volume = "80",
    pages = "012325",
    year = "2009"
}

@article{Wichterich:2008vfx,
    author = "Wichterich, H. and Molina-Vilaplana, J. and Bose, S.",
    title = "{Scaling of entanglement between separated blocks in spin chains at criticality}",
    eprint = "0811.1285",
    archivePrefix = "arXiv",
    primaryClass = "quant-ph",
    doi = "10.1103/PhysRevA.80.010304",
    journal = "Phys. Rev. A",
    volume = "80",
    number = "1",
    pages = "010304",
    year = "2009"
}

@article{Shapourian:2018ozl,
    author = "Shapourian, Hassan and Ryu, Shinsei",
    title = "{Entanglement negativity of fermions: monotonicity, separability criterion, and classification of few-mode states}",
    eprint = "1804.08637",
    archivePrefix = "arXiv",
    primaryClass = "quant-ph",
    doi = "10.1103/PhysRevA.99.022310",
    journal = "Phys. Rev. A",
    volume = "99",
    number = "2",
    pages = "022310",
    year = "2019"
}

@article{Klco:2021cxq,
    author = "Klco, Natalie and Beck, D. H. and Savage, Martin J.",
    title = "{Entanglement structures in quantum field theories: Negativity cores and bound entanglement in the vacuum}",
    eprint = "2110.10736",
    archivePrefix = "arXiv",
    primaryClass = "quant-ph",
    reportNumber = "IQuS@UW-21-012",
    doi = "10.1103/PhysRevA.107.012415",
    journal = "Phys. Rev. A",
    volume = "107",
    number = "1",
    pages = "012415",
    year = "2023"
}

@article{peschel_reduced_2009,
   author = "Eisler, Viktor and Peschel, Ingo",
    title = "{Reduced density matrices and entanglement entropy in free lattice models}",
    eprint = "0906.1663",
    archivePrefix = "arXiv",
    primaryClass = "cond-mat.stat-mech",
    doi = "10.1088/1751-8113/42/50/504003",
    journal = "J. Phys. A",
    volume = "42",
    number = "50",
    pages = "504003",
    year = "2009"
}

@article{Eisert:1998pz,
    author = "Eisert, Jens and Plenio, Martin B.",
    title = "{A Comparison of entanglement measures}",
    eprint = "quant-ph/9807034",
    archivePrefix = "arXiv",
    reportNumber = "J. Mod. Opt. 46, 145-154 (1999)",
    doi = "10.1080/09500349908231260",
    journal = "J. Mod. Opt.",
    volume = "46",
    pages = "145--154",
    year = "1999"
}

@article{Zyczkowski:1999iw,
    author = "Zyczkowski, Karol",
    title = "{On the volume of the set of mixed entangled states. 2.}",
    eprint = "quant-ph/9902050",
    archivePrefix = "arXiv",
    doi = "10.1103/PhysRevA.60.3496",
    journal = "Phys. Rev. A",
    volume = "60",
    pages = "3496",
    year = "1999"
}

@article{Longo:2009mn,
    author = "Longo, Roberto and Martinetti, Pierre and Rehren, Karl-Henning",
    title = "{Geometric modular action for disjoint intervals and boundary conformal field theory}",
    eprint = "0912.1106",
    archivePrefix = "arXiv",
    primaryClass = "math-ph",
    doi = "10.1142/S0129055X10003977",
    journal = "Rev. Math. Phys.",
    volume = "22",
    pages = "331--354",
    year = "2010"
}

@article{Longo:2017mbg,
    author = "Longo, Roberto and Xu, Feng",
    title = "{Relative Entropy in CFT}",
    eprint = "1712.07283",
    archivePrefix = "arXiv",
    primaryClass = "math.OA",
    doi = "10.1016/j.aim.2018.08.015",
    journal = "Adv. Math.",
    volume = "337",
    pages = "139--170",
    year = "2018"
}

\bibliographystyle{nb}

\end{document}